\documentclass[preprint2]{aastex}

\slugcomment{To appear in 2006, ApJ, 642}

\shorttitle{Obscured quasars in the SWIRE/Chandra Survey}
\shortauthors{Polletta et al.}

\def\rp{$r^\prime$}
\def\gp{$g^\prime$}
\def\rp{$r^\prime$}
\def\ip{$i^\prime$}
\def\msun{M$_{\odot}$}
\def\lsun{L$_{\odot}$}
\def\deg{$^{\circ}$}

\def\hr{$^{h}$}
\def\min{$^{m}$}

\def\chandra {{\it Chandra}}

\def\spitzer {{\it Spitzer}}
\def\nh {${\rm N_\mathrm{H}}$}

\def\kms{\ifmmode {\rm\,km\,s^{-1}}\else
    ${\rm\,km\,s^{-1}}$\fi}
\def\kmsMpc{\ifmmode {\rm\,km\,s^{-1}\,Mpc^{-1}}\else
    ${\rm\,km\,s^{-1}\,Mpc^{-1}}$\fi}
\def\ergAcm2{\ifmmode {\rm\,ergs\,cm^{-2}\,{\rm \AA}^{-1}}\else
    ${\rm\,ergs\,cm^{-2}\,\AA^{-1}}$\fi}
\def\cm2{\ifmmode {\rm\,cm^{-2}}\else
    ${\rm\,cm^{-2}}$\fi}
\def\ergcm2s{\ifmmode {\rm\,ergs\,cm^{-2}\,s^{-1}}\else
    ${\rm\,ergs\,cm^{-2}\,s^{-1}}$\fi}
\def\cgsdeg2{\ifmmode {\rm\,ergs\,cm^{-2}\,s^{-1}\,deg^{-2}}\else
    ${\rm\,ergs\,cm^{-2}\,s^{-1}\,deg^{-2}}$\fi}
\def\sqdeg{\ifmmode {\rm\,deg^{2}}\else
    ${\rm\,deg^{2}}$\fi}
\def\ergsHz{\ifmmode {\rm\,ergs\,s^{-1}\,Hz^{-1}}\else
    ${\rm\,ergs\,s^{-1}\,Hz^{-1}}$\fi}
\def\ergs{\ifmmode {\rm\,ergs\,s^{-1}}\else
    ${\rm\,ergs\,s^{-1}}$\fi}
\def\ergsA{\ifmmode {\rm\,ergs\,s^{-1}\,\AA^{-1}}\else
    ${\rm\,ergs\,s^{-1}\,\AA^{-1}}$\fi}
\def\WHz{\ifmmode {\rm\,W\,Hz^{-1}}\else
    ${\rm\,W\,Hz^{-1}}$\fi}
\def\WHzsr{\ifmmode {\rm\,W\,Hz^{-1}\,sr^{-1}}\else
    ${\rm\,W\,Hz^{-1}\,sr^{-1}}$\fi}
\def\ergscm2Hz{\ifmmode {\rm\,ergs\,cm^{-2}\,s^{-1}\,Hz^{-1}}\else
    ${\rm\,ergs\,cm^{-2}\,s^{-1}\,Hz^{-1}}$\fi}
\def\Msun{M$_{\odot}$}

\def\lya{Ly$\alpha$}
\def\nv{\ion{N}{5} $\lambda$1240}

\def\sioiv{\ion{Si}{4}/\ion{O}{4}] $\lambda$1400}
\def\siiv{\ion{Si}{4} $\lambda$1397}
\def\oivp{\ion{O}{4}] $\lambda$1402}

\def\civ{\ion{C}{4} $\lambda$1549}
\def\heii{\ion{He}{2} $\lambda$1640}

\def\nev{[\ion{Ne}{5}] $\lambda$3426}
\def\oii{[\ion{O}{2}] $\lambda$3727}

\def\neiii{[\ion{Ne}{3}] $\lambda$3869}
\def\hbeta{H$\beta$}

\def\oiii{[\ion{O}{3}] $\lambda$5007}

\def\halpha{H$\alpha$}

\begin{document}


\title{Chandra and Spitzer unveil heavily obscured quasars in the SWIRE/Chandra
Survey\altaffilmark{1}}


\author{Maria del Carmen Polletta\altaffilmark{2}, Belinda J. Wilkes\altaffilmark{3},
 Brian Siana\altaffilmark{4,5}, Carol J. Lonsdale\altaffilmark{2,6},
 Roy Kilgard\altaffilmark{3}, Harding E. Smith\altaffilmark{2,5}, 
 Dong-Woo Kim\altaffilmark{3}, Frazer Owen\altaffilmark{5,7}, Andreas
 Efstathiou\altaffilmark{8}, Tom Jarrett\altaffilmark{6}, Gordon Stacey\altaffilmark{9},
 Alberto Franceschini\altaffilmark{10}, Michael Rowan-Robinson\altaffilmark{11},
 Tom S.R. Babbedge\altaffilmark{11}, Stefano Berta\altaffilmark{10}, Fan Fang\altaffilmark{4},
 Duncan Farrah\altaffilmark{9}, Eduardo Gonz\'alez-Solares\altaffilmark{12},
 Glenn Morrison\altaffilmark{5,13}, Jason A. Surace\altaffilmark{4}, Dave L. Shupe\altaffilmark{4}}

\altaffiltext{1}{Some of the data presented herein were obtained at the W.M.
Keck Observatory, which is operated as a scientific partnership among the
California Institute of Technology, the University of California and the
National Aeronautics and Space Administration. The Observatory was made
possible by the generous financial support of the W.M. Keck Foundation.
Based on observations at the Kitt Peak National Observatory, National
Optical Astronomy Observatory, which is operated by the Association of
Universities for Research in Astronomy, Inc. under cooperative agreement
with the National Science Foundation. The National Radio Astronomy
Observatory is a facility of the National Science Foundation operated under
a cooperative agreement by Associated Universities Inc.}
\altaffiltext{2}{Center for Astrophysics \& Space Sciences, University of California, San
      Diego, La Jolla, CA 92093, USA}
\email{mcp@auriga.ucsd.edu}
\altaffiltext{3}{Harvard/Smithsonian Center for Astrophysics, 60 Garden
      Street, Cambridge, MA 02138, USA}   
\altaffiltext{4}{\spitzer\ Science Center, California Institute
      of Technology, 100-22, Pasadena, CA 91125, USA}
\altaffiltext{5}{Visiting astronomer, Kitt Peak National Observatory,
National Optical Astronomy Observatories, operated by AURA Inc, under
cooperative agreement with the National Science Foundation.}
\altaffiltext{6}{Infrared Processing \& Analyis Center, California Institute
      of Technology, 100-22, Pasadena, CA 91125, USA}
\altaffiltext{7}{National Radio Astronomy Observatory$^{\ddagger}$, P.O. Box O, Socorro, 
      NM 87801, USA}
\altaffiltext{8}{School of Computer Science and Engineering, Cyprus College,
     6 Diogenes Street, Engomi, 1516 Nicosia, Cyprus}
\altaffiltext{9}{Department of Astronomy, Cornell University, Ithaca, NY
      14853, USA}
\altaffiltext{10}{Dipartimento di Astronomia, Universit\'a di Padova, Vicolo
      dell'Osservatorio 2, I-35122, Padova, Italy}
\altaffiltext{11}{Astrophysics Group, Blackett Lab., Imperial College,
      Prince Consort Road, London, SW7 2BW, UK}
\altaffiltext{12}{Institute of Astronomy, University of Cambridge, Madingley Road,
     Cambridge, CB3 0HA, UK}
\altaffiltext{13}{Institute for Astronomy, University of Hawaii,
     and Canada-France-Hawaii Telescope, Kamuela, Hawaii 96743-8432, USA}

\begin{abstract}

Using the large multi-wavelength data set in the \chandra/SWIRE Survey (0.6
$deg^2$ in the Lockman Hole), we show evidence for the existence of
highly obscured (Compton-thick) AGN, estimate a lower limit to their surface density and
characterize their multi-wavelength properties. Two independent selection
methods based on the X-ray and infrared spectral properties are
presented. The two selected samples contain 1) 5 X-ray sources with hard
X-ray spectra and column densities $\gtrsim 10^{24}$\cm2, and 2) 120
infrared sources with red and AGN-dominated infrared spectral energy distributions (SEDs).
We estimate a surface density of at least 25 Compton-thick AGN 
deg$^{-2}$ detected in the infrared in the \chandra/SWIRE field of which
$\sim$40\%\ show distinct AGN signatures in their optical/near-infrared
SEDs, the remainings being dominated by the host-galaxy emission. Only
$\sim$33\%\ of all Compton-thick AGN are detected in the X-rays at our
depth (F(0.3-8 keV)$>$10$^{-15}$\ergcm2s). 

We report the discovery of two sources in our sample of Compton-thick AGN,
SWIRE\_J104409.95+585224.8 ($z$=2.54) and SWIRE\_J104406.30+583954.1
($z$=2.43), which are the most luminous Compton-thick AGN at high-$z$
currently known. The properties of these two sources are discussed in detail
with an analysis of their spectra, SEDs, luminosities and black-hole masses.

\end{abstract}


\section{Introduction}

According to the AGN unification models~\citep{antonucci93,krolik99}, all
AGN are intrinsically similar, and the observational differences among
various types are due to the geometry and orientation with respect to the
line of sight of obscuring matter surrounding the central super-massive
black hole (SMBH). According to this model, obscuring matter is ubiquitous
in AGN, but the effects of absorption are only seen when the line of sight
intercepts it. Alternative models attribute the presence of obscuration to a
stage in the early phases of AGN evolution during a merger~\citep{fabian99}.
Mergers of large disk galaxies hosting a SMBH induce the growth of the SMBH
and of the host galaxy spheroid throughout vigorous star-formation episodes.
During this phase the AGN is surrounded by large amounts of gas and dust,
thus it appears as heavily obscured. As the AGN reaches a certain
luminosity, its radiation can cause the expulsion and destruction of the
surrounding material and the system appears as an unobscured
AGN~\citep{silk98,granato04,springel05,hopkins05,dimatteo05,cattaneo05}.

Since AGN that are not affected by obscuration are relatively easy
to detect and identify across the entire wavelength spectrum, they have been
well sampled up to high redshifts and their properties, space density,
luminosity function and redshift distributions are well
measured~\citep{ueda03,richards05}. On the other hand, AGN obscured by
even only moderate column densities (\nh$\simeq$10$^{22}$
\cm2)~\citep{richards03,white03}, are routinely missed in observations at
various wavelengths because of the difficulty of detecting and identifying
them, and, therefore, they are not as well understood. There is evidence
that obscured AGN are numerous and might even outnumber unobscured AGN,
however optically and X-ray selected samples of AGN are still dominated by
unobscured AGN and large and complete samples of obscured AGN are still
missing providing only few observables to constrain models.

\subsection{Indirect evidence for the existence of highly obscured AGN}

In the local Universe and for moderate (Seyfert-like) nuclear luminosities
($<10^{44}$\ergs), the observed ratio between obscured (\nh$>$10$^{22}$\cm2)
and unobscured (\nh$<$10$^{22}$\cm2) AGN is 4:1~\citep{osterbrock88,
madau94, comastri95, risaliti99, piconcelli03} and $\sim$ 45\% of the 
obscured ones are Compton-thick (column density larger than
$10^{24}$ \cm2)~\citep{risaliti99,maiolino95}. Large column densities
($\sim 10^{24}$\cm2) have been directly measured only in a few, $\sim$10, sources (e.g.
NGC1068, Circinus, NGC4945). For most of the known Compton-thick AGN,
$\sim$40~\citep{comastri04}, only indirect evidence to their extreme column
densities is available~\citep{bassani99,risaliti99,maiolino03}, such as the
flatness of the hard X-ray continuum, a large equivalent width (EW) of the
K$_{\alpha}$6.4 keV iron fluorescent emission line, or lower than expected
values of the ratio between the flux in the X-rays and in other wavelengths.
At higher-redshift ($>$1) and luminosities ($\geq 10^{44}$\ergs), the
distribution of absorption in AGN is not as well constrained with only a few
examples of confirmed obscured quasars. There is evidence that the fraction
of obscured AGN decreases with higher intrinsic
luminosities~\citep{ueda03,szokoly04,barger05,treister05}, and possibly
increases with redshift~\citep{lafranca05}.

Other indications for the existence of numerous obscured AGN have been
provided by AGN studies at wavelengths less affected by obscuration, such as
infrared (IR) and radio. \citet{stern05} compared the surface density of an
IR-selected sample of AGN ($F(8\mu m)> 76 \mu Jy$) with that of an
optically-selected (R-mag$<$21)~\citep{wolf03} sample of AGN and found 2.8
times more AGN in the IR than in the optical sample. Since the ratio between
the limiting fluxes of the two samples corresponds to the typical
R-band/8\micron\ flux ratio of unobscured AGN, the higher density of
IR-selected AGN is attributed to an excess of obscured AGN. Similar results
were obtained from an IR and radio selected sample of AGN~\citep{martinez05}
in which the fraction of obscured sources at high-redshift ($z\sim$2) is
estimated to be 50\% (obscured:unobscured=1:1) or as high as 87\%\
(obscured:unobscured=2.6:1) if sources which are not confirmed
spectroscopically are also taken into account. An obscured:unobscured ratio
of 2:1 was also derived from a sample of AGN selected at
24\micron~\citep{alonso05}.

The existence of a large population of obscured AGN is also suggested by the
shape of the X-ray background at high energies. More than 85\%\ of the 2--10
keV Cosmic X-ray Background has been resolved by sources detected in deep
X-ray surveys~\citep{moretti03,deluca04}. However, the resolved fraction
decreases at higher energies, e.g. in the 4-6 keV energy range the resolved
fraction is about 70--90\%, whilst in the 8--12 keV band no more than 50\%\
is resolved. Less than 30\% is resolved above 10 keV, where the bulk of the
CXRB energy density is produced~\citep{worsley04,worsley05}. The spectral
shape of the residual background cannot be produced by a simple
superposition of unobscured AGN spectra, but by an X-ray population with
faint low-energy X-ray fluxes and hard X-ray spectra, as in obscured
AGN~\citep{worsley05}.

\subsection{Searches for obscured AGN}

In spite of the difficulty of finding and identifying highly obscured AGN,
several searches have been conducted combining multiwavelength
data~\citep{webster95,wilkes02,padovani04,donley05,stern05,martinez05,zakamska04,urrutia05},
performing very deep observations~\citep{treister05,vanduyne04} or surveying
large areas of the sky~\citep{cutri02,fiore03,zakamska04,urrutia05}.
Obscured AGN candidates have been selected among X-ray sources with hard
X-ray spectra or with high ($>$10) X-ray over optical flux
ratios~\citep{fiore03,rigby04}, with radio emission in excess compared to
the IR~\citep{donley05,urrutia05} or with narrow emission lines in their
optical spectra~\citep{zakamska04}. These sources are predominantly
characterized by column densities of the order of 10$^{22}$\cm2 and by
various types of optical spectra, with narrow emission lines as expected in
type 2 AGN, but also with broad emission lines, as in unobscured AGN, or
typical of normal, early and late-type, galaxies~\citep{fiore03,perola04}. 

All these studies aimed at finding the obscured AGN population predicted by
models and indirect observations have been successful to varying degrees.
However, most of these sample are affected by selection effects and
characterized by properties too broad to constrain models, and the measured
column densities are only moderate (\nh=10$^{22-23}$\cm2).  A common
property of these moderately obscured AGN is the variety of optical/near-IR
spectral energy distributions (SEDs), with only a minority of sources
showing typical AGN signatures~\citep{franceschini05,donley05,rigby04}.
Consequently, any search in a specific wavelength range will provide
incomplete samples of obscured AGN.

In this work, we aim at identifying and characterizing only the most obscured
AGN, with columns densities of the order of 10$^{24}$ \cm2, the so-called
Compton-thick AGN. By looking at the most obscured AGN which are also the
hardest to find because of their elusive properties, we hope to provide
useful constraints on AGN models. Our analysis is based on the sources
detected by \spitzer\ and \chandra\ in a wide area (0.6 \sqdeg)
multi-wavelength (from radio to X-ray) survey performed in the Lockman Hole,
the \chandra/SWIRE survey. This field was selected for deep follow-up
observations within the \spitzer\ Wide-Area Infrared Extragalactic Survey
(SWIRE)~\citep{lonsdale03,lonsdale04} legacy project. By comparing samples
selected independently in the X-rays and in the IR, the incompleteness level
of each selection method can be estimated. The available multiwavelength
data set is described in Section~\ref{obs}. In Section~\ref{ctagn_sel}, we
present our selection methods for heavily obscured AGN, an X-ray based
selection in Section~\ref{ctagn_x_sec} and an IR based selection in
Section~\ref{ctagn_ir_sec}. The general properties of the two samples, SEDs,
X-ray over optical and mid-IR fluxes are discussed in Sections~\ref{opx_sec}
and~\ref{irx_sec}. A detailed analysis of the properties of two
spectroscopically confirmed Compton-thick quasars present in both samples is
given in sections~\ref{qso2_obs} (data description),~\ref{sed} (SED
analysis), and~\ref{lum_mbh} (bolometric luminosity and black hole mass
estimate). A comparison between our candidates and other samples of heavily
obscured AGN is presented in section~\ref{comp_ctagn}. In
Section~\ref{surf_dens}, we estimate a lower limit to the surface density of
Compton-thick AGN detected in the IR at our sensitivity limits in the
\chandra/SWIRE field and compare our estimates with current models. Our
results are summarized in Section~\ref{conc}.

Throughout the paper, we adopt a flat cosmology with H$_0$ = 71 \kmsMpc,
$\Omega_{M}$=0.27 and $\Omega_{\Lambda}$=0.73~\citep{spergel03}.

\section{Observations} 
\label{obs}

The \chandra/SWIRE field, located in the northern part of the Lockman Hole
Field (10\hr45\min, +59$^{\circ}$), has been selected as the target for the
deepest IR, optical and radio SWIRE observations, and for a moderately deep
and wide \chandra\ survey.  This field has the lowest cirrus sky emission of
all of the SWIRE fields (0.38 MJy/sr at 100\micron) and has no contamination
from moderate brightness radio sources, making it ideally suited for a
radio-IR survey. The neutral Galactic column density toward this field is on
average 6.43$\times 10^{19}$ \cm2~\citep{dickey90}, making it ideal also for
deep X-ray surveys. The data available in this field are summarized in
Table~\ref{obs_log} and details on the observations are given in the
following sections.

\subsection{Optical and Infrared Imaging Observations} 
\label{opt_data}

Optical imaging in $U$, \gp, \rp, and \ip\ was obtained with the Mosaic
Camera at the Kitt Peak National Observatory (KPNO) Mayall 4-mt Telescope on
February 2002 (\gp, \rp, and \ip) and January 2004 ($U$). The coverage of
the field is not uniform, the central 0.3 \sqdeg\ ($\alpha$=161\deg--162\deg\ and
$\delta$=58.75\deg--59.25\deg) were covered with a 3 hour exposure in \gp,
\rp, and \ip, and with a 6 hour exposure in $U$-band to 5$\sigma$
limiting Vega magnitudes of 24.8 ($U$), 25.9 (\gp), 25.2 (\rp), and 24.4
(\ip). The surrounding area at $\delta>$58.6\deg\ was also covered in four
bands with 50 min exposure in \gp, and \rp, 30 min exposure in \ip, and with
a 2 hour exposure in $U$-band to a depth of 24.3 ($U$), 25.2 (\gp), 24.4 (\rp), and
23.5 (\ip). The small region at $\delta<$58.6\deg\ was observed only in the
\gp, \rp, and \ip\ bands with a 30 min exposure to a depth of 23.7 (\gp),
23.5 (\rp), and 22.9 (\ip). The astrometry is good to less than 0.4 arcsec
and the seeing varies between 0.9 and 1.4 arcsec. The data were processed
with the Cambridge Astronomical Survey Unit's reduction pipeline following
the procedures described in~\citet{babbedge04a}. Fluxes were measured within
3\arcsec\ apertures (diameter) and corrected to total fluxes using profiles
measured on bright stars. Total magnitudes, derived by integrating over the
curve of growth, were adopted for sources extended and bright in the three
filters
\gp, \rp, and \ip\ (\gp$<$23.8, \rp$<$22.7, \ip$<$22.0). The optical catalog
contains 77,355 sources (galaxies and stars) of which 45,573 are detected in
at least two bands.

Near-IR $K_s$ imaging observations were carried out with the 200\arcsec\
Hale Telescope of the Palomar Observatory using 
the Wide Infrared Camera (WIRC;~\citet{wilson03}) on
2004 March 29 under photometric conditions. The field was partially
covered, 0.43 \sqdeg, with 24 8\farcm 5$\times$8\farcm 5 pointings of 72 min
exposure. Data reduction for the near-IR imaging consisted of median-sky
removal, flat-fielding using a median ``sky'' image derived from the science
observations, co-addition, and astrometric and flux calibration. The seeing
FWHM ranges between 0.7 and 1.3 arcsec. Sources were extracted using
SExtractor~\citep{bertin96} and MAG\_BEST magnitudes were adopted.
Calibration was carried out using the near-IR 2MASS Point Source
Catalog~\citep{cutri03}. The Ks-band photometric uncertainty, relative to
2MASS, is $\sim$6\% and the 5-sigma sensitivity is 20.5 mag (Vega). The
$K_s$ catalog contains 19,876 sources, of which 17,140 have an optical
counterpart.

Observations with the Infrared Array Camera (IRAC)~\citep{fazio04} were
performed on 2003, December 5 and 2004, April 24--30 and observations with
the Multiband Imaging Photometer (MIPS)~\citep{rieke04} were performed on
2003, December 9 and 2004, May, 4--11. The IRAC depth was 120--480 seconds,
depending on exact field location, with a median depth of 240 seconds and
MIPS depth was 160--360 seconds, with a median depth of 360 seconds. Fluxes
were measured using SExtractor~\citep{bertin96} from mosaics of the flat
fielded images processed by the \spitzer\ Science Center using the S11 data
pipelines. Fluxes were extracted in 5\farcs8 diameter apertures for IRAC
($\sim$2-3$\times$ the FWHM beam) and 12\arcsec\ for MIPS 24\micron\, using
SExtractor~\citep{bertin96} and corrected for aperture to total fluxes using
the IRAC/MIPS point spread functions (PSFs). In the case of extended sources
(SExtractor stellarity index $<$0.8 and ISO\_Area$>$200) in the IRAC images,
Kron fluxes were used. Details of the IRAC and MIPS data processing are
given in~\citet{surace05}. The 5$\sigma$ depths of the
\spitzer\ data are 5, 9, 43, 40 and 230 $\mu$Jy at 3.6, 4.5, 5.8, 8.0 and
24\micron, respectively. The IR catalog contains 41,262 sources, of which
31,106 have an optical counterpart.

\subsection{\chandra\ X-ray Data and Analysis} 
\label{xray_data}

We have obtained \chandra\ Advanced CCD Imaging Spectrometer
(ACIS-I)~\citep{Weisskopf96} observations in a 3$\times$3 raster of a 0.6
$deg^2$ region with center $\alpha$= 10$^h$ 46$^m$ and $\delta$=+59\deg\ 01\arcmin\
and $\sim$2\arcmin\ overlap between contiguous pointings within the Lockman
Hole field of the SWIRE survey. The exposure time for each observation was
$\sim$70 ksecs, reaching broad- (0.3--8 keV), soft- (0.3--2.5 keV) and
hard-band (2.5-8 keV) fluxes of $\sim 10^{-15}$, 5$\times 10^{-16}$, and
10$^{-14}$
\ergcm2s, respectively. The observations were obtained on 2004, September 12-26
and processed using the XPIPE pipeline developed for analysis of
\chandra\ data for the ChaMP project~\citep{kim04}. XPIPE screens bad data,
corrects instrumental effects remaining after the standard pipeline
processing, detects the X-ray sources (using {\sc
WAVDETECT}~\citep{freeman02} in the CIAO 3.2 software
package\footnote{\chandra\ Interactive Analysis of Observations (CIAO),
http://cxc.harvard.edu/ciao/}) and determines counts in the soft
(0.3$-$2.5keV), hard (2.5$-$8.0keV) and broad (0.3$-$8.0keV) bands. A false-positive threshold
of 10$^{-6}$ in {\sc WAVDETECT} is used to accept a source, corresponding to
$\sim 1$ spurious source per ACIS-I chip, or 4 per field. The background and
exposure-corrected count-rates for each source were converted to fluxes
using conversion factors computed using XSPEC and assuming a power-law model
($F(E) \propto E^{-(\Gamma-1)} e^{-\sigma(E)\cdot N_{\mathrm{H}}}$, where
$E$ is the energy, $F(E)$ the flux density, $\sigma (E)$ is the
photo-electric cross-section~\citep{morrison83}, $\Gamma$ is the photon
index, and \nh\ is the column density associated with the absorbing material)
with $\Gamma$=1.7 and \nh=6$\times$10$^{19}$ \cm2. The
analysis of the X-ray data and details on the observations will be presented
in a future publication.

A total of 812 sources were detected in the initial analysis. The X-ray
source list was cross-correlated with the \spitzer\ source list using a
search radius corresponding to the quadratic sum of the \chandra\ positional
uncertainty (2\arcsec\ minimum) and of the IR positional uncertainty that
was fixed to 2\arcsec. Based on the visual inspection of the images and on
the low detection reliability, 20 sources are considered to be spurious.
This is roughly the number expected since 1 source per ACIS chip would lead
to 27 spurious sources. These X-ray sources were not included in the X-ray
catalog for further analysis, reducing the total number of X-ray sources to
792. An IR counterpart is matched to 766 sources, 631 of which are also
detected in the optical images. Sixteen sources are detected only in the
optical and 10 X-ray sources do not have either an IR or optical
counterpart. The majority of the X-ray sources (561 out of 792 sources or
71\%) has a unique optical or IR counterpart within the positional
uncertainty and 213 sources have multiple matches (130 sources have 2
matches, 49 have 3 matches and the remaining 34 have more than 3 matches).
In the case of multiple matches the closest source was chosen as the
counterpart, unless there was another candidate at similar distance that was
a brighter and redder IR source (10 cases).  To estimate the reality of the
associations we calculated the probability of random matches between the
X-ray sources and the possible Spitzer counterparts following the same
procedure discussed in~\citet{fadda02} and in~\citet{franceschini05} which
assumes that the IR population follows a Poisson spatial distribution. Most
of the \chandra\ sources with multiple associations are unambiguously
identified as the X-ray positional uncertainty is small ($\sim$2\arcsec) and
one counterpart is at less than 0\farcs 5 from the X-ray source and the
others are at more 2\arcsec. In eight cases, the positional uncertainties
were so large, i.e. sources at large off-axis angles, to have several
possible matches. The closest source was selected, however the reliability
of these matches is very low. Approximately 88\%
have a probability of random matches $P<$5\%. Summing the probabilities we
expect about 19 false associations in the 774 matched sources. Eighteen of
the 774 matched sources are close to bright sources, mostly stars, and,
therefore, do not have reliable optical and IR photometric measurements.

\subsection{Radio Imaging} 
\label{radio_data}

A deep, 1.4~GHz radio map centered at $\alpha$=10\hr46\min,
$\delta$=+59\deg01\arcmin, and covering 40\arcmin$\times$40\arcmin\ in
the \chandra/SWIRE field, was obtained at the Very Large Array (VLA) during
multiple dates, on 2001, December 15, on 2002, January--March and on 2003,
January 6 (Owen, F. et al., in preparation). VLA configurations A/B/C and D were
used. The total integration time spent on source was 500~ks.  The
root-mean-squared (rms) noise in the center of the radio image is
2.7~$\mu$Jy. The source density decreases at larger distances from the
center of the field, from about 9400 sources deg$^{-2}$ within 10\arcmin\
from the center ($\sim$4.5$\mu$Jy rms), to about 4300 sources deg$^{-2}$
at a distance between 10\arcmin\ and 20\arcmin, and less than 1000 sources
sources deg$^{-2}$ at a distance greater than 20\arcmin ($\sim$16$\mu$Jy
rms). There are 2052 radio sources in the entire field and 2000 are also
detected in the IR with IRAC. The fraction of IR sources that are detected
at radio wavelengths varies from 13\% within 10\arcmin\ from the center of
the radio field to about 7\% in the entire radio field. The fraction of
X-ray sources that are radio detected is almost four times higher than the fraction
of radio detected IR sources, with 50\% of all X-ray sources within
10\arcmin\ from the center and 27\% in the whole field.

\subsection{Optical Spectroscopy}

Spectroscopic observations were carried out using various facilities, with
Hydra on the WIYN Observatory on 2004, February 11--15; with the Low
Resolution Imaging Spectrometer (LIRS)~\citep{oke95} on the Keck~I telescope
on February 24--25, 2004 and March 3--4, 2005; and with the Gemini
Multi-Object Spectrograph (GMOS) on the Gemini Observatory on February
21--23, 2004. Details on these observations will be published in a future
publication (Smith, H. et al., in preparation). Details on the Keck observations
are given in section~\ref{opt_spec}. Spectroscopic redshifts from the Sloan
Digital Sky Survey\footnote{http://www.sdss.org} (SDSS) are also available.
In total, spectroscopic redshifts are available for 574 IR sources of which
74 are also X-ray sources, 48 from Keck, 81 from Gemini, 412 from WIYN, and
35 from SDSS.

The IR, X-ray and radio source lists and the matched multi-wavelength
catalogs with be presented in a future publications (Polletta, M. et al.; Owen,
F. et al., in preparation).

\section{Selection of obscured AGN candidates} 
\label{ctagn_sel}

In the following sections, we present two methods for selecting AGN with
extreme column densities (\nh$\gtrsim 10^{24} cm^{-2}$), the so called
Compton-thick AGN, one based on the X-ray properties and one based on the
optical-IR properties. Since the methods require knowing the redshift, and
spectroscopic redshifts are available for only a small fraction of sources,
we supplement the spectroscopic redshifts with photometric redshifts.
Photometric redshifts were derived using the code HYPERZ~\citep{hyperz}.
HYPERZ measures photometric redshifts by finding the best-fit, defined by
the $\chi^2$-statistic, to the observed SED with a library of galaxy templates.
We use 24 galaxy templates that represent normal galaxies (9), starbursts
(3) and AGN (12) and cover the wavelength range from 1000\AA\ to 500$\mu m$
(Polletta et al., in preparation). The same method and template library has
been used to fit the SEDs of a sample of X-ray detected AGN in the ELAIS-N1
field~\citep{franceschini05}. 

\subsection{X-ray selected Compton-thick AGN} 
\label{ctagn_x_sec}

Compton-thick AGN can show a variety of X-ray spectra, soft and hard,
according to the amount of absorption and reflection
components~\citep{smith96,matt00}. Therefore, soft X-ray spectra do not
necessarily imply low or lack of absorption. However, hard X-ray spectra and
X-ray luminosities greater than 10$^{42}$ \ergs\ can be only explained by
presence of absorption and can, therefore, be used to identify obscured AGN.

In this work, we select sources with hard X-ray spectra and estimated column
densities of the order of 10$^{24}$ \cm2. This method is biased against
Compton-thick AGN in which the primary radiation is completely obscured at
the observed energies ($<$8 keV) and only the warm scattered component is
observed which produces soft X-ray spectra. This selection effect is less
important for high-$z$ sources where the observed radiation is emitted at
higher energies in the source rest-frame, which are less affected by
obscuration, e.g.; for a column density of 10$^{24}$ \cm2\, the observed
flux between 0.3 and 8 keV is reduced by about 98\%\ at $z\sim$0, and by
71\%\ at $z$=2 assumig a photon index $\Gamma$=1.7. Since most of the X-ray
sources in the sample are too faint (75\% of the sample has less than 50
broad-band X-ray counts) to perform full spectral fitting, the amount of
absorption is estimated by comparing the counts in the hard- and soft-X-ray
bands. Previous studies~\citep{dwelly05,mainieri02,dellaceca04,perola04}
have shown that colour based analysis are effective in deriving the
properties of X-ray sources with few counts.

Hardness ratios, HR, defined as (H-S)/(H+S), where H corresponds to the X-ray
counts in the hard band (2.5-8 keV) and S to the X-ray counts in the soft
band (0.3-2.5 keV) were derived from the observed counts for all of the
sources. In order to derive the corresponding absorption,
Sherpa~\citep{freeman01} simulations were performed for each source assuming
an absorbed power-law model. The power-law slope was fixed to a conservative
(in terms of \nh\ estimate) $\Gamma = 1.7$ value, corresponding to the
observed mean for AGN~\citep{nandra94} and the column density varied from
\nh= 10$^{19}$ \cm2\ to 10$^{24.5}$ \cm2. The Bayesian method~\citep{dyk04}
was applied to take into account the differences in effective area across
the detector by estimating the local background for each source. Each
spectrum was used as input to MARX\footnote{http://space.mit.edu/CXC/MARX}
to create a simulated data set. Hardness ratios were calculated from the
simulations of each source and the corresponding \nh\ were tabulated. The
tabulated values were then used to determine the \nh\ by comparison with the
observed HR. In order to derive the effective hydrogen column density,
\nh$^{rest}$, the measured \nh\ was corrected for the redshift of the source
taking into account the energy dependence of the photo-electric cross
section~\citep{morrison83}, \nh$^{rest}$=\nh$^{obs}\times
(1+z)^{2.6}$~\citep{barger02,longair92}. Photometric redshifts were used
when spectroscopic redshifts were not available.

\begin{figure}[!ht]
\plotone{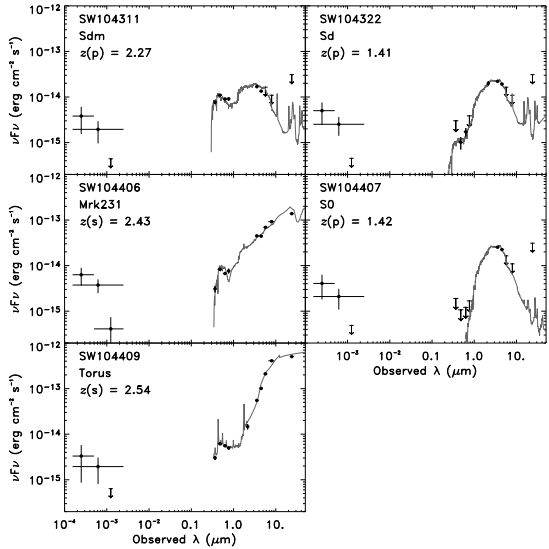}
\caption{\footnotesize SED in $\nu F\nu$ of the 5 X-ray selected Compton-thick AGN (full circles).
Abbreviated source names are reported on the upper-left corner of each
panel. Downward arrows indicate 5$\sigma$ upper limits. The broad-, hard-,
and soft-X-ray fluxes are shown as crosses. The X-ray flux energy range
is indicated by the length of the horizontal line. The X-ray fluxes are
derived assuming an absorbed power-law model with photon index, $\Gamma$
equal to 1.7 and Galactic absorption, \nh=6$\times$10$^{19}$\cm2. A downward
arrow at 2$\sigma$ limit is used in the X-rays when the counts are less than
1$\sigma$. The best-fit template for each object is shown as a grey curve
and the template name is reported in each panel. The spectroscopic ($z$(s))
or photometric ($z$(p)) redshift of each object is also reported.}
\label{ctagn_x_sed}
\end{figure}

We found 10 X-ray sources with an intrinsic \nh$^{rest}\gtrsim 10^{24}$
\cm2\ as derived from the observed HR, assuming spectroscopic redshifts when
available and photometric redshifts for the others. Photometric redshifts
are still preliminary as more spectroscopic redshifts are being
collected and improvements are being implemented. Currently, the total $rms$
for the whole sample of 574 sources with spectroscopic redshifts is 0.26 in
1+$z$, the rate of outliers (defined by
$|z_{phot}-z_{spec}|/(1+z_{spec})\geq 0.2$) is 8\%, and the
$rms$ obtained after removing the outliers is 0.08. Although these values
are consistent with those derived in other samples of galaxies and
AGN~\citep{babbedge04b}, the uncertainties are still quite large. Therefore,
after the selection based on the best photometric redshift, we examined all
of the possible solutions obtained by fitting the observed SED with various
templates and redshifts. After comparing the reduced
$\chi^2_{\nu}$ of all solutions with the $\chi^2_{\nu}$ given by the
best-fit solution, $\chi^2_{\nu}(Best)$, we removed four sources with
secondary solutions with $\chi^2_{\nu}$ $<$ $2\times \chi^2_{\nu}(Best)$ at
a redshift below the value required to have an \nh$^{rest}$=10$^{24}$ \cm2.
Another source was removed because of poor optical and IR photometric data
due to a bright nearby star which did not allow a reliable fit. The final
X-ray selected sample of obscured AGN candidates contains 5 sources. Their
basic properties (coordinates, \rp\-band magnitudes, IR and radio fluxes)
are reported in Table~\ref{ctagn_x}, their X-ray properties (broad-, soft-
and hard-band X-ray fluxes, redshifts, column densities and absorption
corrected broad band luminosities) are listed in Table~\ref{ctagn_x_xprop},
and their SEDs are shown in Figure~\ref{ctagn_x_sed}.

Two sources show SEDs dominated by AGN emission. Spectroscopic data are also
available for these two sources and show narrow emission lines typical of
type 2 AGN. A detailed analysis of the SEDs of these two sources is presented in
Sections~\ref{qso2_obs}, and~\ref{sed}. The SEDs of the other three sources
are dominated by starlight in the optical and near-IR and are best-fitted
with spiral galaxy templates. The estimated redshifts range from 1.4 to 2.5,
the column densities range from 1.0 to 9$\times$10$^{24}$ and the
absorption-corrected X-ray (0.3-8 keV) luminosities from
2.5$\times$10$^{45}$ to 9.5$\times 10^{45}$ \ergs\ implying that these are
all Compton-thick quasars.

\subsection{IR-selected obscured AGN} 
\label{ctagn_ir_sec}

Since Compton-thick AGN might be too faint in the X-rays to be detected
at our sensitivity level, we need a complementary selection method that can
be applied to the entire IR sample independently of their X-ray data.

In obscured AGN, the X-ray radiation is absorbed by gas and dust surrounding
the nuclear region and re-emitted in the mid- and far-IR after being
thermally reprocessed. Since IR radiation is less affected by obscuration
than optical and X-ray emission, AGN selection at these wavelengths is less
biased against obscured AGN. However, AGN usually represent only a small
fraction of all of the sources detected in IR surveys compared to the far
more numerous and strong IR emitters, such as galactic sources, normal and
starburst galaxies. Moreover, their IR spectral properties can be
indistinguishable from those of non AGN sources when only a few IR bands are
available. IR-selected AGN can thus be less reliable than X-ray and optical
selected samples. The {\it Spitzer Space Telescope}~\citep{werner04}, thanks
to the wavelength range accessible to IRAC~\citep{fazio04}, offers the
capability to easily identify a large fraction of the AGN
population~\citep{lacy04,stern05} through a simple color analysis that
efficiently removes the majority of non-AGN sources.

In the mid-IR, AGN are characterized by red and almost featureless
spectra~\citep{houck05,hao05,weedman05}. These properties make their IRAC
(3.6, 4.5, 5.8, and 8.0 $\mu$m) colors unique among other IR sources, such
as galaxies and stars, providing a powerful tool to identify
them~\citep{lacy04,stern05,evanthia05}. However, this diagnostic is
effective only when the AGN is the dominant energy source. In cases where
thermal radiation produced by dust associated with the AGN is self-absorbed
and/or thermal radiation from dust in star-forming regions is more luminous
than that produced by the AGN, e.g. as in the Seyfert 2 galaxies NGC 4945
and NGC 6240~\citep{maiolino03,rigopoulou99}, the AGN IR emission is fainter
or negligible compared to that produced by star-formation processes in the
host-galaxy~\citep{peeters04}. The IR SED and IRAC colors of these AGN may
be indistinguishable from those of starburst and normal star-forming
galaxies (i.e. dominated by cool/warm dust and PAH features at $z<$0.6 and
by stellar light at $z>$0.6)~\citep{franceschini05,alonso04,rigby04}. In
these sources, the AGN might manifest itself at other wavelengths, in the
X-rays, if not completely obscured, as in the 3 stellar-dominated
Compton-thick X-ray AGN reported in Section~\ref{ctagn_x_sec}, in the radio
or in optical and IR spectra, if the light is not diluted by the host galaxy
starlight~\citep{moran02}.

In order to select heavily obscured AGN candidates among the IR population
independently of their X-ray properties, we require a red and featureless IR
SED, and red optical SEDs or red optical-IR colors to remove unobscured AGN.
This requirement would be satisfied by only a subset of all obscured AGN for
the reasons given above.  We first selected all of the extragalactic IR
sources that are detected at a 5$\sigma$ level in at least 3 IR bands over
the wavelength range 3.6-24$\mu$m.  This reduced the IR sample from 41,262
to 4493 sources of which 2726 (60\%) are detected at 24\micron. Note that
about 46\%\ of the extragalactic X-ray sources with an IR counterpart in the
SWIRE/\chandra\ field do not satisfy this selection criterion. The next step
in the selection procedure is done automatically through an algorithm that
calculates the spectral slope and goodness of a power-law model fit to the
observed SED ($\nu F_{\nu}$) between 2.15 and 24$\mu$m, which includes the
$K_s$-band data, the four IRAC bands and the MIPS 24$\mu$m band. All of the
sources with a monotonically rising IR SED, a spectral slope $\alpha_{IR}$
(defined as $F_{\nu}\propto \lambda^{\alpha_{IR}}$) larger than 1.0 and reduced
$\chi^2_{\nu}<13.3(\alpha_{IR}-1)\leq 20$ are then selected. The slope
threshold is defined to reject sources with blue SEDs typical of early-type
galaxies. The $\chi^2$-test is used to select sources with smooth red SEDs,
not necessarily power-law like, and to remove sources with variations in
their SEDs due to the presence of spectral features, e.g. PAHs. AGN may show
IR SEDs with a convex round shape in Log(F$_{\nu}$) instead of a straight
power-law spectrum, or a dip at 24$\mu$m due to the 9.7$\mu$m Silicate
absorption feature at
$z\sim$1.5~\citep{franceschini05,alonso04,rigby04,rigby05,houck05,weedman05,hao05}.
For these AGN a power-law model fit does not provide a good fit, but it is a
reasonable first-order approximation to their IR SEDs. The expected
$\alpha_{IR}$ and $\chi^2_{\nu}$ values for AGN and normal galaxies were
derived from a simulated catalog of sources of different types and redshifts
and several combinations of detections in the same filter systems as
available for the SWIRE catalog. The simulated catalog was created from the
same template library used to derive photometric redshifts and uncertainties
to the observed fluxes were added in order to resemble those of the SWIRE
catalog. The SEDs of a sub-set of randomly selected objects with a broad
range of $\chi^2$ and $\alpha_{IR}$ were also visually inspected for
verification and parameters tuning. Non-detections were taken into account
by requiring the power-law model to be consistent with the upper limits.
This selection reduces the sample from 4493 to 248 sources of which 93\%\
are detected at 24\micron. Although the selection does not require a
detection at 24\micron, the large fraction of 24\micron\ sources is not
surprising due to the combination of the minimum required power-law slope
($\alpha_{IR}$=1) and the SWIRE sensitivity. The faintest source with
$\alpha_{IR}$=1 and detected in 3 IRAC bands will have a 24\micron\ flux of 120$\mu$Jy, only a factor
of 2 lower than our 5$\sigma$ limit. Sources with brighter fluxes and redder
SED will then be easily detected at 24\micron. Only 23\%\ (90
sources) of the X-ray detected sources with 3 IR detections have IR SEDs
that satisfy the above criterion. This is consistent with previous SED
analysis of X-ray selected AGN which find typical AGN SEDs for only
$\sim$30\% of all X-ray selected AGN~\citep{franceschini05}. In order
to remove likely unobscured quasars, we then selected all of the sources
with optical-IR colors redder than those typical of unobscured quasars or
red optical SEDs. These requirements are satisfied if a power-law
($F_{\nu}\propto \lambda^{\alpha_{opt}}$) fit to the optical SED of a source detected in \gp,
\rp, \ip\ has a slope, $\alpha_{opt}$, greater than 2, or if at least two of
the following conditions are verified, $F(3.6\mu m)/F($\gp$)\geq 15$,
$F(3.6\mu m)/F($\rp$)\geq 13$, and $F(3.6\mu m)/F($\ip$)\geq 10$. This final
criterion is satisfied by 181 sources, of which 78 do not have an optical
counterpart at the survey limits (see Section~\ref{opt_data}). Note that for
blank sources fainter than 16$\mu$Jy at 3.6\micron\ and 
located where the optical observations are the least sensitive
($\delta<$58.6\deg) this criterion can not verified, but at this stage of the
selection, they are kept in the sample for completeness. This selection
removes 67 sources, of which 42 are X-ray sources. Among the rejected
sample, 61 sources have optical-IR SEDs consistent with those of unobscured
AGN like optically selected quasars.

\begin{figure}[!ht]
\plotone{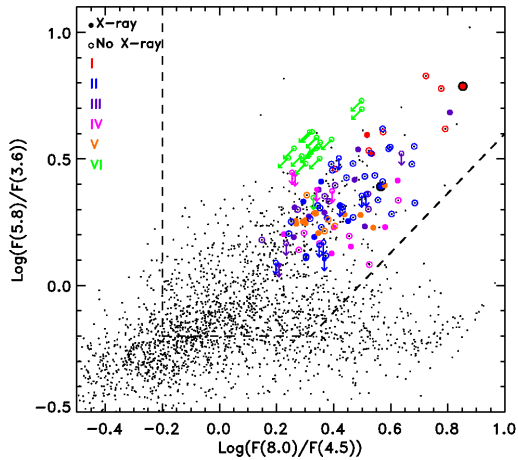}
\caption{\footnotesize IRAC color-color diagram (F(5.8\micron)/F(3.6\micron) $versus$
F(8.0\micron)/F(4.5\micron)) of all of the IR sources detected in four IRAC
bands in the \chandra/SWIRE field (black dots). IR-selected obscured AGN
candidates are shown as open circles, or full circles if they are also X-ray
sources. Downward arrows indicate sources that are not detected at
5.8\micron, and arrows pointing toward the bottom-left corner indicate
sources that are not detected at neither 5.8\micron\ nor 8.0\micron. The
colors correspond to different SED types (I:red, II:blue, III:purple,
IV:magenta, V:orange and VI:green) (see Section~\ref{ctagn_ir_sec}). The two
confirmed Compton-thick quasars, SW104406 and SW104409, are shown with large
black circles (see section~\ref{sed}). The dashed line shows the region
preferentially occupied by AGN, identified by~\citet{lacy04}.}
\label{irac_cols}
\end{figure}

The SEDs of the remaining 181 obscured AGN candidates were fitted using
HYPERZ and the library of 12 normal galaxy and 12 AGN templates. Only the
sources that did not have any acceptable solutions (a minimum in the
$\chi^{2}_{\nu}$ distribution) with normal galaxy templates were kept for a final sample
of 120 sources. Among the rejected sources, several can be high redshift galaxies
whose optical and near-IR emission is due to stellar light. Some of the
rejected sources might host an AGN, but in order to increase our sample
reliability we decided to remove any dubious cases. This selection removes 7
X-ray sources from the sample, yielding a total of 41 X-ray sources among
the IR-selected obscured AGN. The X-ray properties of the 41 X-ray sources
out of the 120 IR-selected sample of AGN candidates are reported in
Table~\ref{ctagn_ir_xprop} and described in Section~\ref{ctagn_ir_x}. For
all of the selected sources, even for those that are not detected in the
optical images, the 3.6\micron\ over optical flux ratio is constrained to be
higher than the limits given above. The IRAC colors of the selected sample
compared to the rest of IR sources in the
\chandra/SWIRE field are shown in Figure~\ref{irac_cols}. The dashed line
delimits the region where AGN are most likely to be found~\citep{lacy04}
(see also~\citet{stern05,evanthia05}).


In the initial sample of 4493 sources, less than 9\%\ are X-ray sources, but
34\%\ of the selected sample of obscured AGN candidates are X-ray sources.
Among the IR-selected AGN that were rejected because consistent with
unobscured AGN, 63\% are X-ray sources. These values are consistent with a
selection that favors AGN. Only two sources are not detected at 24\micron\,
and four have a 24\micron\ flux below the nominal 5$\sigma$ limit. A
24\micron\ detection for these four sources was confirmed after a visual
inspections of the images. Note that the sensitivity of the 24\micron\ data
varies across the field due to a variation in the number of coverages. The
adopted 5$\sigma$ limit of 230$\mu$Jy is valid for the areas with median
coverage in the SWIRE/\chandra\ field.

\begin{figure}[!ht]
 \plotone{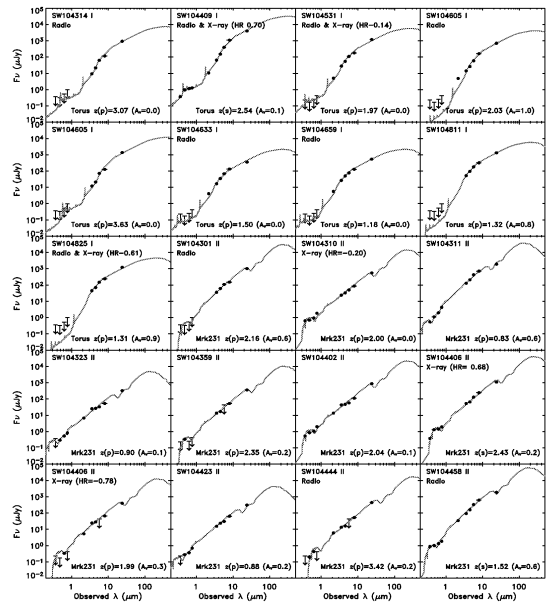}
 \caption{\footnotesize SEDs of the 120 IR-selected obscured AGN candidates. Five $\sigma$
 upper limits are reported as downward arrows. The sources are ordered by
 SED class, from I to VI (see text) and by right ascension within each SED
 class. The abbreviated source name and the SED class are given on the
 upper-left corner of each panel. Each SED class is fitted with an AGN
 template (I: Torus, II: Mrk 231, III: QSO1, IV: IRAS 19254$-$7245 South, V:
 QSO2, and VI with any of the previous templates). The best-fit template for
 each source at the spectroscopic redshift, $z$(s), if available, or at the
 photometric redshift, $z$(p), is shown as a grey curve. The amount of
 extinction applied to the template and the redshift are reported on the
 bottom-right corner in each panel. X-ray sources are identified by the note
 ``X-ray'' on the upper-left corner and the HR value is also given and radio
 sources are identified by the note ``Radio''.}
\label{ctagn_ir_sed}
\end{figure}
\addtocounter{figure}{-1}
\begin{figure}[!ht]
 \plotone{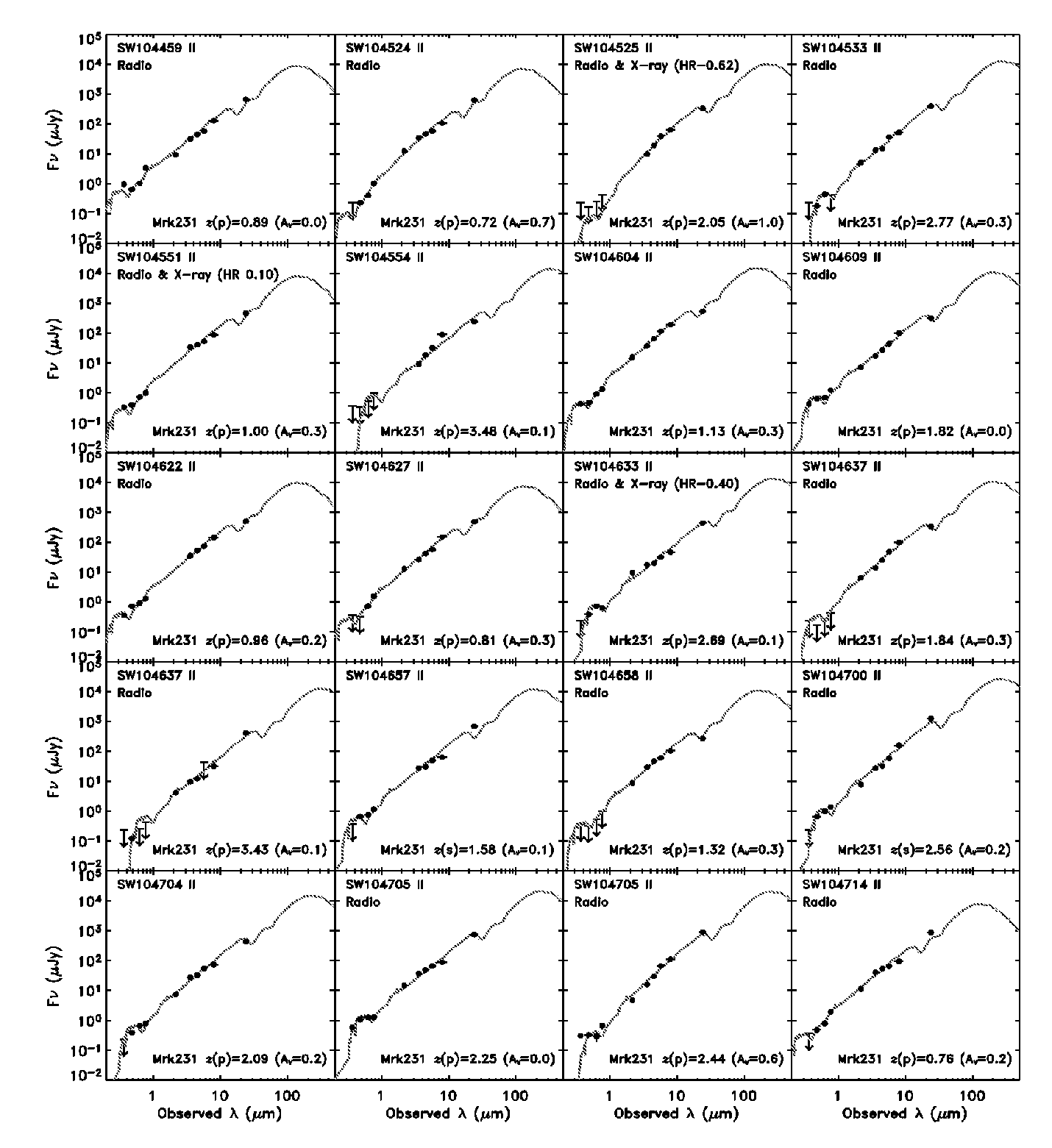}
 \caption{\footnotesize \it Continued}
\end{figure}
\addtocounter{figure}{-1}
\begin{figure}[!ht]
 \plotone{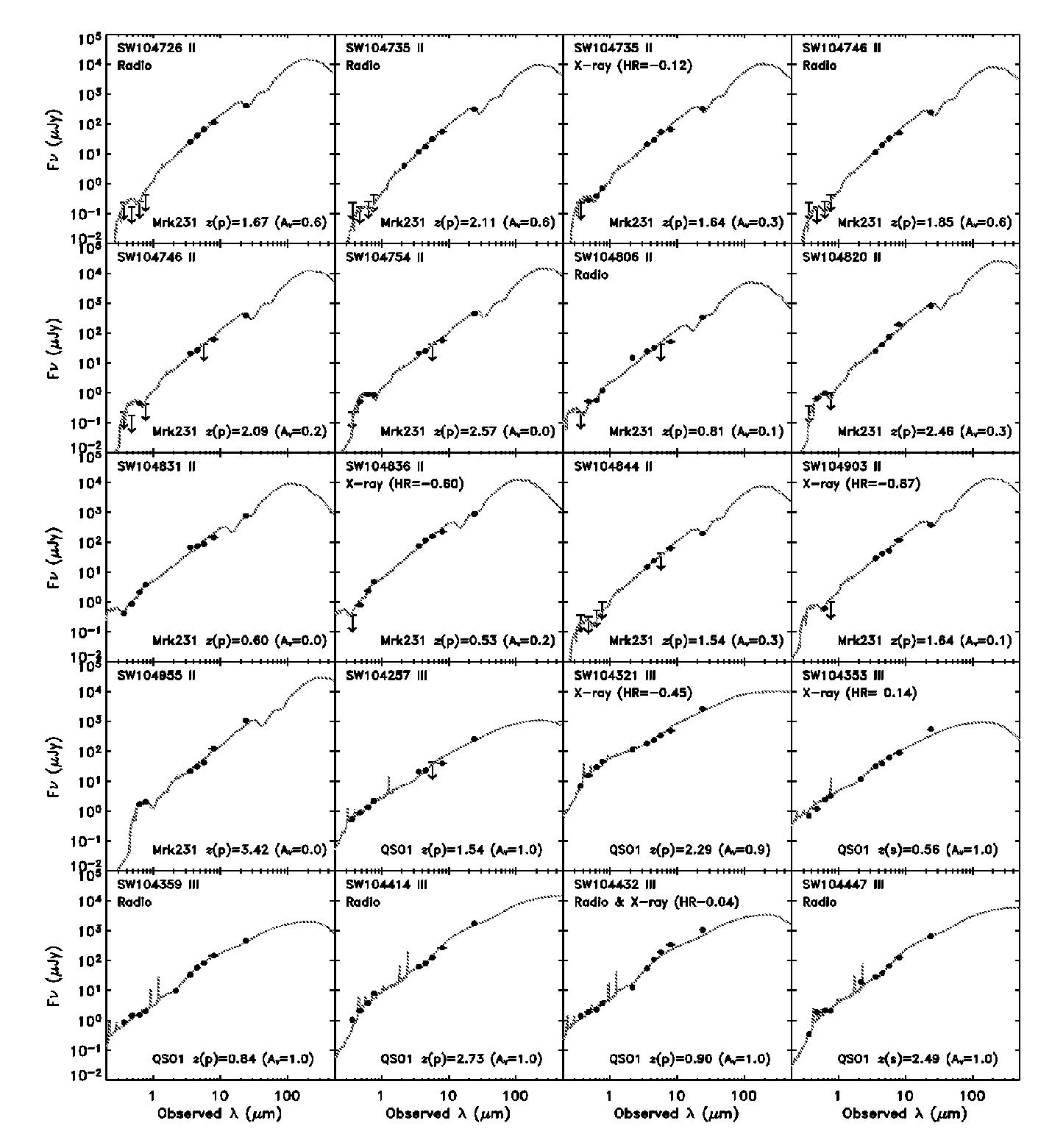}
 \caption{\footnotesize \it Continued}
\end{figure}
\addtocounter{figure}{-1}
\begin{figure}[!ht]
 \plotone{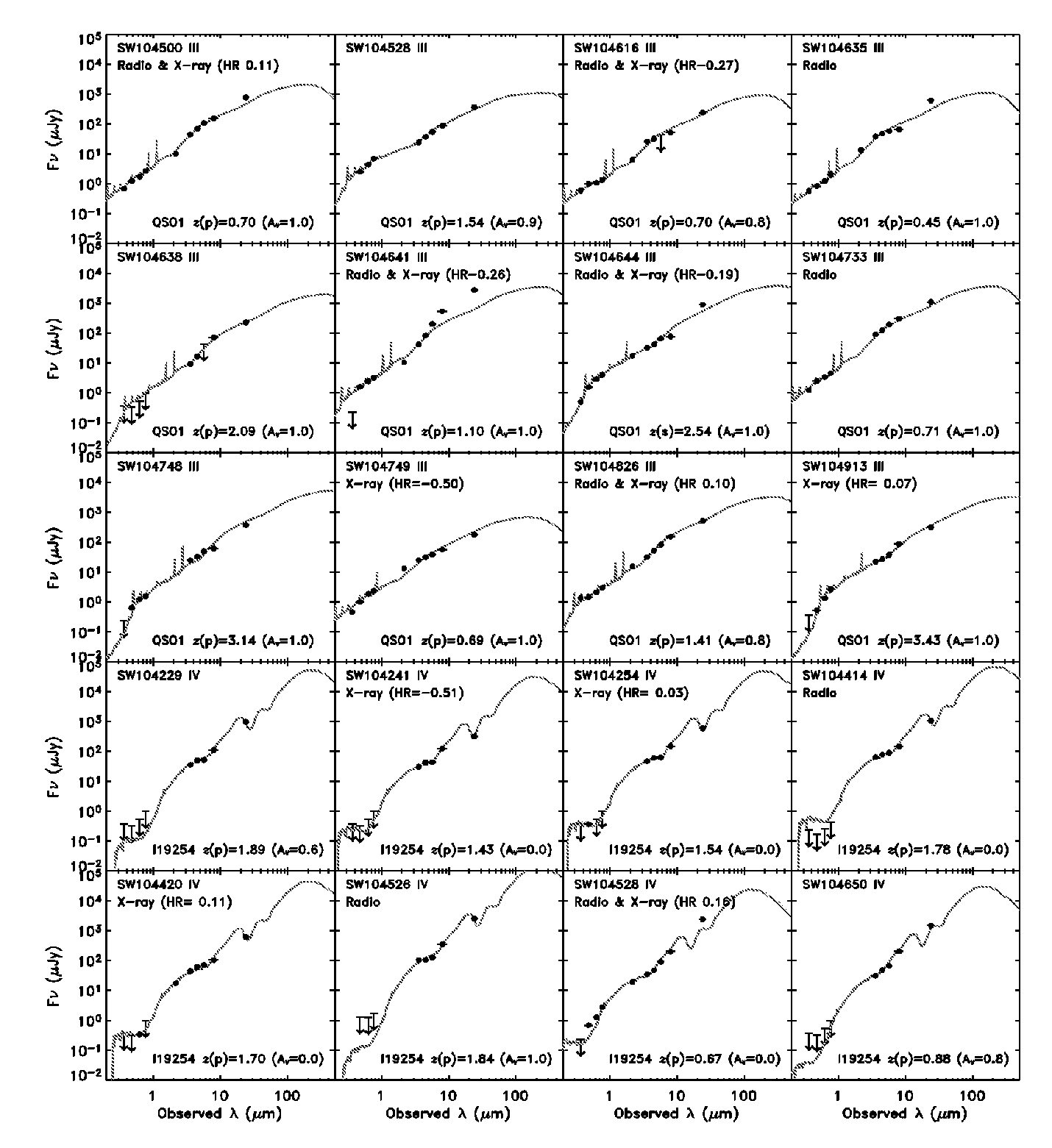}
 \caption{\footnotesize \it Continued}
\end{figure}
\addtocounter{figure}{-1}
\begin{figure}[!ht]
 \plotone{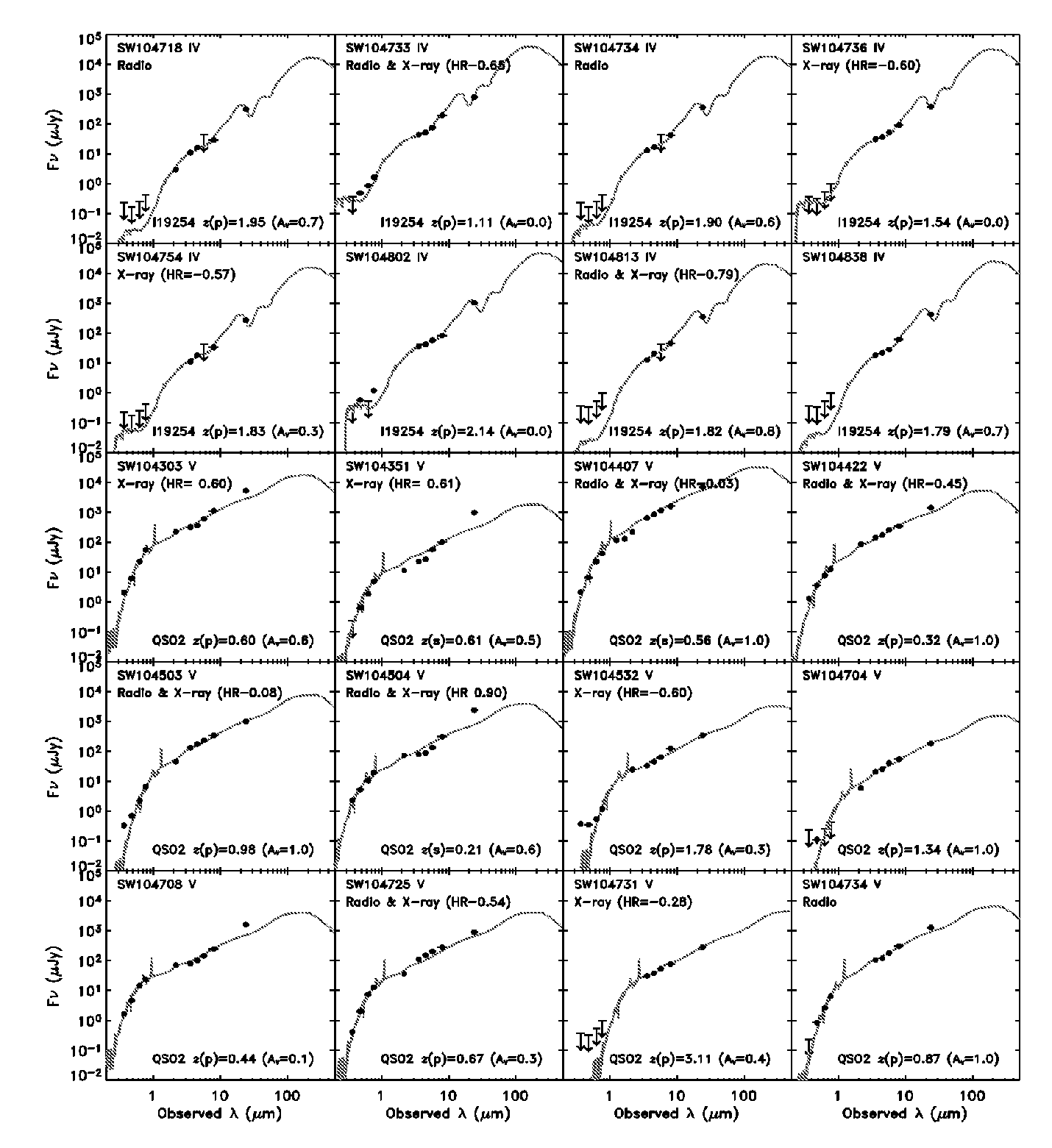}
 \caption{\footnotesize \it Continued}
\end{figure}
\addtocounter{figure}{-1}
\begin{figure}[!ht]
 \plotone{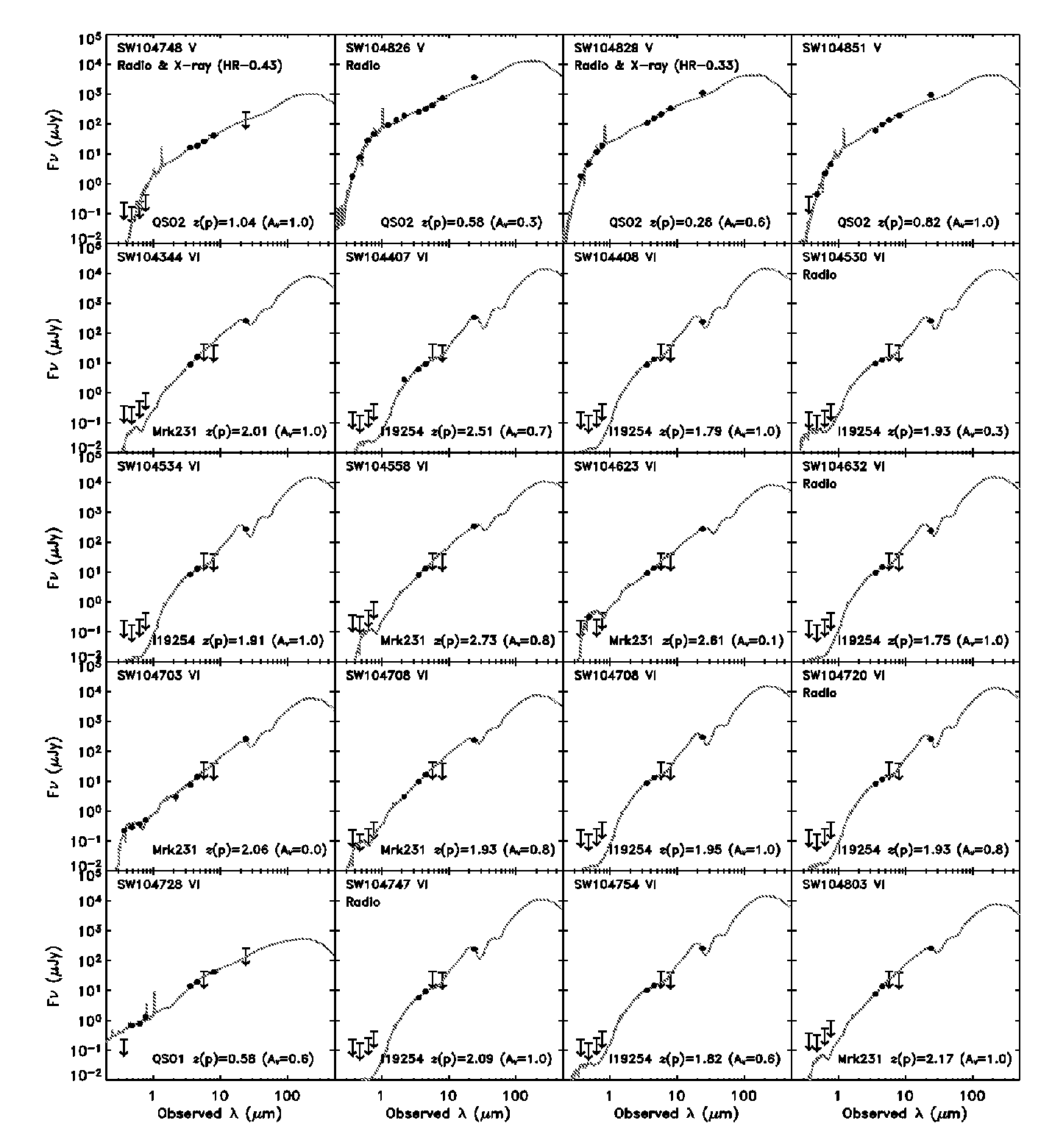}
 \caption{\footnotesize \it Continued}
\end{figure}

The sample was divided in six categories, from I to VI, five categories
(I--V) were defined based on their best-fit template one category (VI) was
defined based on the low number (3) of detections in the IR. Class I sources
(9) have IR SEDs characterized by a convex shape that are well fitted with a
``Torus'' template (see Section~\ref{sed_27114}). Class II sources (44) show
power-law like optical-IR SEDs, similar to Mrk 231 or slightly redder
(A$_{\rm V}<$1). Similarly, class III sources (19) have power-law like
optical-IR SEDs, but not as red as Mrk 231; a reddened QSO template (A$_{\rm
V}=$0.6--1.0) provides a better fit. Class IV sources (16) show SEDs
consistent with templates of composite sources with contributions from both
star-forming regions and an AGN component. This interpretation is supported
by an excess at 24$\mu$m with the respect to the extrapolation of the
power-law representing the hot dust continuum associated with the AGN, this
excess is likely associated to emission from the PAH bands. Class V objects
(16) are characterized by a turnover in the SED with optical SEDs much
redder than the IR SED. These sources might be composite objects where the
optical emission is dominated by stellar light, from either an old or a
reddened stellar population, and the IR emission is associated to the AGN.
Class VI objects (16) are fainter than the rest of the sample and detected
only in three bands from 3.6 to 24$\mu$m, their SEDs show a distinct
signature of AGN in their extreme red F(4.5\micron)/F(3.6\micron) flux
ratios. Upper limits to the flux in the bands are consistent with the models
used to describe the other classes. Names, coordinates, \rp\ magnitudes, IR
fluxes and classification of the IR-selected obscured AGN candidates are
listed in Table~\ref{ctagn_ir} and their SEDs are shown in
Figure~\ref{ctagn_ir_sed}. Photometric redshifts were used to over-plot the
best-fit template on each SED when spectroscopic redshifts were not
available.  The redshifts are used only to visually characterize the SED
shape of these sources by comparing them with AGN templates.


Only eleven sources have spectroscopic
redshifts. Spectroscopic redshifts are indicated as $z$(s) and photometric
redshifts as $z$(p) in Figure~\ref{ctagn_ir_sed} and listed in 
Table~\ref{ctagn_ir}. The optical spectra of seven sources (SW104351,
$z$=0.609; SW104409, $z$=2.540; SW104406, $z$=2.430; SW104447, $z$=2.488; SW104644,
$z$=2.542; SW104657, $z$=1.579; and SW104700, $z$=2.562), show emission lines from
an AGN, e.g. \lya, \civ, and their width ranges from 1200 to $\sim$1700 km/s,
typical of narrow-line or type 2 AGN. The optical spectrum of SW104407
($z$=0.555) is dominated by stellar light typical of a post-starburst galaxy
(Balmer absorption line) and high-ionization narrow emission lines, \nev,
\neiii, from the AGN. Two sources, SW104353 ($z$=0.563) and SW104458
($z$=1.52), show only one emission line, \oii, consistent with emission from
new-born stars. No AGN signatures are observed, but the optical spectra
are characterized by low signal-to-noise. The spectrum of SW104504
($z$=0.214) shows broad Balmer emission lines, \halpha, \hbeta, and the
\oiii\ forbidden line, typical of a starburst galaxy.

\subsubsection{X-ray properties of the IR-selected obscured AGN candidates}
 \label{ctagn_ir_x}

A small fraction, 41 sources, corresponding to 34\%\ of the IR-selected
obscured AGN candidates are also X-ray detected. Five of these 41 X-ray
sources have multiple IR counterparts. However, in all five cases the
matched source is at less than 1\arcsec\ from the X-ray source with a
probability of random association $<$1.3\%, and the neighbour sources are
more than 2\arcsec\ distant. Therefore, we consider all these associations
reliable. The 41 X-ray sources show all types of optical-IR SEDs, except
class VI. Their X-ray spectra (see Table~\ref{ctagn_ir_xprop}) show a wide
range of hardness ratios, from $-$0.87 to 0.90, with a median value of
$-$0.26.  Assuming our redshift estimates to derive the intrinsic column
density and a photon index
$\Gamma$=1.7, 26 (63\%) sources have column densities larger than 10$^{22}$
cm$^{-2}$ and of these, 12 have \nh$>$10$^{23}$ \cm2. Only two 
sources are also present in the X-ray selected sample of Compton-thick
(\nh$>$10$^{24}$ \cm2) AGN, SWIRE\_J104409.95+585224.8 (SW104409 hereafter)
and SWIRE\_J104406.30+583954.1 (SW104406 hereafter).

\subsubsection{Optical $versus$ X-ray flux} 
\label{opx_sec}

\begin{figure}[!ht]
\vspace{-8mm}
\plotone{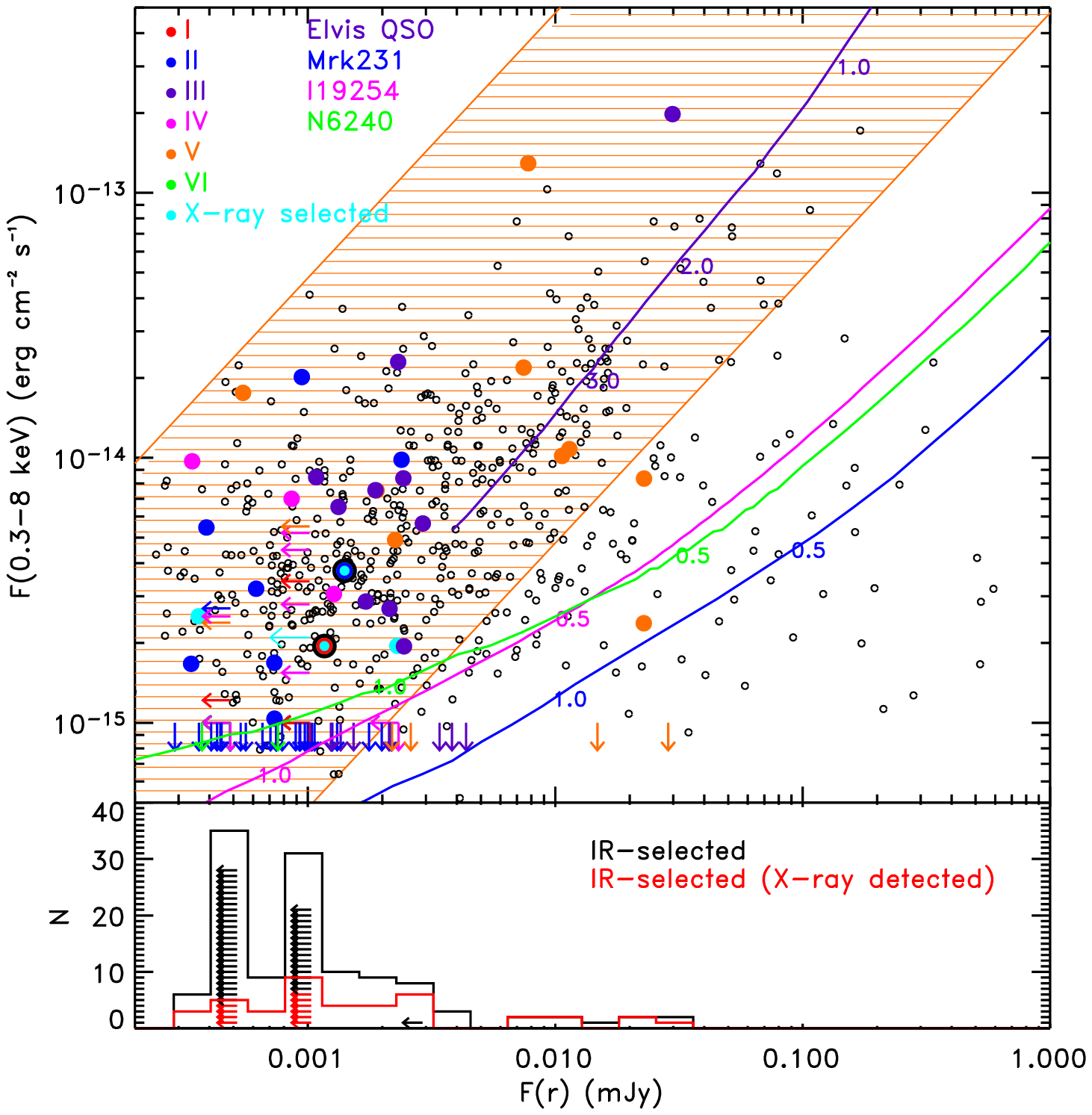}
\vspace{-8mm}
\caption{\footnotesize Top panel: Broad (0.3-8 keV) X-ray flux $versus$ the \rp-band flux of all of the
X-ray sources in the \chandra/SWIRE field (black open symbols), of X-ray
selected obscured AGN candidates (cyan) and of IR-selected obscured AGN
candidates (I:red, II:blue, III:purple, IV:magenta, V:orange and VI:green). Downward
arrows represent sources not detected in the X-rays and leftward arrows
represent sources not detected in the optical \rp-band. The X-ray fluxes are
derived assuming an absorbed power-law model with photon index, $\Gamma$
equal to 1.7 and Galactic absorption, \nh=6$\times$10$^{19}$\cm2. The two
Compton-thick quasars, SW104406 and SW104409, are shown with large black
circles. The orange shaded area delimits the region of $Log(F(X)/\nu F($\rp$)) =
\pm$1, typical of ``classical'' AGN. The purple, blue, magenta and green curves
represent the tracks of known AGN at various redshifts as annotated, the
Elvis QSO template (purple), Mrk 231 (blue), IRAS 19254$-$7245 South (magenta)
and NGC6240 (green). Bottom panel: Distribution of the \rp-band flux of the
IR-selected sample of obscured AGN candidates. The black line represents the
entire IR-selected sample, the red curves shows only the X-ray detected sub-sample.
Left-pointing arrows represent \rp-band 5$\sigma$ upper limits, each arrows
refers to a single source.}
\label{optx}
\end{figure}

The optical and broad-band (0.3-8 keV) X-ray fluxes of all of the X-ray
sources in the \chandra/SWIRE field and of the IR-selected obscured AGN
candidates are compared in the top panel of Figure~\ref{optx} where the
X-ray sample is shown as black open circles and the IR-selected sample as
full circles. We use the full band flux to minimize the uncertainty
associated with the factors used to convert counts to flux. The distribution
of \rp-band fluxes for the IR-selected sample is shown in the bottom panel
of Figure~\ref{optx}. The sample of IR-selected obscured AGN candidates is
shown as full circles of different colors according to their classification
(I in red, II in blue, III in purple, IV in magenta, V in orange and VI in green). The
X-ray-selected sources are shown as cyan full circles. In case of no optical
detection a left pointing arrow at the 5$\sigma$ \rp-band value is
shown and in case of no X-ray detection a downward arrow at 10$^{-15}$
\ergcm2s\ is shown. We also plot the expected observed fluxes of four AGN
templates at various redshifts from 0.1 to 3. The AGN templates correspond
to the median of a sample of optically selected quasars~\citep{elvis94}:
``Elvis QSO'', the Compton-thick BAL QSO/Seyfert 1 galaxy Mrk 231
($z$=0.042; ~\citet{braito04}), and the Compton-thick Seyfert 2 galaxies
IRAS 19254$-$7245 South (I19254) ($z$=0.0617; ~\citet{berta03,braito03}) and
NGC 6240 (N6240) ($z$=0.0244; ~\citet{vignati99,iwasawa01}). All templates
other then the ``Elvis QSO'' are characterized by extreme (Compton-thick)
absorption in the X-rays. The hatched area represents the locus of $F(X)/\nu F$(\rp)
between 0.1 and 10, which is traditionally considered the locus where
``classical'' AGN lie~\citep{akiyama03}. Sources with $F(X)/\nu F$(\rp)$>$10 are
expected to be mostly obscured AGN at high-$z$~\citep{perola04}, while
sources with $F(X)/\nu F$(\rp)$<$0.1 are expected to be mostly star-forming
galaxies whose X-ray emission is not powered by an AGN. However, this simple
picture becomes more complex for fainter AGN~\citep{comastri03}. 


The three Compton-thick AGN templates are characterized by low X-ray over
optical flux ratios ($F(X)/\nu F$(\rp)$<$ 0.1) at low-$z$ and as the redshift
increases, they move into the locus of ``classical'' AGN. Thus, extreme
$F(X)/\nu F$(\rp) ratios are not expected for local Compton-thick AGN even at high
redshifts. Lower flux ratios ($F(X)/\nu F$(\rp)=0.1--10), consistent with those of
classical AGN, are also observed in the X-ray selected Compton-thick AGN
sample. The majority (108) of the IR-selected obscured AGN candidates also
has $F(X)/\nu F$(\rp)=0.1--10, 91 of which show $F(X)/\nu F$(\rp) between 0.1 and 1 and
12 sources have $F(X)/\nu F$(\rp)$<$0.1. It is clear that obscured AGN do not have
unique $F(X)/\nu F$(\rp) ratios. Therefore, a selection based on large ($>$10)
$F(X)/\nu F$(\rp) values would miss a large fraction of obscured AGN.

\subsubsection{Mid-Infrared $versus$ X-ray flux} 
\label{irx_sec}

\begin{figure}[!ht]
\plotone{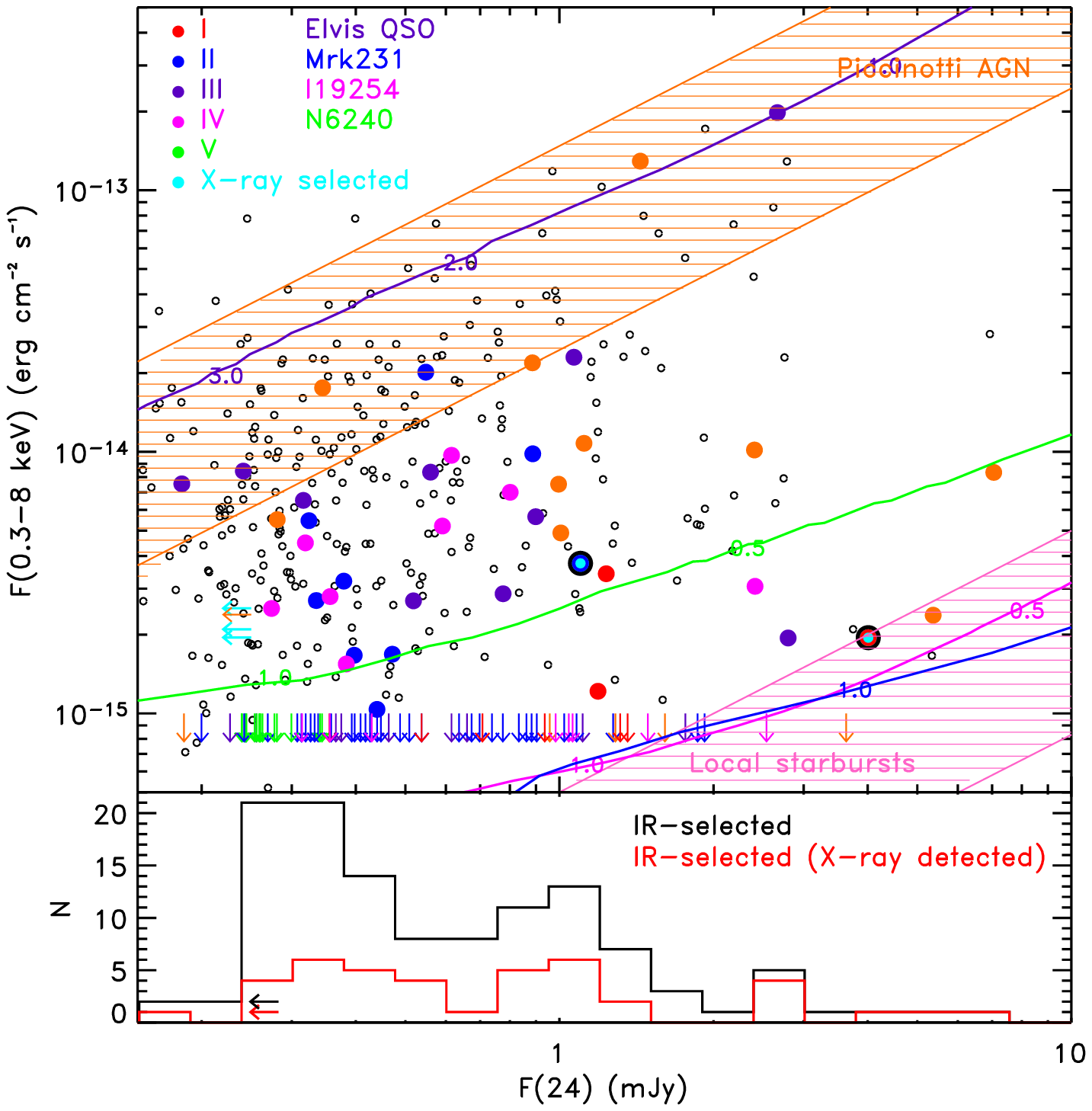}
\caption{\footnotesize Top panel: Broad (0.3-8 keV) X-ray flux $versus$ the 24$\mu$m flux of
all of the X-ray sources in the \chandra/SWIRE field (black open circles).
Symbols as in Figure~\ref{optx}. The orange dashed area represents the area
that the AGN in the Piccinotti's sample~\citep{piccinotti82} would occupy
and the pink area represents the area that local starburst galaxy would
occupy (adopted from~\citet{alonso04}). The X-ray fluxes are derived assuming
an absorbed power-law model with photon index, $\Gamma$ equal to 1.7 and
Galactic absorption, \nh=6$\times$10$^{19}$\cm2. Bottom panel: Distribution
of the 24\micron\ flux of the IR-selected sample of obscured AGN candidates.
The black line represents the entire IR-selected sample, the red curves
refers only to the X-ray detected sub-sample. Left-pointing arrows represent 24\micron\ 5$\sigma$
upper limits.}
\label{irx}
\end{figure}

The observed mid-IR flux at 24$\mu$m and the broad-band (0.3-8 keV) X-ray
flux for the X-ray sample in the \chandra/SWIRE field are compared in the
top panel of Figure~\ref{irx} where the X-ray sample is shown as black open
circles. Downward arrows indicate the full band (0.3-8 keV) flux upper limit
of 1$\times 10^{-15}$ \ergcm2s\ and leftward arrows indicate the 230$\mu$Jy
5$\sigma$ limit at 24$\mu$m. The sample of X-ray and IR-selected obscured
AGN shown in Figure~\ref{optx} is also shown in Figure~\ref{irx} as full
circles or downward arrows in colors corresponding to different classes
(from I to VI). The curves represent the expected observed fluxes of the
four AGN templates shown in Figure~\ref{optx} at various redshifts from 0.1 to 3.
In the bottom panel of Figure~\ref{irx}, the distribution of 24\micron\
fluxes of the IR-selected sample of obscured AGN candidates is shown.


Hard X-rays to mid-IR flux ratios in the local Universe range from 10$^{-3}$
for starburst galaxies to $\simeq$1 for unobscured
AGN~\citep{alexander01,manners04,lutz04}. The locus occupied by hard
X-ray-selected AGN from the~\citet{piccinotti82} sample with detected mid-IR
emission and
$z<$0.12 corresponds to the hatched area in orange
($F(X)/(\nu_{24}F(24\micron))$=0.19--1.17). The locus occupied by local
starburst galaxies corresponds to the hatched pink area
($F(X)/(\nu_{24}F(24\micron))$ = 5$\times$10$^{-4}$ -- 3$\times$10$^{-3}$). The two loci have been
adapted from~\citet{alonso04} after correcting the X-ray flux from 2-10 keV
to 0.3-8 keV assuming an absorbed power-law model with
\nh=10$^{21}$ cm$^{-2}$ and spectral index $\Gamma$=1.7 for the Piccinotti's
sample and $\Gamma$=2.0 for the starbursts. Sources with low or no
absorption, such as the Elvis QSO template, show large X-ray over mid-IR
flux ratios which are almost constant up to $z$=4 (0.6--0.9). Sources
characterized by large column-densities are characterized by a broader range
of $F(X)/\nu_{24}F(24\micron))$ values, from 6$\times$10$^{-4}$ to 0.6 for
$z<$4. Their X-ray over mid-IR flux ratios increase at larger redshifts,
moving them into the region occupied by unobscured AGN. As absorption
increases the X-ray flux decreases, while the 24\micron\ flux is only
slightly affected. At higher-$z$, however, the IR flux decreases more
rapidly since the observed wavelengths correspond to shorter wavelengths in
the rest-frame and the X-ray flux becomes less affected by obscuration as
higher energy photons are collected, resulting in larger X-ray/24\micron\
flux ratios~\citep{alexander01,fadda02,manners04}.

All of the X-ray selected Compton-thick AGN show X-ray/mid-IR flux ratios
lower than those of unobscured AGN
($F(X)/(\nu_{24}F(24\micron))$=0.001--0.03). Among the IR-selected obscured
AGN candidates only 7 sources (SW104310, SW104321, SW104422, SW104532,
SW104616, SW104725, SW104749) have $F(X)/(\nu_{24}F(24\micron))>$0.19 as the
Piccinotti's sample. Three of these 7 sources are obscured in the X-rays
(\nh$>10^{22}$ \cm2). Thus, obscured AGN do not show unique
$F(X)/(\nu_{24}F(24\micron)$ values, however low values ($<$0.2) are more
likely. This results is in agreement with a recent study of X-ray and
24$\mu$m-selected AGN, which shows that there is no correlation
between $F(X)/\nu_{24}F(24\micron)$ and the amount of absorption in the X-ray or their
optical properties~\citep{rigby05}.

The range of IR and optical fluxes of the X-ray detected sources are very
similar to the values observed in the entire IR-selected sample of obscured
AGN candidates (see bottom panels of Figures~\ref{optx} and~\ref{irx}).
However, there is a higher fraction of faint sources in the whole sample
compared to the X-ray detected sub-sample, and the majority of sources has
smaller X-ray over 24\micron\ flux ratios. Smaller ratios suggest that the
fraction of obscured sources and/or the amount of absorption is higher in
the non X-ray detected sub-sample. Since we can not quantify the amount of
obscuration in the non X-ray detected sources, we assume that the
distribution of absorption in the entire IR-selected sample is similar to
that observed in the X-ray detected sub-sample (63\% with \nh$>$10$^{22}$
\cm2, 29\%\ with \nh$>$10$^{23}$ \cm2, and 5\%\ with \nh$>$10$^{24}$ \cm2).
However, it is plausible that the estimated fractions of obscured sources
are only lower limits to the real distribution for the reasons given above.

Although the ratios between the X-ray flux and the optical or the mid-IR
flux are affected by absorption, they cannot be used as an effective
method to select obscured AGN. As shown in the two previous sections and in
previous works~\citep{rigby04}, these flux ratios also depend on the
AGN luminosity, the host galaxy contribution and redshift, and, therefore,
they are not unique for AGN with large amounts of absorption.

\subsubsection{Radio properties of the IR-selected obscured AGN candidates}
 \label{ctagn_ir_radio}

Although a detailed discussion on the radio properties of the AGN in this
field and on the radio population in general will be presented in future
works, here we give a brief summary of the radio detection rate of the
IR-selected obscured AGN candidates. Half (60 sources) of the sample is
detected in the radio (see Table~\ref{ctagn_ir}). The fraction of radio
sources per class, I, II, III, IV, V, and VI, is, respectively, 77, 50, 57,
43, 56 and 25\%. The fraction of radio sources among the IR-selected
obscured AGN candidates is much higher than the fraction of radio sources
among either all IR sources or all X-ray sources. More specifically, within
12\arcmin\ from the center of the radio field, there are 26 IR-selected
obscured AGN and 24 (=92\%) are radio sources. The fraction of radio sources
in this area per class is 100\%\ (2/2) of class I, 92\%\ (11/12) of class
II, 100\%\ (6/6) of class III, 100\% (1/1) of class IV, 75\%\ (3/4) of class
V, and 100\%\ (1/1) of class V. Since large ratio powers are usually
associated with the presence of an AGN, the high fraction of radio detection
of this sample is consistent with the hypothesis that these sources are
mainly powered by an AGN.

\section{Photometric and spectroscopic data of SW104409 and SW104406}
\label{qso2_obs}

Having discussed some general properties of the X-ray and IR-selected
obscured AGN candidates in the SWIRE/\chandra\ field, we now focus on the
only two sources that are selected by both selection methods, SW104409 and
SW104406. These two sources are the only ones among the X-ray sample of
Compton-thick AGN for which a spectroscopic redshift is available,
providing a confirmation on their Compton-tick nature. In this section we
review the available multi-wavelength data for the two sources and in the
next section we construct their SEDs, compare them to those of other
Compton-thick AGN and present a simple model to explain the observed
properties.

Optical and IR photometric fluxes as measured during the observations
described in section~\ref{opt_data} for SW104409 and SW104406 are listed in
Table~\ref{phot_tab1} and~\ref{phot_tab2}.

\subsection{Optical Spectroscopy} 
\label{opt_spec}

\begin{figure}[!ht]
\plotone{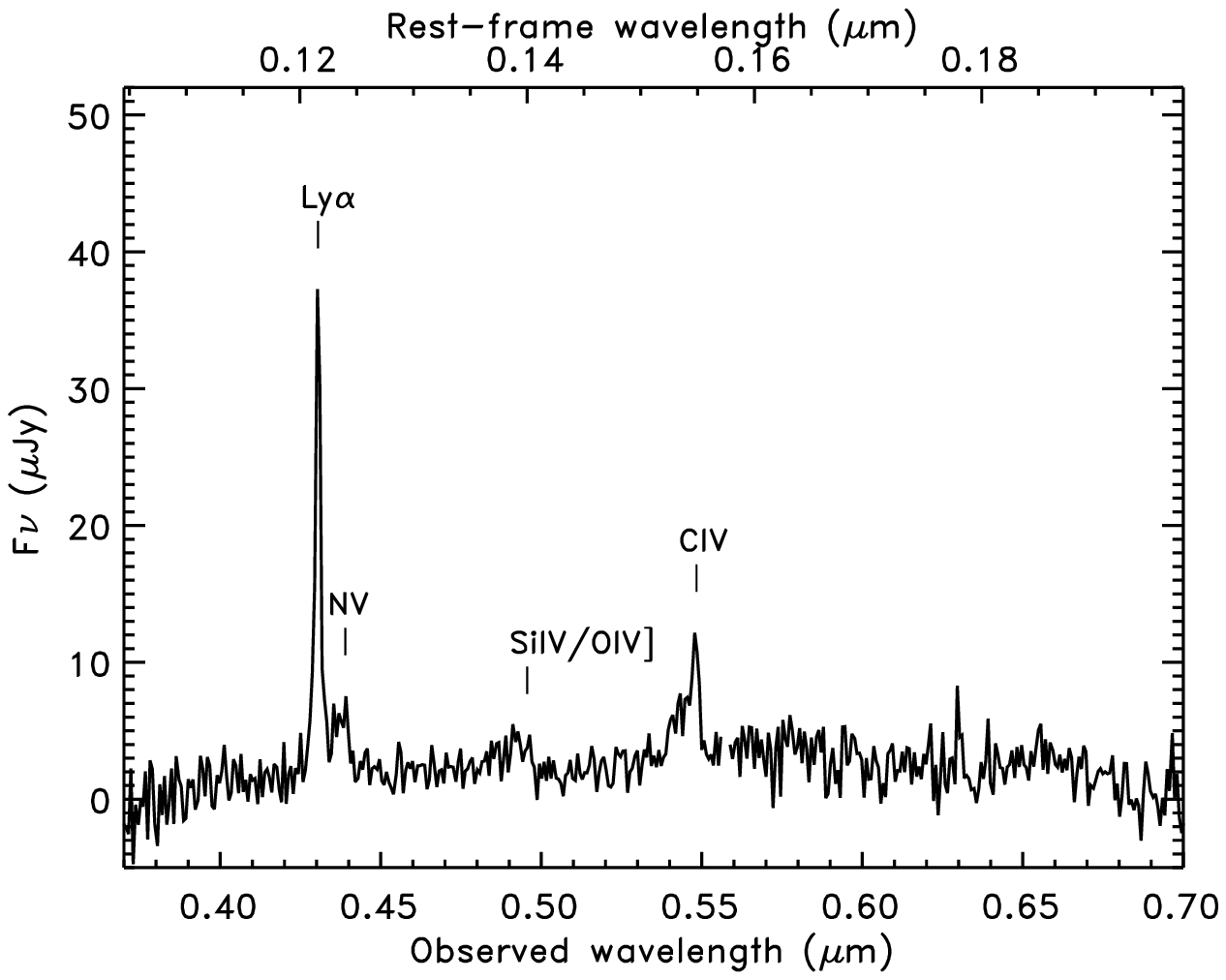}
\caption{\footnotesize Optical spectrum of SW104409 obtained with the Keck~I telescope.
Detected emission features are labeled.}
\label{figSpec1}
\end{figure}

\begin{figure}[!ht]
\plotone{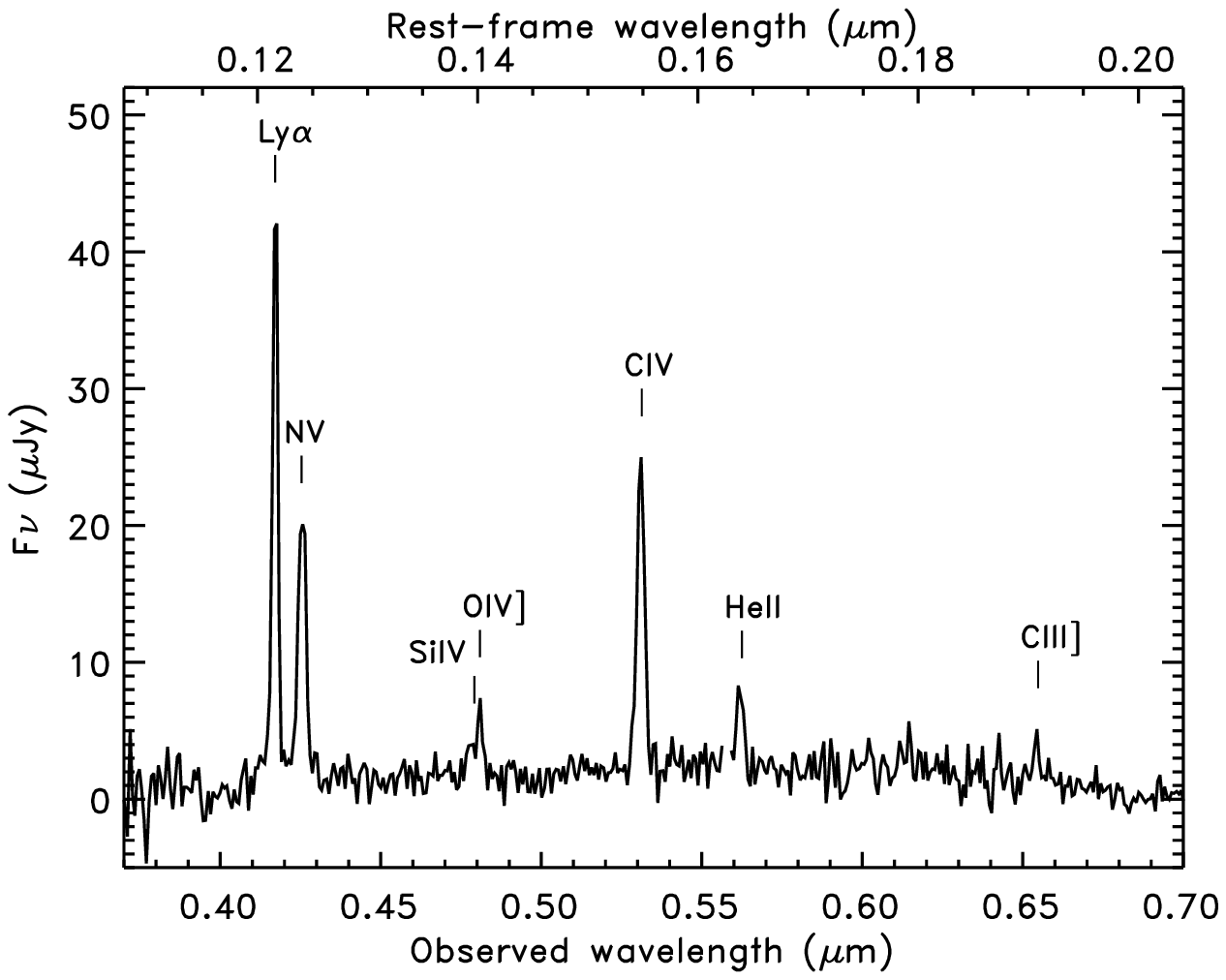}
\caption{\footnotesize Optical spectrum of SW104406 obtained with the Keck~I telescope.
Detected emission features are labeled}
\label{figSpec2}
\end{figure}

Spectroscopic observations for SW104409 and SW104406 were carried out on
2005 March 03 and 04, respectively, with the Low Resolution Imaging
Spectrometer (LIRS)~\citep{oke95} on the Keck~I telescope. The observations
were taken in multi-object mode and three equal integrations were performed
for a total exposure time of 1.5 hour. The observations were taken with
1\farcs 5 wide slitlets aligned near the parallactic angle. The effective
wavelength range of the blue spectrograph of the instrument is
3500$-$6700\AA.  A 300 lines/$mm$ grism with a blaze wavelength of 5000\AA\
was used resulting in 1.43 \AA$/pixel$ dispersion.  We used a HgAr lamp
spectrum obtained at the same position angle immediately following the
observations for wavelength calibration. The night was photometric with
$\sim$ 1\farcs 3 seeing and the spectra were calibrated using the
observations of the standard star G191B2B from~\citet{massey88}. The spectra
of SW104409 and SW104406 are shown in Figures~\ref{figSpec1} and
~\ref{figSpec2}, respectively. The optical flux and spectral shape of
SW104409 and SW104406 at the time of the Keck observations are consistent,
within 1$\sigma$, with the earlier broad-band \gp\ and \rp\ photometric
measurements.


The spectrum of SW104409 shows a narrow (FWHM = 1324$\pm$80 \kms) \lya\
emission line and two asymmetric emission lines with a broad blue-ward
component, \civ\ and \nv. The spectrum of SW104406 shows several narrow
emission lines, e.g.; \lya\ with FWHM = 1360$\pm$20 \kms and \civ\ with FWHM
= 1485$\pm$60 \kms. The line identifications and parameters (FWHM, fluxes
and the rest-frame equivalent width ($W_{\lambda,rest}$)), calculated for
single Gaussian fits using IRAF~\citep{kriss94}, are listed in
Table~\ref{spec_tab1} and~\ref{spec_tab2}. In the case of SW104409, since
the signal-to-noise is too low to constrain a Gaussian fit for the broad
component of the asymmetric emission lines, the total flux and equivalent
width were derived by integrating the total spectrum around the lines and
subtracting the continuum (see Table~\ref{spec_tab1}). The estimated mean
redshifts are $2.54 \pm 0.02$ for SW104409, and $2.430 \pm 0.003$ for
SW104406.

According to the properties of its optical spectrum, SW104406 is a classical
type 2 QSO, while SW104409 could be considered a peculiar type 2 QSO because
of the superposition of weak asymmetric broad components on top of the
stronger narrow emission lines.

A more detailed analysis and an in depth investigation of the spectrum of
SW104409 is beyond the scope of this work. The blue-shifted broad
components of the \civ\ and \nv\ emission lines suggest the presence of
highly ionized gas moving toward the observer as an outflow or wind from the
accretion disk as observed in other quasars~\citep{gallagher05}. However,
this interpretation is difficult to reconcile with a geometry in which the
AGN is obscured.

\subsection{X-ray Data and Spectral Analysis} 
\label{xray_spec}

Details (seq. N., OBSID and exposure time) on the X-ray observations of
SW104409 and SW104406 are listed in Table~\ref{xobs_log}. SW104409, located
at an off-axis angle of 3\farcm 6 has 11 broad-band counts, only 2 of which
fall in the soft band. SW104406, at an off-axis angle of 5\farcm 8, is
brighter in the X-ray than SW104409 with 22 broad-band counts and 3 in the
soft band. With so few counts, the error on the source's hardness ratio is
large and spectral modeling would not constrain any parameter. Therefore, in
order to constrain the amount of absorption in these sources we applied the
Bayesian method described in section~\ref{xray_data}~\citep{dyk04}. We
derive an HR of 0.85$^{+0.06}_{-0.39}$ for SW104409 and of
0.61$^{+0.21}_{-0.23}$ for SW104406, which correspond to intrinsic column
densities of 2.0$^{+0.5}_{-1.3}\times 10^{24}$ \cm2 for SW104409 and
1.0$^{+0.7}_{-0.3}\times 10^{24}$ \cm2 for SW104406. These extreme column
densities indicate that both sources are borderline Compton-thick AGN.
Alternative models, such as a reflection component due to ionized or neutral
gas, can not be ruled out, but they would also indicate large column
densities ($\sim$10$^{24}$\cm2). Assuming that the X-ray spectra of SW104409
and SW104406 are due to transmitted components through column densities of
\nh=2$\times$10$^{24}$ cm$^{-2}$, and
\nh=1$\times$10$^{24}$ cm$^{-2}$, respectively, then the
absorption-corrected X-ray luminosities in the rest-frame, assuming a photon
index $\Gamma$=1.7, are 4$\times$10$^{45}$\ergs\ and 5$\times$10$^{45}$
\ergs, respectively.

\subsection{Radio Data}

SW104409, located at 17\arcmin\ from the center of the radio field, is
unresolved, but clearly detected. Its radio flux, measured on an image
convolved to 3\arcsec\ to reduce instrumental effects this far off-axis, is
273$\pm$15$\mu$Jy. SW104406, located at the edge of the radio field
(26\arcmin\ off-axis), is not detected in the radio (there is only an
apparent 2.4$\sigma$ detection), therefore we assume a 3$\sigma$ upper limit
to its radio flux of 162$\mu$Jy.

\section{Spectral Energy Distributions of SW104409 and SW104406} 
\label{sed}

In this section, we analyze the spectral energy distributions (SEDs) from
X-ray to radio wavelengths of the two obscured quasars discussed above,
SW104409 and SW104406. The SED shapes and luminosities are compared to those
of known AGN: Elvis QSO template and the BAL QSO/Seyfert 1 galaxy Mrk 231.
The SEDs are interpreted assuming the unification scenario for which the
absorbing material is distributed around the central source in a toroidal
shape. However, alternative models can not be ruled out by the available
data.

\subsection{SW104409\label{sed_27114}}

\begin{figure}[!ht]
\plotone{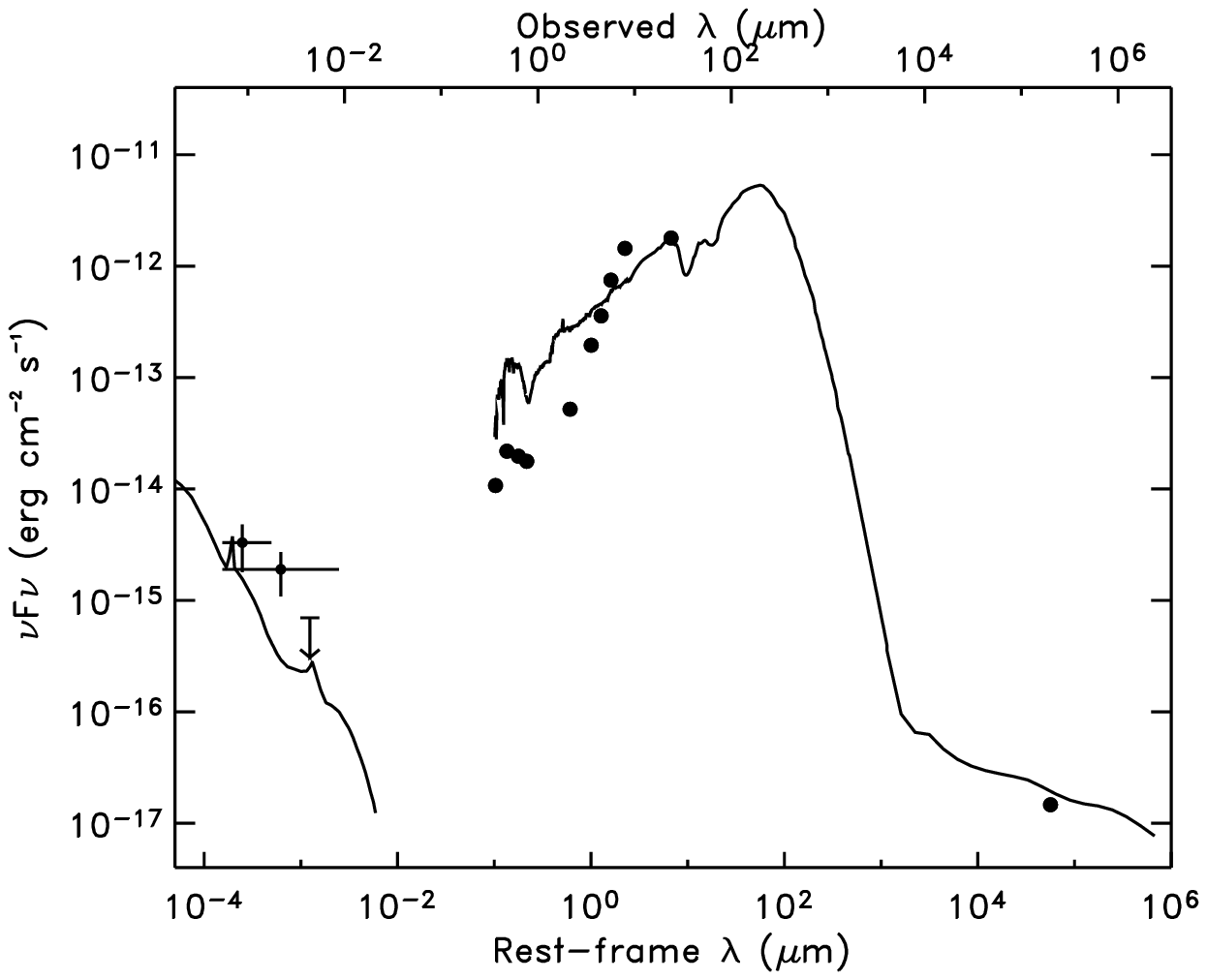}
\caption{\footnotesize SED of SW104409 (black circles) compared to the SEDs of the
Compton-thick AGN (black curve), Mrk 231, normalized at the observed 24$\mu$m flux. 
The crosses correspond to the observed X-ray flux in the broad, and hard
X-ray bands. The downward arrow is the 2$\sigma$ upper limit to the
soft-X-ray flux. The X-ray flux energy range is indicated by the length of
the horizontal line. The X-ray fluxes are derived assuming an absorbed
power-law model with photon index, $\Gamma$ equal to 1.7 and Galactic
absorption, \nh=6$\times$10$^{19}$\cm2.}
\label{figSED1}
\end{figure}

The SED of SW104409, from X-ray to radio wavelengths, is shown in
Figure~\ref{figSED1} and compared to the SED of Mrk 231. SW104409 is characterized by a blue
optical spectrum up to 2200\AA\ in the source rest-frame, followed by a
rapid rise at longer wavelengths with an observed \rp-$K_s$=4.13 (Vega),
fitting the conventional definition of extremely red object (ERO;
$R - K > 4$)~\citep{elston88}. The U-band drop-out is probably due to
intergalactic medium (IGM) attenuation~\citep{madau95}, as expected for such
a high redshift object. The comparison with Mrk 231 shows how extreme the IR
SED of SW104409 is ($F^{rest}\nu(2.2\mu m)/F^{rest}\nu(0.6\mu m)$=94
compared to 9.7 for Mrk 231). However, the mid-IR (10\micron) over radio
flux ratios are very similar and the mid-IR (10\micron) over X-ray flux is
only 3 times higher in SW104409 than in Mrk 231.


\begin{figure}[!ht]
\plotone{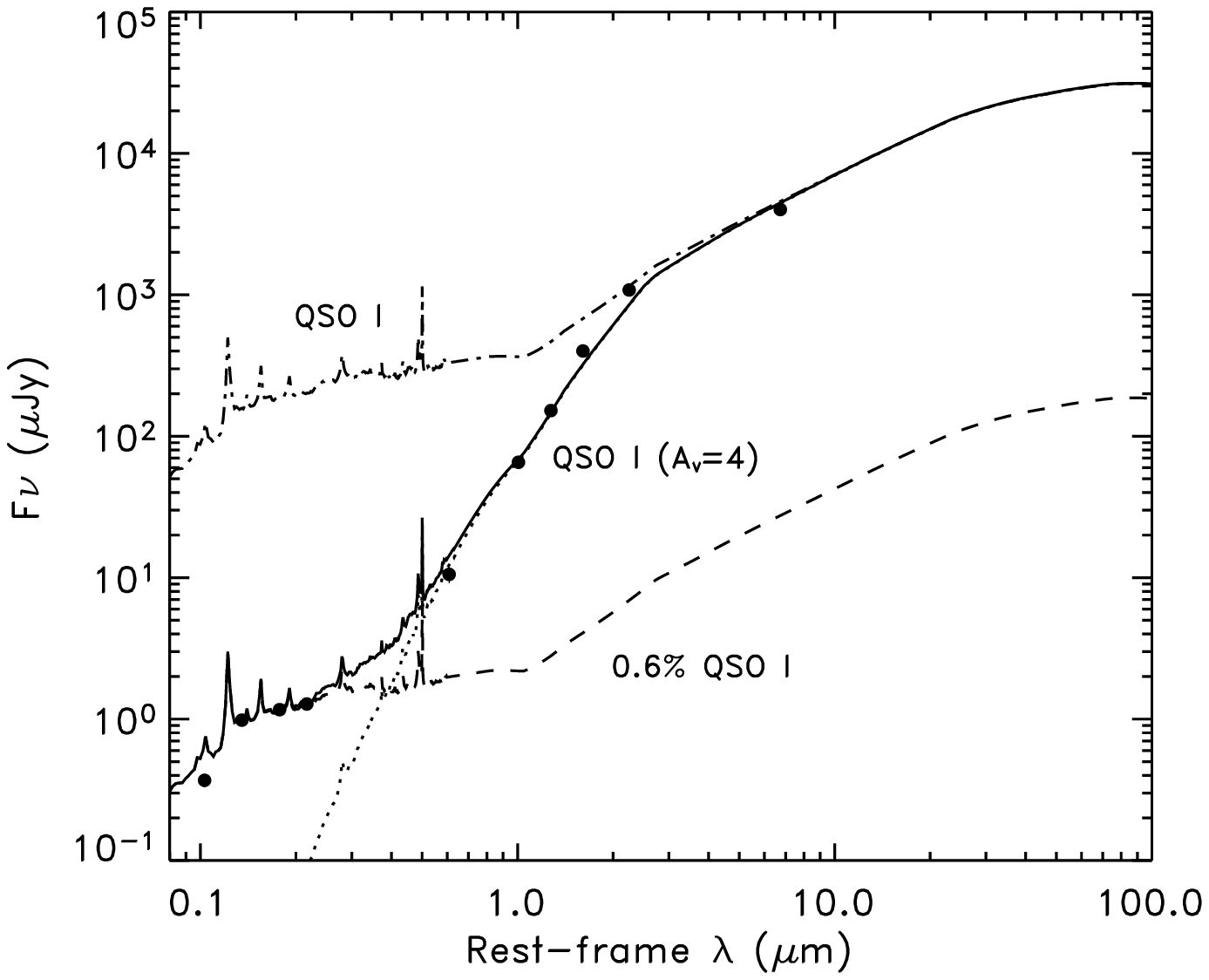}
\caption{\footnotesize Observed SED of SW104409 (black circles) compared to an
unobscured QSO template: 1) normalized to the mid-IR flux of SW104409
(dashed-dotted curve), 2) scaled to match the optical flux of SW104409
(dashed line), and 3) extinguished by A$_\mathrm{V}$=4 to fit the IR data points
(solid curve) (see section~\ref{opt_spec}). The solid curve corresponds to
the sum of the extinguished (3) and the scattered components (2).}
\label{extinction}
\end{figure}

The X-ray and near-IR properties of SW104409 indicate that the source is
heavily obscured. However, the observed optical continuum is blue and the
spectrum shows emission lines with broad components. A plausible explanation
for the observed optical spectrum could be scattering. The scattered light
preserves the spectral shape of the intrinsic component, but its flux
corresponds to a fraction of the primary component which depends on the
covering factor of the scattering medium~\citep{smith03}. The fraction of
scattered radiation can be estimated by comparing the observed and the
intrinsic (unabsorbed) optical flux. Since the observed optical spectrum is
similar to that observed in optically selected quasars and scattering does
not modify the spectral shape of the intrinsic spectrum, we estimated the
intrinsic optical flux by assuming that the intrinsic, before absorption,
SED of SW104409 is similar to that of an unobscured QSO normalized at the
observed mid-IR observed flux. In Figure~\ref{extinction}, we show the SED
of SW104409 and an unobscured QSO template in three cases: 1) normalized at
the observed 24\micron\ flux of SW104409 to represent the intrinsic, before
absorption, emission of SW104409; 2) scaled to match the observed optical
data, to represent the scattered component, and 3) reddened by an extinction
A$_\mathrm{V}$=4 mag (E(B$-$V)$\simeq$1). Reddening was applied as
prescribed in~\citet{calzetti99} assuming a foreground screen of dust at the
redshift of the source. According to this scenario, SW104409 resembles an
optically selected quasar whose light is reddened by an extinction
A$_\mathrm{V}$=4 mag and, therefore, completely suppressed at the observed
optical wavelengths (ultraviolet in the rest-frame) and reddened
in the near-IR (optical in the rest-frame). Assuming that the observed
optical blue spectrum is due to scattering, the scattered fraction
corresponds to 0.6\%\ of the intrinsic optical emission.
 

Note that throughout this work, we apply reddening as prescribed
in~\citet{calzetti99}. We also investigated the use of the Small Magellanic
Cloud (SMC) extinction curve~\citep{prevot84,bouchet85} that well reproduces
the optical spectra of dust reddened quasars in SDSS~\citep{richards03}. The
two prescriptions produce similar reddening at $\lambda>$5000\AA, but the
SMC law produces redder spectra at shorter wavelengths for the same amount
of extinction.

\begin{figure}[!ht]
\plotone{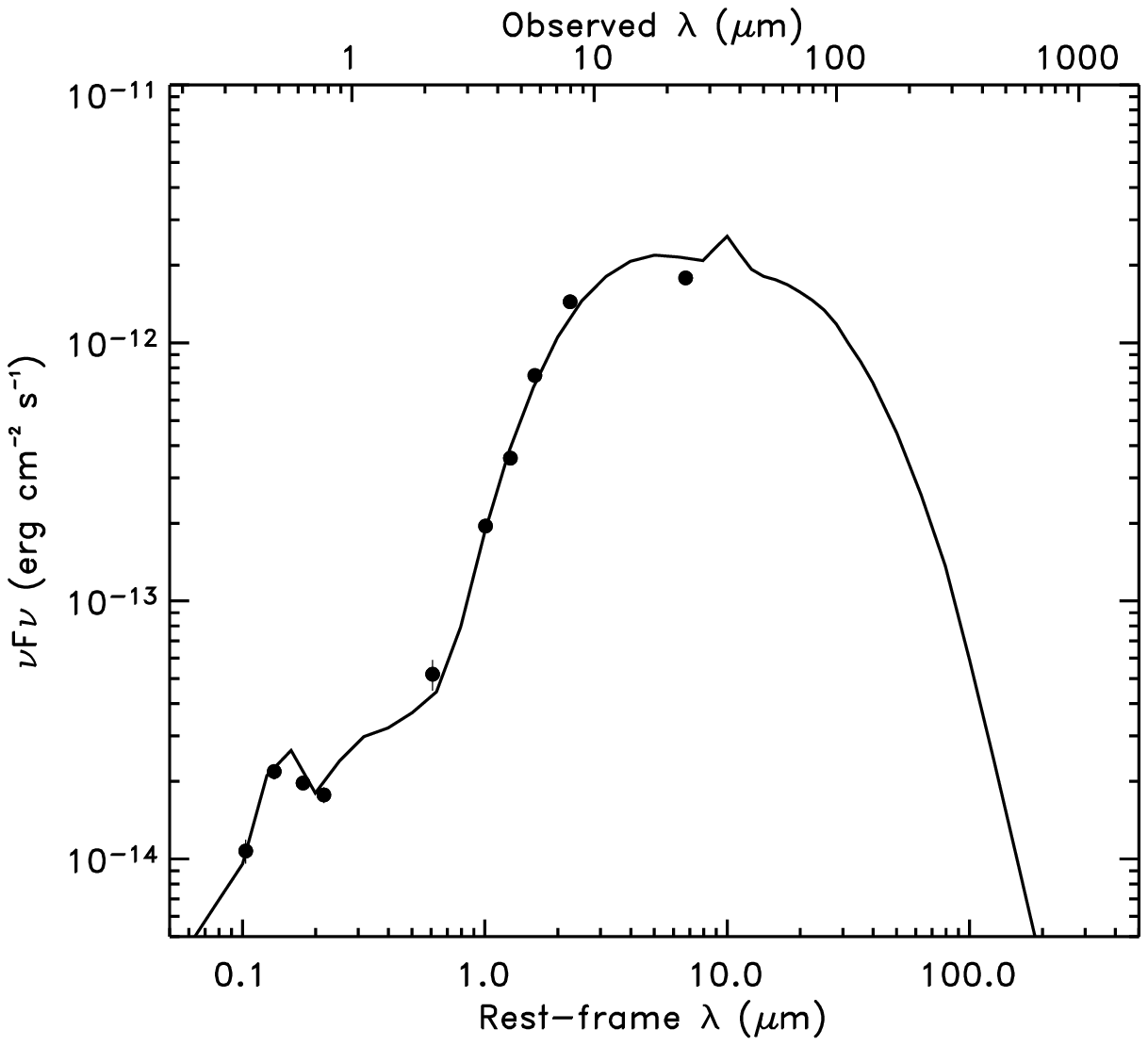}
\caption{\footnotesize Observed SED of SW104409 (black circles) compared to a model
of tapered disk with opening angle of 60\deg, optical depth $\tau_V$=129 and line of sight
inclination of 61\deg (black solid curve).}
\label{ae_model}
\end{figure}

Different geometries of the obscuring material were investigated by
comparing the observed SED of SW104409 with predictions from radiative
transfer models within the AGN unification scenarios. The models assume that
the absorbing material is distributed in a toroidal shape around the central
heating source. Two geometries were taken into account, flared and tapered
disks, as described in detail in~\citet{andreas95}. In flared disks the
thickness of the disk increases linearly with distance from the central
source. In tapered disks the thickness of the disk in the inner part also
increases linearly with distance from the source but stays constant in the
outer part. We find that a good fit to the rest-frame ultraviolet to mid-IR
spectrum of SW104409 can be obtained with a tapered disk with an opening
angle of 60 degrees. The predicted SED for this model is shown in
Figure~\ref{ae_model} and compared to the observed SED of SW104409. The
predicted inclination of the line of sight with respect to the disk axis (61
degrees) implies that it is almost grazing the boundaries of the torus. The
line of sight optical depth at 1000 \AA\ rest-frame through the torus is 700
which corresponds to an optical depth $\tau_V$=129 or A$_\mathrm{V}$=140.
The best-fit model also assumes that the emission from the torus suffers
additional extinction of 0.5 magnitudes by dust which is located at some
distance from the nucleus; e.g. in the host galaxy. The predicted
optical emission, which well agrees with the observed optical data, is
produced by light scattered by the torus.


Due to the high optical depth to the central source, the optical emission
must be scattered light. According to this model, scattering takes place in
the inner walls of the disk and its surface, therefore, in order to observe
scattered light the inclination must be slightly larger than the opening
angle of the disk. In case of larger inclination angles, the disk would
obscure the scattered light. For inclination angles smaller than the opening
angle, the scattered light would still be visible, however this geometry
would be inconsistent with the obscured nature of this source. The
similarity between the inclination and opening angles is also suggested
by the observed 1--10\micron\ SED, characteristic of emission from hot
dust. Since the scattering region and the hot dust region are almost
co-spatial, the inclination must be similar to the opening angle,
otherwise the disk would obscure the emission from this region. An
alternative to this model is scattering produced by electrons or dust in the
opening cone of the accretion disk, instead than from the walls of the disk.
In this case the similarity between the opening and inclination angles is
not required, however it would still be required to explain the shape of the
near- to mid-IR continuum. This model is consistent with the simple scenario
described above. The large difference in the amount of extinction is likely
due to the different assumptions made for the dust distribution. In the
simple model of scattering+obscured QSO, we assume a foreground screen of
dust absorbing the emitted radiation. This assumption does not take into
account re-emission from dust and produces an underestimation of the dust
opacity. Radiative transfer models are more accurate in estimating the dust
opacity since they take into account the transmission of the intrinsic
radiation through the dusty material. According to the results from the
radiative transfer model and assuming a Galactic dust to gas ratio, the
estimated gas column density surrounding the central regions is
$\sim$2.6$\times$10$^{23}$ \cm2, consistent with what is observed in the
X-ray. However, since the X-ray emission is produced in the
nucleus and the optical and near-IR radiation are emitted in the outer
regions, the absorbing gas intercepted by the X-ray photons is expected to
be closer to the disk plane and have a larger column density than the
material intercepted by the optical/near-IR radiation. In spite of its
simplicity, the proposed picture reconciles the multiwavelength observations
of SW104409 and agrees well with the unification model.

\subsection{SW104406}

\begin{figure}[!ht]
\plotone{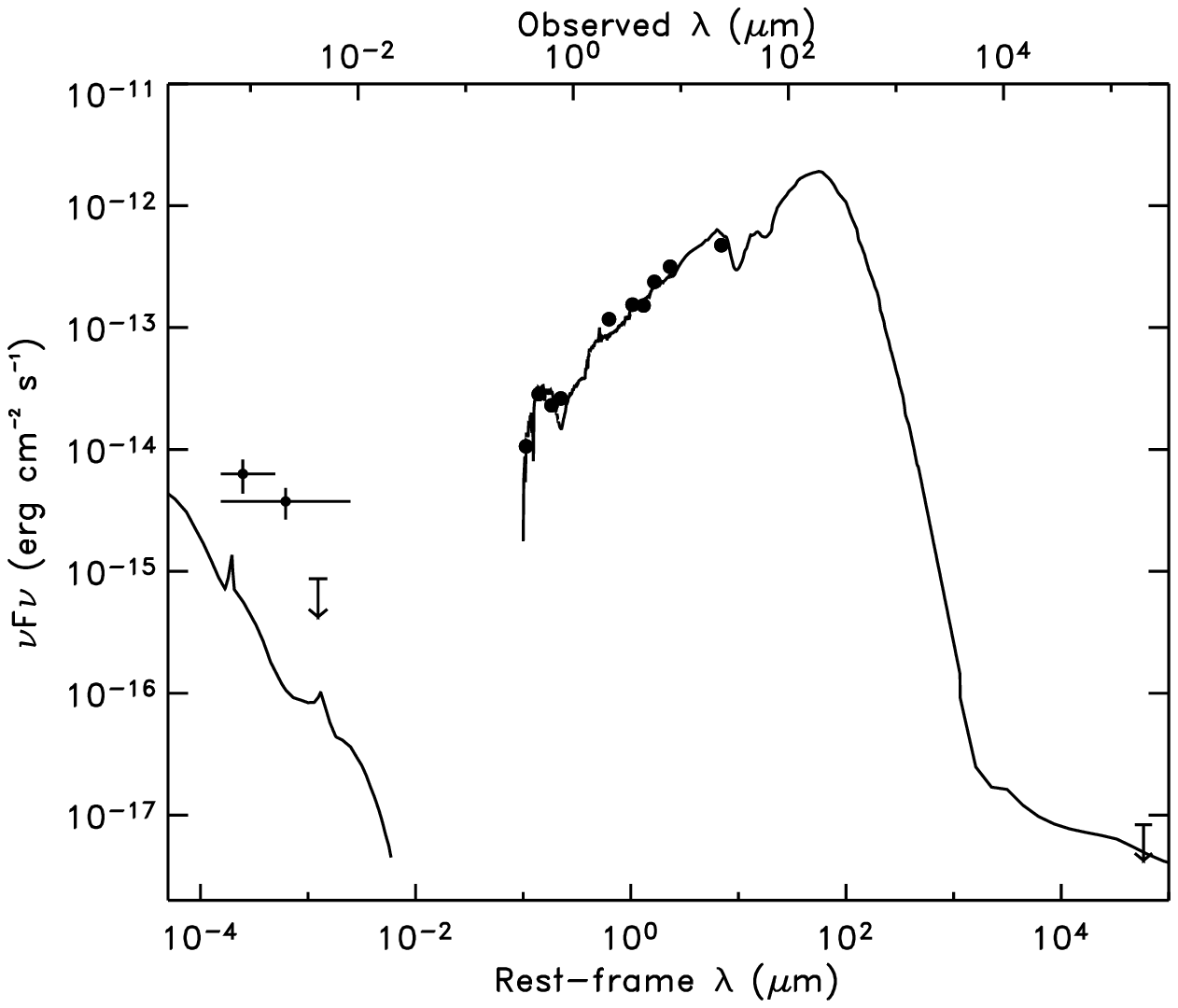}
\caption{\footnotesize SED of SW104406 (black circles) compared to the SED of the
Compton-thick Seyfert 1 Mrk 231 normalized at 24\micron\ with
an A$_{V}$=0.2 mag additional extinction (black curve).  The crosses correspond to the observed X-ray
flux in the broad, and hard X-ray bands. The downward arrow is the 2$\sigma$
upper limit to the soft-X-ray flux. The X-ray flux energy range is indicated
by the length of the horizontal line. The X-ray fluxes are derived assuming
an absorbed power-law model with photon index, $\Gamma$ equal to 1.7 and
Galactic absorption, \nh=6$\times$10$^{19}$\cm2.}
\label{figSED2}
\end{figure}

The spectral energy distribution (SED) of SW104406 is shown in
Figure~\ref{figSED2}. The optical spectrum is characterized by a blue
continuum as in the case of SW104409 with no indication for intrinsic
reddening. As in the case of SW104409, the optical data are characterized by
a U-band drop-out which is probably caused by IGM attenuation. SW104406 also
fits within the conventional definition of EROs with an observed
\rp-$K_s$=4.82 (Vega). Its SED is very similar to the SED of Mrk 231 after
applying an additional extinction of A$_{\rm V}$=0.2 as shown in
Figure~\ref{figLUM2}. The optical spectrum can be fitted with this reddened
template and an additional component, as scattered light, is not required by
the data. The X-ray over IR ratio of SW104406 is about 10 times higher than
for Mrk 231, but the X-ray spectrum is similarly hard or even harder.
Contamination from stellar light emission from the host galaxy or a
starburst, which would appear as a broad bump with a peak at
$\lambda^{rest}\sim$1.6\micron\ in not observed.


In summary, both sources show SEDs where the AGN dominates at all observed
wavelengths and any contribution from other energy source is negligible. The
observed SED are both consistent with a scenario in which the optical and
near-IR sources are obscured. In the X-rays, the observed spectrum is likely
due to the transmitted direct component seen through Compton-thick matter.

\section{Luminosity, Black-Hole Mass and Accretion Rate of SW104409 and
SW104406} 
\label{lum_mbh}

\begin{figure}[!ht]
\plotone{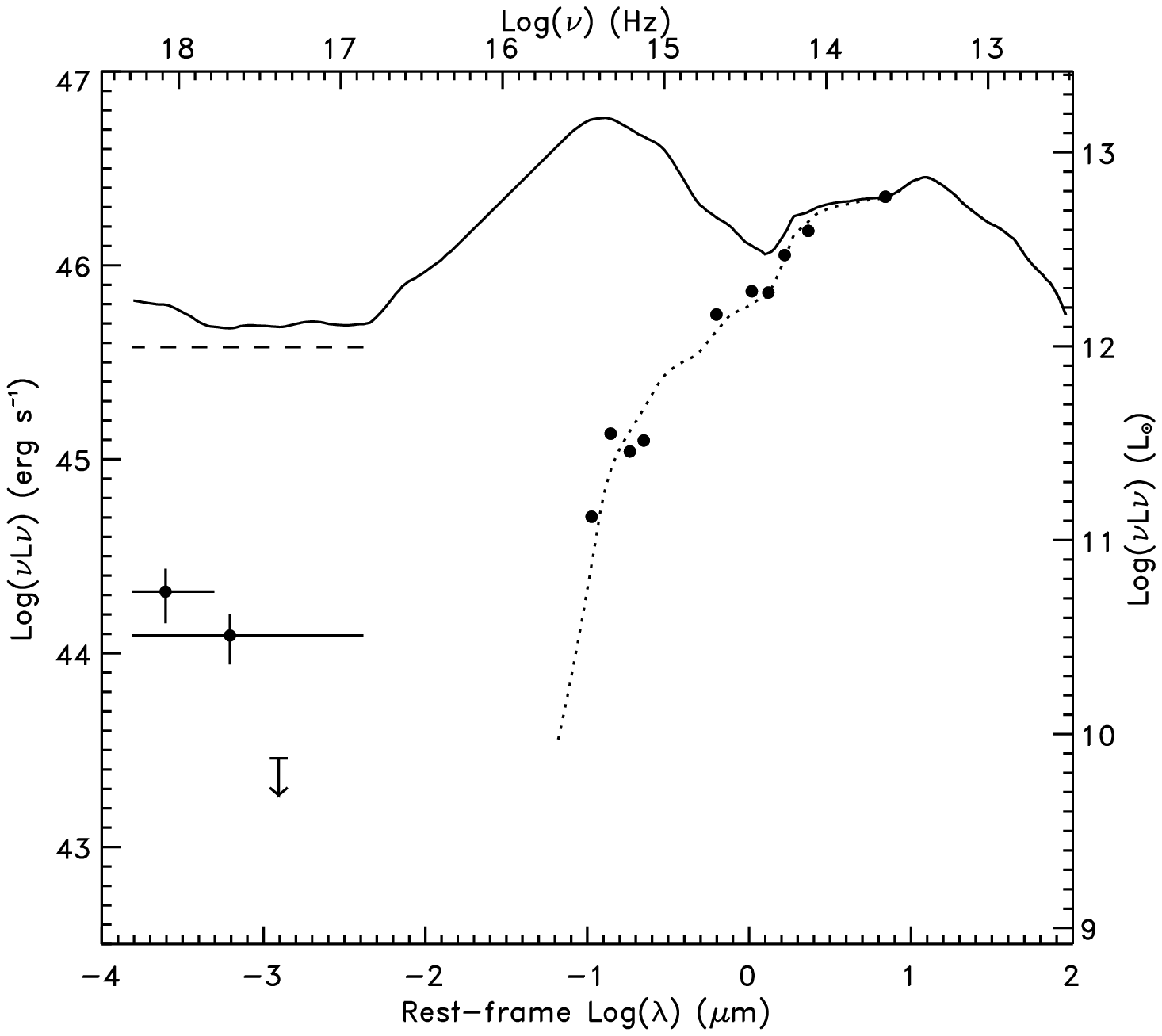}
\caption{\footnotesize SED in $\nu L\nu$ of SW104406 (black full circles) compared to
Elvis QSO template normalized at 24\micron\ in two cases: 1) with no
additional extinction (black solid curve) and 2) with 1.7 mag additional extinction
(black dotted curve). The dashed line corresponds to the absorption-corrected broad
band X-ray luminosity. The crosses correspond to the X-ray
luminosities in the broad, and hard X-ray bands. The downward arrow is the 2$\sigma$
upper limit to the soft-X-ray luminosity. The X-ray energy range is indicated
by the length of the horizontal lines.}
\label{figLUM2}
\end{figure}

\begin{figure}[!ht]
\plotone{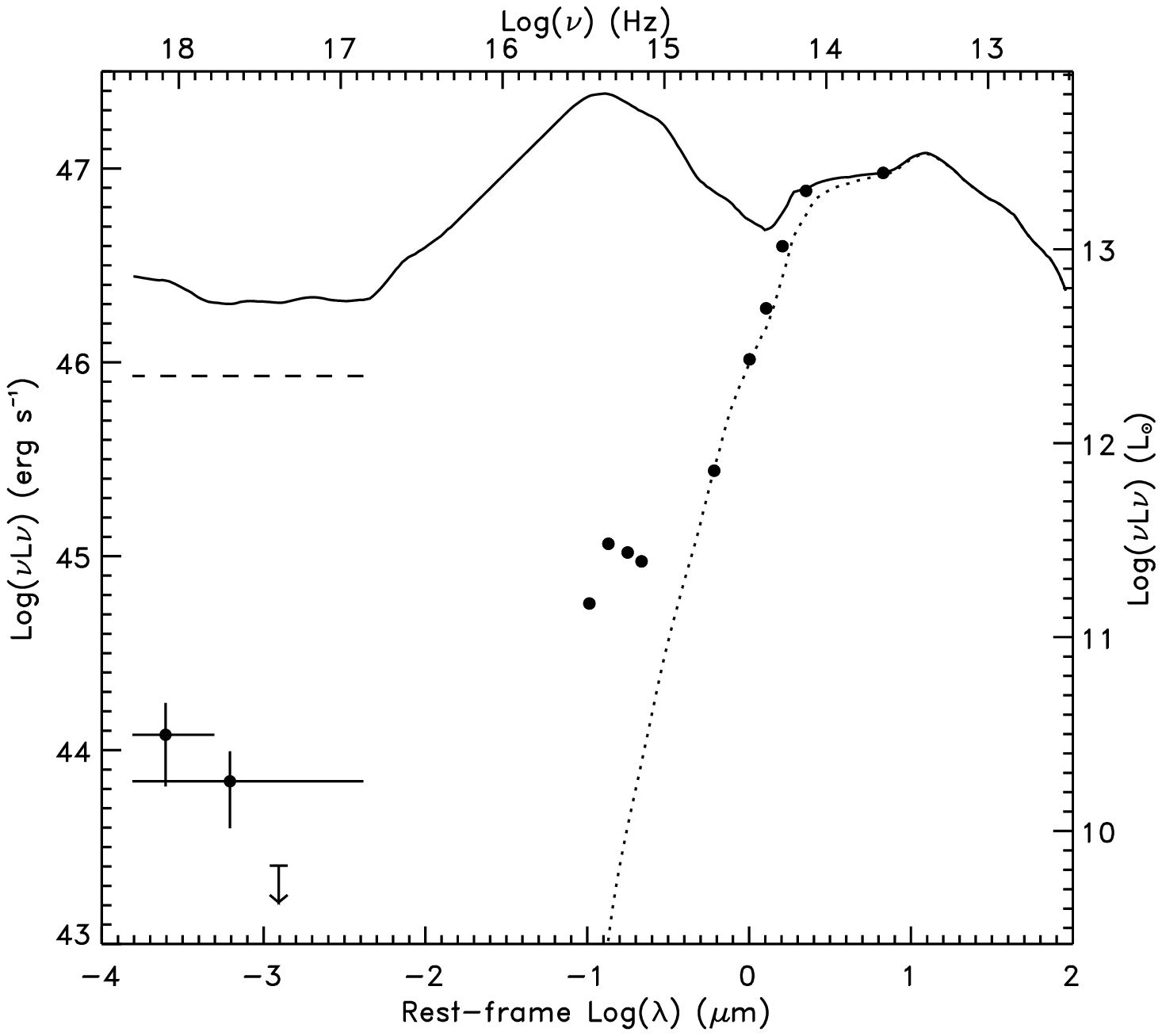}
\caption{\footnotesize SED in $\nu L\nu$ of SW104409 (black full circles) compared to
Elvis QSO template normalized at 24\micron\ in two cases: 1) with no
additional extinction (black solid curve) and 2) with 4.0 mag additional
extinction (black dotted curve). The dashed line corresponds to the absorption-corrected
broad band X-ray luminosity.}
\label{figLUM1}
\end{figure}

The luminosities as a function of wavelength of SW104406 and SW104409 are
shown in Figures~\ref{figLUM2} and~\ref{figLUM1}, respectively. The
luminosity distribution of the unobscured quasar template, ``Elvis
QSO''~\citep{elvis94} normalized to the mid-IR flux of the two quasars is
also shown for comparison. The template is also shown after applying enough
extinction to reproduce the red IR SEDs of SW104409 and SW104406.


Due to the lack of data at $\lambda_{rest}>$7$\mu m$ we cannot directly
measure the total IR luminosity, but we can derive it making some
assumptions about the SED shape. Assuming the model shown in
Figure~\ref{extinction} for SW104409 and a reddened (A$\mathrm{V}$=0.2)
Mrk 231 for SW104406 at $\lambda= 7-1000 \mu m$, the IR luminosities, L(3-1000
$\mu m$), are 3.4$\times 10^{47}$~\ergs\ (=8.8$\times 10^{13}$~\lsun) and 1.2$\times 10^{46}$~\ergs\ (=3.1$\times
10^{12}$~\lsun), respectively. SW104409 is a hyper-luminous IR galaxy
(HYLIRG) and SW104406 an ultra-luminous IR galaxy (ULIRG)~\citep{sanders96}.
The integrated luminosities in different wavelength ranges, as well as the
bolometric luminosity of both sources are listed in Table~\ref{lum_tab}. The
radio luminosity is derived assuming the radio spectral index typically
observed in AGN, $\alpha=-0.8$~\citep{condon88}, where $F_\nu\propto
\nu^\alpha$. The implied rest-frame 1.4~GHz radio luminosity density of
SW104409 is $L_{\rm 1.4 GHz} = 5.3 \times 10^{31}$~\ergsHz, and less than
3.0$ \times 10^{31}$~\ergsHz\ for SW104406. These high radio powers are a
signature of the presence of an AGN.

The bolometric luminosity can be derived by adding the absorption-corrected
X-ray luminosity, the estimated radio luminosity (only in the case of
SW104409) and the optical-IR luminosity. The derived bolometric luminosities
are 3.9$\times 10^{47}$~\ergs\ (1.0$\times 10^{14}$ \lsun) for SW104409 and
1.9$\times 10^{46}$~\ergs\ (4.6$\times 10^{12}$ \lsun) for SW104406. Because
of the uncertainty associated with the lack of data above
$\lambda^{rest}$=7$\mu$m, we derived a lower limit to the bolometric
luminosity by replacing the optical-IR luminosity $L(0.03-1000\mu m)$ with
$L(0.03-10\mu m)$. The obtained values are a factor of 2 lower than the
estimates obtained by extrapolating the model to $\lambda>$10\micron\ (see
values in Table~\ref{lum_tab}). The absorption-corrected X-ray luminosities
of SW104409 and SW104406 correspond, respectively, to $\sim$1\% and 30\%\ of
their bolometric luminosity.

Assuming that the AGN is the main source of the observed bolometric
luminosity and that it is accreting at the Eddington limit
($L_{bol}=L_{Edd}$), the estimated black-hole mass for SW104409 and SW104406
is $\simeq 2.9\times 10^9$ \Msun\ and 1.4$\times 10^8$ \Msun, respectively,
or 1.2$\times 10^9$ \Msun\ and 6.6$\times 10^7$ \Msun, if we do not include
the energy emitted at $\lambda>$10\micron\ in the estimate of the AGN
bolometric luminosity. If we assume an accretion efficiency of 10\%, the
derived accretion rates are 68~\Msun\ yr$^{-1}$ for SW104409 and 3.4~\Msun\
yr$^{-1}$ for SW104406, or 28~\Msun\ yr$^{-1}$ for SW104409 and 1.5~\Msun\
yr$^{-1}$ for SW104406 if we do not include the energy emitted at
$\lambda>$10\micron\ in the estimate of the AGN bolometric luminosity. The
black-hole mass of SW104409 is comparable to the highest measured values in
the local Universe, i.e. M87~\citep{ford94} and Cyg A~\citep{tadhunter03}
having M$_{\rm BH} \simeq$3$\times 10^9$ \Msun. Its accretion rate is also
among the highest observed in quasars at $z\leq2$~\citep{mclure04}. SW104406
is also characterized by a large black-hole mass and accretion rate, but it
is not as extreme as SW104409. Since most of the quasars accrete below their
Eddington limit~\citep{mclure04}, it would be more realistic to assume a
lower Eddington ratio, however, this would imply even higher black hole
masses.

\section{Comparison with other Compton-thick AGN at $z\gtrsim$2}
\label{comp_ctagn}

Only a few Compton-thick quasars at $z\gtrsim$2 are currently known. Here,
we compare their properties with those of SW104409 and SW104406
and investigate whether they represent the same population or if the different
selection methods are finding objects with different properties. The largest
and probably best sampled sample of Compton-thick AGN currently known
contains 4 sources~\citep{alexander05a,alexander05b}. These
sources were drawn from an X-ray detected sub-millimeter selected sample in
a 0.12 $deg^2$ field. Optical data from HST, IR data from \spitzer, radio data
and optical high-resolution spectra are available for all of the sources. 
Spectroscopic redshifts range from 2 to 2.5 and broad-band (0.5-8 keV) X-ray
fluxes range from 0.7 to 1.3$\times$10$^{-15}$~\ergcm2s. These sources host
both an AGN and a powerful starburst, and their optical-near-IR SEDs are
dominated by stellar light~\citep{borys05}. The optical spectra of all
sources are also dominated by a starburst component~\citep{chapman05}. The
AGN bolometric luminosities, estimated from the observed X-ray flux after
correcting it for absorption, range from 2.2 to 4.4 $\times 10^{11}$ \lsun,
the derived black-hole masses range from 0.6 to 1.6 $\times 10^{7}$ \msun\
and the accretion rates vary from 0.13 to 0.35
\msun/$yr$. The main differences between this sample and the two obscured
quasars SW1044090 and SW104406 is in the AGN bolometric luminosity, and thus
in the accretion rates and the black-hole masses which are, on average, two
order of magnitudes lower. Another important difference is in the presence
of a dominant starburst component which is absent or negligible in SW104409
and SW104406. It is clear that the two quasars presented in this work and
these AGN/sub-millimeter galaxies show some major differences and might be
very different objects. However, before deriving any conclusions on the
differences between these two samples, we should evaluate how the different
methods applied to estimate the AGN bolometric luminosity affect these
results.

We tried to reproduce~\citet{alexander05b}'s estimates using their energy
ranges and found that their correction for absorption to the rest-frame
X-ray luminosity is about 4 times smaller than ours for \nh=10$^{24}$\cm2.
The most likely explanation for the observed difference is the difference
in the assumed X-ray model used to derive the K-correction.
~\citet{alexander05a,alexander05b} adopt a model which includes an absorbed
power-law component with $\Gamma$=1.8, a neutral reflection component, a
scattered component of ionized gas, and a Fe K$\alpha$ mission line at 6.4
keV. We also notice a discrepancy in the column density estimates, their
values being on average 2.5 times larger than what we would predict with our
method. Although this would yield larger luminosities, their correction factor
due to the different spectral model is much smaller than ours, thus their
absorption-corrected rest-frame luminosities are lower that what we would
estimate. If we apply our method to their sample, the derived
absorption-corrected rest-frame X-ray luminosities are, on average, higher
by a factor of 8 (from 3 to 12.5 times higher). Their smaller X-ray
luminosities imply smaller AGN bolometric luminosities, black-hole masses
and accretion rates by the same factors. A third difference between our
methods is in the derivation of bolometric luminosity. They assume a
constant factor between the absorption-corrected rest-frame X-ray luminosity
and the bolometric luminosity of 6\%, while we add the luminosity measured
throughout the whole spectrum (30\%\ for SW104409 and 1\%\ for SW104406; see
Section~\ref{lum_mbh}). Even after correcting by the factors described
above, the differences in AGN luminosity, SMBH mass and accretion rate
between the two quasars presented in this work and the sample
in~\citet{alexander05a,alexander05b} are still significant, more than one
order of magnitude compared to almost two order of magnitude initially
measured. Thus, we conclude that the four sources in~\citet{alexander05b}
and our two quasars are different. 

Other examples of Compton-thick AGN at high-$z$ are the type 2 quasars,
CXO-52 ($z$=3.288; ~\citet{stern02}), CDFS-202 ($z$=3.700; ~\citet{norman02})
and CDFS-263 ($z$=3.660; ~\citet{mainieri05}).  The former two sources show
similar characteristics to SW104409 and SW104406. The X-ray selected
type 2 QSO detected at sub-millimeter wavelengths, CDFS-263, is instead more
similar to the sub-millimeter selected AGN discussed above. The
SEDs of these quasars at $\lambda>$2.5\micron\ are not currently available,
but at shorter wavelengths CXO-52 and CDFS-202 show very similar SEDs to
those of SW104409 and SW104406 with AGN-dominated optical spectra and red
optical-near-IR colors. Assuming that the absorption-corrected X-ray
luminosity corresponds to 10\% of the AGN bolometric
luminosity~\citep{elvis94}, we derive a bolometric luminosity,
$L_{bol}$, of 3.3$\times$10$^{45}$ \ergs\ for CXO-52 and of
2$\times$10$^{46}$ \ergs\ for CDFS-202. Assuming accretion at the Eddington
limit, the black-hole masses are 2.5$\times$10$^{7}$ \msun\ and
1.1$\times$10$^{8}$ \msun\, respectively. These values are also lower than
what we derive for SW104409 and similar to those derived for SW104406, but
higher that those measured in~\citep{alexander05a}'s sub-millimeter selected
Compton-thick AGN. In CXO-52 and CDFS-202, as in SW104409 and SW104406, the
AGN dominates over the host galaxy and the associated luminosity and SMBH
mass are one order of magnitude higher than observed in sub-millimeter
selected Compton-thick AGN.

\begin{figure}[!ht]
\plotone{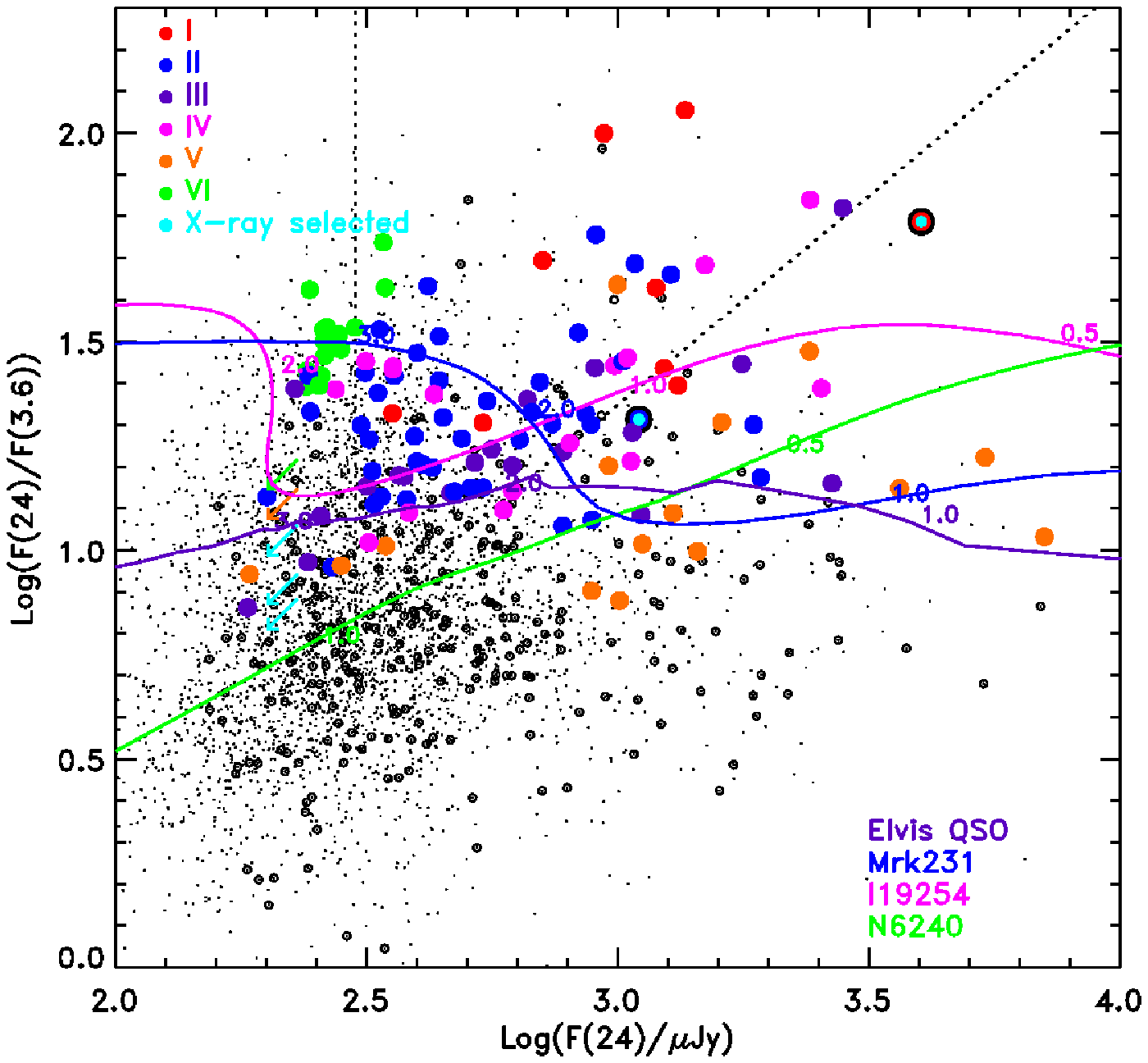}
\caption{\footnotesize IR color F(24$\mu$m)/F(3.6$\mu$m) $versus$ the IR flux at
24$\mu$m for the IR (black dots) and X-ray (black open circles) sources detected 
at 3.6 and 24 $\mu$m in the \chandra/SWIRE field. IR-selected obscured AGN
candidates are shown as full circles with colors corresponding to
different SED types (I:red, II:blue, III:purple, IV:magenta, V:orange, and
VI:green). Symbols as in Figure~\ref{optx}. X-ray selected obscured AGN  are shown
as cyan full circles and as cyan arrows if non-detected at 24$\mu$m. 
The dotted line defines the IR selection criteria used
in~\citet{martinez05} for obscured AGN at high-$z$. }
\label{f51_f5}
\end{figure}

The two Compton-thick AGN discussed in this work also differ from a sample
of high-$z$, heavily obscured, and luminous AGN candidates selected at IR
and radio wavelength~\citep{martinez05}. We applied the same selection
criteria to the complete IR and X-ray samples, to the IR-selected
obscured AGN candidates and to the X-ray selected Compton-thick AGN in
Figure~\ref{f51_f5}. Only one source among the IR-selected obscured AGN
candidates satisfies their selection criteria, SWIRE\_J104641.38+585213.9
and none of the X-ray selected Compton-thick AGN. Although the two
Compton-thick AGN, SW104409 and SW104406, are luminous and obscured AGN at
$z>$2 (see section~\ref{qso2_obs}), they do not pass either the 3.6$\mu$m
or the radio limit (see large black circles in Figure~\ref{f51_f5}) required
by~\citet{martinez05}. In the entire field only 6 sources satisfy their
selection criteria, of which 2 show power-law like IR SEDs and are
also detected in the X-ray. One is SWIRE\_J104641.38+585213.9 which we classify
as class III source (see table~\ref{ctagn_ir}), the other source is
characterized by an SED consistent with an unobscured QSO. The remaining 4
sources are characterized by SEDs more similar to those in class IV.
Although both selection methods are based on IR-colors derived from
\spitzer\, there is little overlap among the two samples. 


A compilation of all heavily obscured AGN selected by different methods and
a systematic analysis of their multi-wavelength properties is necessary to
understand the differences among these sources and their origin, e.g. a
different evolutionary stage, SMBH mass, accretion rate or environment.

\section{Statistical predictions for Compton-thick AGN} 
\label{surf_dens}

The two selection methods described in Sections~\ref{ctagn_x_sec}
and~\ref{ctagn_ir_sec} were defined for samples of obscured AGN based on
their X-ray and optical-IR properties, respectively. The X-ray selected
sample includes X-ray detected AGN with hard X-ray spectra and estimated
column densities $\geq$10$^{24}$ \cm2 (5 sources). The IR-selected sample
contains AGN with red optical-IR colors and whose emission dominates in the
near- and mid-IR over the host galaxy light (120 sources). The first
selection method is biased against Compton-thick AGN whose primary X-ray
radiation is completely absorbed in the observed energy range, i.e. sources
at low-$z$ or with column densities $\geq$10$^{25}$ \cm2, and against
sources that are fainter than 10$^{-15}$ \ergcm2s\ at 0.3-8 keV. The second
method is biased against AGN that are too faint to be detected in 3 IR bands
at our sensitivity limits or whose host galaxy is brighter than the AGN in
the optical-IR wavelength range. Because of these selection effects, we can
only estimate a lower limit to the surface density of IR detected
Compton-thick AGN in the \chandra/SWIRE field.

In the sub-sample of 41 IR-selected obscured AGN candidates which are
detected in the X-rays, only two sources, $\sim$5\%, are confirmed
Compton-thick AGN, SW104409 and SW104406. If we assume that this fraction
does not depend on the X-ray flux and is the same for the entire IR selected
sample of obscured AGN candidates (but see discussion in
Section~\ref{irx_sec}), we estimate about 6 sources (5\%\ of 120 sources) to
be Compton-thick AGN. Since only two out of five X-ray selected
Compton-thick AGN are also selected by the IR-selected criteria, we assume
that our IR selection identifies only 40\%\ (2 out of 5) of all
Compton-thick AGN in the field. Thus, the estimated total number of Compton-thick
AGN in the field is about 15 ($\simeq$6/0.4) or 25 sources deg$^{-2}$. The
estimated number of sources (15) should be considered a lower limit to the
total number of Compton-thick AGN detected in the IR at our sensitivity
limits in the \chandra/SWIRE field. Due to the lack of X-ray data for most
of the sources, the identification of these 15 Compton-thick AGN is not
possible because of lack of constraints on their column densities. The only
exceptions are the 5 X-ray selected Compton-thick AGN presented in
Section~\ref{ctagn_x_sec}. The fraction of Compton-thick AGN detected in the
X-ray at our depth (F(0.3-8 keV)$\geq$10$^{-15}$\ergcm2s) is 33\% (5 out of
15). This fraction is consistent with the value estimated by
\citet{treister04} of 30\%\ of all Compton-thick AGN detected at the X-ray
limit of the \chandra\ deep surveys.

We compared our results with the number of obscured AGN detected in the
IR and in the X-rays predicted by various
models~\citep{silva04,treister05,xu03,polletta03}. Since our selection is
very similar to a 24\micron\ flux selection, we compared our results with
those predicted for AGN with a 24\micron\ flux greater than our 5$\sigma$
limit, 230$\mu$Jy. However, our selection is more restricted due to the
additional requirements on the properties of the IR SED and, therefore, we
expect our numbers to be lower than those expected from a limited-flux
sample of AGN. \citet{silva04} predicts about 670 AGN deg$^{-2}$, of which 500 are
obscured (\nh$\geq$10$^{22}$\cm2) (obscured:unobscured=2.9:1) and about 300
are Compton-thick.
\citet{treister05} predicts about 1000 AGN deg$^{-2}$, of which 765 are
obscured (\nh$\geq$10$^{22}$\cm2) (obscured:unobscured=3:1) and no
prediction for Compton-thick AGN is given.
\citet{xu03} and~\citet{polletta03} predict about 1100 AGN deg$^{-2}$, of
which 810 are obscured (\nh$\geq$10$^{22}$\cm2) (obscured:unobscured=2.8:1).
Although these models are different and have not been fully tested with
respect to the available IR and X-ray observables, they are all consistent
with a surface density of about 700--1000 AGN deg$^{-2}$, a ratio between
obscured and unobscured AGN of 3:1, and about 300 Compton-thick AGN. Our IR
selection of AGN yields 120 ($\sim$200 deg$^{-2}$) obscured AGN candidates
and 61 ($\sim$100 deg$^{-2}$) unobscured AGN candidates, corresponding to
only about 20\% of the expected number of AGN, and to an obscured to
unobscured ratio of $\sim$ 2:1. The total number of AGN and the ratio
between obscured and unobscured AGN would likely increase if the selection
was not limited to sources with AGN-dominated and red IR SEDs. Our lower
limit to the surface density of Compton-thick AGN is less than 10\%\ the
expected value, indicating that even in the mid-IR a large fraction of these
sources are elusive.

\section{Conclusions} 
\label{conc}

Theoretical models, simulations and indirect observations predict the
presence of a large population of highly obscured luminous AGN at moderate
and high redshifts. The quest for these objects has been hindered by the
difficulty of detecting and identifying them and, thus, very few examples
are currently known. In this work, we show how the combined high sensitivity
and spatial resolution of the \spitzer\ Space Telescope and of the \chandra\
X-ray Observatory are overcoming these difficulties, offering the capability
of detecting and identifying heavily obscured AGN up to high redshifts,
providing constraints on their surface density and characterizing their
properties.

Using the dataset in the \chandra/SWIRE field (0.6 deg$^2$ in the Lockman
Hole), we conducted a search for Compton-thick AGN. We selected,
independently, two samples of Compton-thick AGN candidates based on their
X-ray spectral properties and optical-IR SED. The X-ray and IR selected
samples contain, respectively, 5 and 120 sources. Based on the properties of
the X-ray selected Compton-thick AGN, we estimate that only 40\%\ of the
population of heavily obscured AGN shows distinct AGN signatures in their
optical-IR SEDs, the rest being dominated by the host-galaxy emission. The
number of Compton-thick AGN detectable in the \chandra/SWIRE field is of at
least 25 sources deg$^{-2}$ and only 33\% of them are detected in the X-ray
down to a flux limit of 10$^{-15}$\ergcm2s. The X-ray over optical flux
ratios of the selected obscured AGN cover a similar range as the rest of the
X-ray sample and well overlap with those observed for ``classical'' AGN,
indicating that searches for heavily obscured AGN among X-ray sources with
extreme X-ray over optical flux ratios might miss a large fraction of these
sources.

The complete sample of AGN selected on the basis of a red power-law like IR
SED contains 181 sources, of which 61 show SEDs similar to those of
unobscured AGN, and 120 are obscured AGN candidates. The complete sample
corresponds to only about 20\%\ of the total number of AGN expected to be
detected in the mid-IR at our sensitivity. The observed ratio between
obscured and unobscured AGN is 2:1, still lower than the models predictions,
$\sim$3:1. The estimated surface density of Compton-thick AGN is only 10\%
the expected value. These results suggest that even in the mid-IR, heavily
obscured AGN are elusive.

Optical spectroscopy of two obscured AGN candidates confirmed that they are
high redshift, SW104409 at $z$=2.54 and SW104406 at $z$=2.43, Compton-thick
QSOs. These sources, characterized by type 2 AGN spectra, very red
optical-IR SEDs and high X-ray luminosities ($\sim 10^{45}$ \ergs), have
properties (SEDs, luminosities and SMBH masses) similar to those observed in
the high-$z$ Compton-thick quasars CXO-52 ($z$=3.288;~\citet{stern02}), and
CDFS-202 ($z$=3.700;~\citet{norman02}). However, they differ from
sub-millimeter selected Compton-thick AGN~\citep{alexander05b} where
emission from a starburst dominates at most wavelengths and black-hole
masses and accretion rates are one order of magnitude lower. SW104409
represents the most luminous ($\sim 10^{14}$ \lsun) Compton-thick QSOs at
high-$z$ currently known with a SMBH mass comparable to the most massive
local SMBH ($\sim$3$\times 10^{9}$\msun). Such a rare object can be found
when large volumes are sampled as in the SWIRE/\chandra\ field
($\sim$0.6\sqdeg).

\acknowledgments

M.P. thanks Alain Omont and Dan Weedman for useful discussions. This work is
based on observations made with the {\it Spitzer Space Telescope}, which is
operated by the Jet Propulsion Laboratory, California Institute of
Technology under NASA contract 1407. Support for this work, part of the {\it
Spitzer Space Telescope} Legacy Science Program, was provided by NASA
through an award issued by the Jet Propulsion Laboratory, California
Institute of Technology under NASA contract 1407. MP, BW and RK are grateful
for the financial support of NASA grant G04-5158A (\chandra). BW is grateful
for the financial support of NASA contract NAS8-39073 (\chandra X-ray
Center).
                                                                                                        
This research makes use of the NASA/IPAC Extragalactic Database (NED) which
is operated by the Jet Propulsion Laboratory, California Institute of
Technology, under contract with the National Aeronautics and Space
Administration.
Based on observations obtained at the Hale Telescope, Palomar Observatory as
part of a continuing collaboration between the California Institute of
Technology, NASA/JPL, and Cornell University.
This publication makes use of data products from the Two Micron All Sky
Survey, which is a joint project of the University of Massachusetts and the
Infrared Processing and Analysis Center/California Institute of Technology,
funded by the National Aeronautics and Space Administration and the National
Science Foundation.

Facilities: \facility{Spitzer(IRAC,MIPS)}, \facility{CXO(ACIS)}.






%
\clearpage

\begin{deluxetable}{lcccllrl}
\tabletypesize{\scriptsize}
\rotate
\tablecaption{Summary of observations in the \chandra/SWIRE Survey\label{obs_log}}
\tablewidth{0pt}
\tablehead{
\colhead{Telescope/Instrument} & \colhead{$\alpha_{2000}$}& \colhead{$\delta_{2000}$} & \colhead{Area}& \colhead{Observing date} & \colhead{Band}& \colhead{Exp. Time} & \colhead{5$\sigma$ limit}  \\
\colhead{                    } &  \colhead{(\hr $\;\;\;$\min)} & \colhead{(\deg $\;\;\;$\arcmin)} &  \colhead{  }  & \colhead{} & \colhead{ }                & \colhead{(sec)    } & \colhead{$\mu$Jy/Vega mag}
}
\startdata
KPNO: Mayall 4mt/Mosaic        & 10 46                & +59 00       &30\arcmin$\times$30\arcmin & 2004 Jan                 & U                            & 21,600              & 24.8                \\
KPNO: Mayall 4mt/Mosaic        & 10 46                & +59 03       &60\arcmin$\times$54\arcmin & 2004 Jan                 & U                            &   7200              & 24.3                \\
KPNO: Mayall 4mt/Mosaic        & 10 46                & +59 00       &30\arcmin$\times$30\arcmin & 2002 Feb                 & \gp, \rp, \ip                & 10,800              & 25.9, 25.2, 24.4    \\
KPNO: Mayall 4mt/Mosaic        & 10 46                & +59 03       &60\arcmin$\times$54\arcmin & 2002 Feb                 & \gp, \rp                     &   3000              & 25.2, 24.4          \\
KPNO: Mayall 4mt/Mosaic        & 10 46                & +59 03       &60\arcmin$\times$54\arcmin & 2002 Feb                 & \ip                          &   1800              & 23.5                \\
KPNO: Mayall 4mt/Mosaic        & 10 46                & +58 33       &60\arcmin$\times$6\arcmin  & 2002 Feb                 & \gp, \rp, \ip                &   1800              & 23.7, 23.5, 22.9    \\
Palomar: 200\arcsec\ Hale/WIRC & 10 46                & +59 00       &36\arcmin$\times$43\arcmin & 2004 Mar 29              & K$_s$                        &   4320              & 20.5                \\
VLA                            & 10 46                & +59 01       &40\arcmin$\times$40\arcmin & 2001, 2002 \& 2003       & 20 $cm$                      & 500,000             & 13.5                \\
\chandra/ACIS-I                & 10 46                & +59 01       &47\arcmin$\times$47\arcmin & 2004 Sep 12-26           & 0.3--8 keV                   & 70,000              & 10$^{-15}$ \ergcm2s \\
\spitzer/IRAC                  & 10 45                & +58 00       & 3.69\deg$\times$3.01\deg  & 2003 Dec \& 2004 Apr     & 3.6, 4.5, 5.8, 8.0 \micron   & 120--480            & 5, 9, 43, 40        \\
\spitzer/MIPS                  & 10 45                & +58 00       & 3.66\deg$\times$3.00\deg  & 2003 Dec \& 2004 May     & 24 \micron                   & 160--360            & 230                 \\
\enddata
\end{deluxetable}


\begin{deluxetable}{c cc c rrrrr c}
\tabletypesize{\scriptsize}
\rotate
\tablecaption{Properties of X-ray selected Compton-thick AGN \label{ctagn_x}}
\tablewidth{0pt}
\tablehead{
\colhead{Source name}  & 
\colhead{$\alpha_{2000}$\tablenotemark{a}} & 
\colhead{$\delta_{2000}$\tablenotemark{a}} & 
\colhead{mag(\rp)} & 
\colhead{$F_{3.6\micron}$} & 
\colhead{$F_{4.5\micron}$} & 
\colhead{$F_{5.8\micron}$} & 
\colhead{$F_{8.0\micron}$} & 
\colhead{$F_{24\micron}$} & 
\colhead{$F_{20cm}$\tablenotemark{b}} \\
\colhead{}      & 
\colhead{(degrees)} & 
\colhead{(degrees)} & 
\colhead{(Vega)}    &
\colhead{($\mu$Jy)} &
\colhead{($\mu$Jy)} &
\colhead{($\mu$Jy)} &
\colhead{($\mu$Jy)} &
\colhead{($\mu$Jy)} & 
\colhead{($\mu$Jy)}
} 
\startdata
 SWIRE\_J104311.22+591128.2  &    160.79675  &    59.191170  &  22.81  &  20  &  20  &  $<$43  &  $<$40  &   $<$230  & $<$87     \\
 SWIRE\_J104322.07+590648.7  &    160.84196  &    59.113541  &  24.83  &  26  &  29  &  $<$43  &  $<$40  &   $<$230  & 64$\pm$17 \\
 SWIRE\_J104406.30+583954.1  &    161.02626  &    58.665039  &  23.35  &  53  &  67  &    131  &    244  &     1099  & $<$162    \\
 SWIRE\_J104407.67+584011.3  &    161.03194  &    58.669800  & $>$25.2 &  30  &  34  &  $<$43  &  $<$40  &   $<$230  & $<$154    \\
 SWIRE\_J104409.95+585224.8  &    161.04146  &    58.873550  &  23.55  &  65  & 152  &    401  &   1082  &     4011  & 273$\pm$15\\
\enddata
\tablecomments{Typical uncertainties to the IR fluxes are around 10\% of the
               measured fluxed and to the optical magnitudes are around 0.04 mag.}
\tablenotetext{a}{IR coordinates}
\tablenotetext{b}{Upper limits correspond to 5$\sigma$.}
\end{deluxetable}

\clearpage
\topmargin=2cm
\footskip=0in
                                                                                          
\begin{deluxetable}{cccccccclllc}
\tabletypesize{\scriptsize}
\rotate
\tablecaption{X-ray properties of X-ray selected Compton-thick AGN \label{ctagn_x_xprop}}
\tablewidth{0pt}
\tablehead{
 \colhead{Source name}&
 \colhead{$\alpha_{2000}$\tablenotemark{a}} &
 \colhead{$\delta_{2000}$\tablenotemark{a}} &
 \colhead{Total counts} &
 \colhead{Flux\tablenotemark{b}}   &
 \colhead{Flux\tablenotemark{b}}   &
 \colhead{Flux\tablenotemark{b}}   &
 \colhead{HR} &
 \colhead{z$_{phot}$} &
 \colhead{\nh$^{obs}$\tablenotemark{c}} &
 \colhead{\nh$^{rest}$\tablenotemark{d}} &
 \colhead{Log(L)\tablenotemark{e}}  \\
 \colhead{}  &
 \colhead{(degrees)} &
 \colhead{(degrees)} &
 \colhead{(0.3-8 keV)} &
 \colhead{(0.3-8 keV)} &
 \colhead{(0.3-2.5 keV)} &
 \colhead{(2.5-8 keV)} &
 \colhead{} &
 \colhead{} &
 \colhead{} &
 \colhead{} &
 \colhead{}
} 
\startdata
 SWIRE\_J104311.22+591128.2  &    160.79686  &    59.191322  &   12$\pm$5    &   20$\pm$8  &  0$\pm$3  &   38$\pm$15  &  1.00$^{+0.0}_{-0.15}$   &   2.27                  & $>$30$^{+40}_{-23}$    &    863$^{+1151}_{-662}$  & 45.9 \\
 SWIRE\_J104322.07+590648.7  &    160.84201  &    59.113667  &   16$\pm$5    &   25$\pm$9  &  0$\pm$3  &   50$\pm$17  &  1.00$^{+0.0}_{-0.04}$   &   1.41                  & $>$30$^{+\inf}_{-10}$  &    312$^{+\infty}_{-104}$& 45.5 \\
 SWIRE\_J104406.30+583954.1  &    161.02638  &    58.665276  &   22$\pm$6    &   37$\pm$11 &  4$\pm$4  &   63$\pm$20  &  0.61$^{+0.21}_{-0.23}$  &   2.43\tablenotemark{f} &     4$^{+3}_{-1}$      &     99$^{+74}_{-25}$     & 45.7 \\
 SWIRE\_J104407.67+584011.3  &    161.03212  &    58.670094  &   12$\pm$5    &   21$\pm$9  &  0$\pm$3  &   41$\pm$17  &  1.00$^{+0.0}_{-0.13}$   &   1.42                  & $>$20$^{+\inf}_{-11}$  & $>$200$^{+\infty}_{-109}$& 45.4 \\
 SWIRE\_J104409.95+585224.8  &    161.04143  &    58.873802  &   11$\pm$5    &   19$\pm$8  &  2$\pm$3  &   33$\pm$15  &  0.85$^{+0.06}_{-0.39}$  &   2.54\tablenotemark{f} &     8$^{+2}_{-5}$      &    214$^{+54}_{-134}$    & 45.6 \\
\enddata
\tablenotetext{a}{X-ray coordinates}
\tablenotetext{b}{X-ray flux in 10$^{-16}$\ergcm2s\ derived assuming an
 absorbed power-law model with photon index, $\Gamma$ equal to 1.7 and
 Galactic \nh\ (6$\times$10$^{19}$ \cm2). Uncertainties reflect only the
 statistical errors from the observed counts and do not include 
 uncertainties in the spectral model.}
\tablenotetext{c}{\nh\ in the observer rest-frame in 10$^{22}$ cm$^{-2}$.}
\tablenotetext{d}{\nh\ in the source rest-frame in 10$^{22}$ cm$^{-2}$.}
\tablenotetext{e}{Logarithm of the 0.3-8 keV absorption-corrected rest-frame luminosity in \ergs.}
\tablenotetext{f}{Spectroscopic $z$. Photometric $z$ are reported in
Table~\ref{ctagn_ir}.}
\end{deluxetable}

\clearpage
\topmargin=0cm
\footskip=1cm

\begin{deluxetable}{c cc r rrrrr r ccc}
\tabletypesize{\scriptsize}
\setlength{\tabcolsep}{0.02in}
\tablecaption{Properties IR-selected obscured AGN candidates \label{ctagn_ir}}
\tablewidth{0pt}
\tablehead{
\colhead{Source name}  & \colhead{$\alpha$\tablenotemark{a}}  & \colhead{$\delta$\tablenotemark{a}}& \colhead{mag(\rp)} & \colhead{$F_{3.6\micron}$} & \colhead{$F_{4.5\micron}$} & \colhead{$F_{5.8\micron}$} & \colhead{$F_{8.0\micron}$} & \colhead{$F_{24\micron}$}  & \colhead{$F_{20cm}$}& \colhead{$z_{spec}$} & \colhead{$z_{phot}$} &\colhead{Class\tablenotemark{b}}  \\
\colhead{    }    & \colhead{(degrees)}             & \colhead{(degrees)}              & \colhead{(Vega)}         &\colhead{($\mu$Jy)}&\colhead{($\mu$Jy)}&\colhead{($\mu$Jy)}&\colhead{($\mu$Jy)}&\colhead{($\mu$Jy)}&\colhead{($\mu$Jy)}& \colhead{} & \colhead{}& \colhead{}
} 
\startdata
 SWIRE\_J104314.93+585606.3  &    160.81221  &    58.935070  & $>$24.4  &   9  &  22  &  63  & 116  &   936  &  90$\pm$23  & \nodata  &  3.07  &       I \\
 SWIRE\_J104409.95+585224.8  &    161.04146  &    58.873550  &  23.55   &  65  & 152  & 401  &1082  &  4011  & 273$\pm$15  &   2.540  &  2.67  &       I\tablenotemark{c} \\
 SWIRE\_J104531.45+591027.0  &    161.38103  &    59.174160  & $>$25.2  &  28  &  54  & 110  & 179  &  1189  & 103$\pm$ 5  & \nodata  &  1.97  &       I\tablenotemark{c} \\
 SWIRE\_J104605.11+584310.8  &    161.52130  &    58.719662  & $>$25.2  &  14  &  26  &  59  & 162  &   707  &  58$\pm$11  & \nodata  &  2.03  &       I \\
 SWIRE\_J104605.54+583742.5  &    161.52309  &    58.628460  & $>$24.4  &  12  &  21  &  72  & 127  &  1359  & $<$91       & \nodata  &  3.63  &       I \\
 SWIRE\_J104633.45+590016.7  &    161.63937  &    59.004631  & $>$25.2  &  17  &  35  &  67  & 131  &   355  &  33$\pm$ 3  & \nodata  &  1.50  &       I \\
 SWIRE\_J104659.42+584624.1  &    161.74757  &    58.773350  & $>$25.2  &  27  &  50  &  77  & 128  &   538  & 223$\pm$16  & \nodata  &  1.18  &       I \\
 SWIRE\_J104811.23+592206.8  &    162.04680  &    59.368561  & $>$24.4  &  53  &  96  & 181  & 319  &  1315  & $<$198      & \nodata  &  1.32  &       I \\
 SWIRE\_J104825.76+591338.9  &    162.10732  &    59.227470  & $>$24.4  &  45  &  70  & 150  & 241  &  1234  & 154$\pm$23  & \nodata  &  1.31  &       I\tablenotemark{c} \\
\\
 SWIRE\_J104301.13+591214.4  &    160.75470  &    59.204010  & $>$24.4  &  36  &  61  & 109  & 150  &  1022  & 185$\pm$44  & \nodata  &  2.16  &      II \\
 SWIRE\_J104310.29+585916.0  &    160.79289  &    58.987789  &  23.78   &  24  &  33  &  50  &  87  &   548  &  $<$68      & \nodata  &  2.00  &      II\tablenotemark{c} \\
 SWIRE\_J104311.81+590649.9  &    160.79923  &    59.113850  &  22.97   & 129  & 210  & 428  & 714  &  1921  &  $<$217     & \nodata  &  0.83  &      II \\
 SWIRE\_J104323.61+590218.3  &    160.84839  &    59.038422  &  24.40   &  25  &  27  &  33  &  54  &   326  &  $<$146     & \nodata  &  0.90  &      II \\
 SWIRE\_J104359.77+585811.3  &    160.99904  &    58.969810  & $>$25.2  &  14  &  17  & $<$43&  53  &   358  & $<$27       & \nodata  &  2.35  &      II \\
 SWIRE\_J104402.02+584504.9  &    161.00841  &    58.751362  &  23.65   &  44  &  48  &  58  & 112  &   882  & $<$30       & \nodata  &  2.04  &      II \\
 SWIRE\_J104406.30+583954.1  &    161.02626  &    58.665039  &  23.35   &  53  &  67  & 131  & 244  &  1099  & $<$162      &   2.430  &  2.14  &      II\tablenotemark{c} \\
 SWIRE\_J104406.71+585130.8  &    161.02797  &    58.858551  &  24.90   &  24  &  28  & $<$43&  66  &   397  & $<$94       & \nodata  &  1.99  &      II\tablenotemark{c} \\
 SWIRE\_J104423.99+584638.3  &    161.09996  &    58.777302  &  24.82   &  15  &  24  &  30  &  64  &   307  & $<$119      & \nodata  &  0.88  &      II \\
 SWIRE\_J104444.84+585019.5  &    161.18681  &    58.838749  &  24.63   &  14  &  16  & $<$43&  52  &   440  &  55$\pm$ 8  & \nodata  &  3.42  &      II \\
 SWIRE\_J104458.45+590314.0  &    161.24353  &    59.053890  &  23.42   &  93  & 157  & 328  & 623  &  1861  &  33$\pm$ 4  &   1.520  &  1.58  &      II \\
 SWIRE\_J104459.07+590705.3  &    161.24612  &    59.118141  &  23.68   &  31  &  45  &  58  & 132  &   674  &  40$\pm$ 4  & \nodata  &  0.89  &      II \\
 SWIRE\_J104524.91+591626.7  &    161.35379  &    59.274071  &  24.69   &  35  &  47  &  59  & 106  &   636  & 140$\pm$15  & \nodata  &  0.72  &      II \\
 SWIRE\_J104525.21+585949.3  &    161.35506  &    58.997040  & $>$25.2  &  10  &  19  &  39  &  64  &   335  &  24$\pm$ 3  & \nodata  &  2.05  &      II\tablenotemark{c} \\
 SWIRE\_J104533.47+590541.1  &    161.38947  &    59.094749  &  24.60   &  13  &  15  &  37  &  52  &   398  &  33$\pm$ 3  & \nodata  &  2.77  &      II \\
 SWIRE\_J104551.81+590345.3  &    161.46588  &    59.062592  &  24.06   &  34  &  41  &  53  &  88  &   472  &  21$\pm$ 5  & \nodata  &  1.00  &      II\tablenotemark{c} \\
 SWIRE\_J104554.47+592322.2  &    161.47697  &    59.389488  & $>$24.4  &   9  &  19  &  33  &  89  &   242  & $<$216      & \nodata  &  3.48  &      II \\
 SWIRE\_J104604.78+591625.3  &    161.51993  &    59.273701  &  23.80   &  38  &  66  & 115  & 193  &   538  & $<$72       & \nodata  &  1.13  &      II \\
 SWIRE\_J104609.55+590849.0  &    161.53979  &    59.146938  &  24.09   &  17  &  27  &  45  & 101  &   319  &  44$\pm$ 3  & \nodata  &  1.82  &      II \\
 SWIRE\_J104622.31+585036.0  &    161.59294  &    58.843330  &  23.79   &  36  &  53  &  74  & 143  &   509  &  79$\pm$12  & \nodata  &  0.96  &      II \\
 SWIRE\_J104627.46+584049.2  &    161.61443  &    58.680328  &  24.04   &  26  &  42  &  57  & 150  &   488  & $<$154      & \nodata  &  0.81  &      II \\
 SWIRE\_J104633.29+584820.3  &    161.63870  &    58.805641  &  24.05   &  17  &  20  &  32  &  47  &   440  &  47$\pm$ 6  & \nodata  &  2.69  &      II\tablenotemark{c} \\
 SWIRE\_J104637.25+585252.4  &    161.65520  &    58.881222  & $>$25.2  &  14  &  25  &  48  &  99  &   332  &  25$\pm$ 4  & \nodata  &  1.84  &      II \\
 SWIRE\_J104637.57+590014.6  &    161.65654  &    59.004059  & $>$25.2  &  10  &  12  & $<$43&  32  &   418  &  24$\pm$ 7  & \nodata  &  3.43  &      II \\
 SWIRE\_J104657.15+592152.8  &    161.73814  &    59.364658  &  24.04   &  28  &  31  &  51  &  64  &   698  & $<$89       &   1.579  &  2.56  &      II \\
 SWIRE\_J104658.67+584042.4  &    161.74446  &    58.678440  & $>$24.4  &  30  &  47  &  61  & 105  &   269  & 192$\pm$43  & \nodata  &  1.32  &      II \\
 SWIRE\_J104700.21+590107.6  &    161.75087  &    59.018780  &  23.72   &  28  &  33  &  59  & 158  &  1273  & 279$\pm$ 6  &   2.562  &  2.85  &      II \\
 SWIRE\_J104704.98+585849.7  &    161.77074  &    58.980469  &  24.18   &  27  &  33  &  55  &  74  &   427  & $<$35       & \nodata  &  2.09  &      II \\
 SWIRE\_J104705.31+584420.3  &    161.77211  &    58.738960  &  23.47   &  37  &  49  &  66  &  87  &   738  & 219$\pm$23  & \nodata  &  2.25  &      II \\
 SWIRE\_J104705.49+590919.4  &    161.77287  &    59.155380  &  25.06   &  16  &  30  &  66  & 111  &   903  &  53$\pm$ 6  & \nodata  &  2.44  &      II \\
 SWIRE\_J104714.84+591916.5  &    161.81181  &    59.321259  &  23.98   &  41  &  54  &  65  &  94  &   862  & 262$\pm$36  & \nodata  &  0.76  &      II \\
 SWIRE\_J104726.03+591305.5  &    161.85844  &    59.218189  & $>$25.2  &  26  &  41  &  68  & 114  &   409  & 169$\pm$16  & \nodata  &  1.67  &      II \\
 SWIRE\_J104735.76+584707.3  &    161.89902  &    58.785370  & $>$25.2  &  12  &  17  &  32  &  56  &   313  &  89$\pm$12  & \nodata  &  2.11  &      II \\
 SWIRE\_J104735.93+590549.7  &    161.89970  &    59.097149  &  24.74   &  21  &  29  &  54  &  67  &   324  & $<$55       & \nodata  &  1.64  &      II\tablenotemark{c} \\
 SWIRE\_J104746.64+585113.5  &    161.94432  &    58.853760  & $>$25.2  &  11  &  20  &  32  &  50  &   244  &  41$\pm$10  & \nodata  &  1.85  &      II \\
 SWIRE\_J104746.79+591759.2  &    161.94496  &    59.299770  &  24.59   &  21  &  27  & $<$43&  62  &   393  & $<$196      & \nodata  &  2.09  &      II \\
 SWIRE\_J104754.20+590956.3  &    161.97582  &    59.165630  &  23.84   &  22  &  26  & $<$43&  58  &   448  & $<$91       & \nodata  &  2.57  &      II \\
 SWIRE\_J104806.02+584456.4  &    162.02510  &    58.749001  &  24.34   &  25  &  33  & $<$43&  52  &   338  & 126$\pm$24  & \nodata  &  0.81  &      II \\
 SWIRE\_J104820.64+592436.6  &    162.08600  &    59.410160  &  23.75   &  25  &  42  &  76  & 196  &   833  & $<$1064     & \nodata  &  2.46  &      II \\
 SWIRE\_J104831.65+592004.6  &    162.13188  &    59.334621  &  22.87   &  68  &  73  &  87  & 147  &   775  & $<$596      & \nodata  &  0.60  &      II \\
 SWIRE\_J104836.72+585414.7  &    162.15300  &    58.904072  &  22.77   &  75  & 117  & 161  & 231  &   887  & $<$176      & \nodata  &  0.53  &      II\tablenotemark{c} \\
 SWIRE\_J104844.25+585139.4  &    162.18439  &    58.860931  & $>$24.4  &  15  &  24  & $<$43&  64  &   200  & $<$248      & \nodata  &  1.54  &      II \\
 SWIRE\_J104903.66+590006.9  &    162.26524  &    59.001930  &  24.24   &  29  &  42  &  51  & 119  &   379  & $<$276      & \nodata  &  1.64  &      II\tablenotemark{c} \\
 SWIRE\_J104955.03+584742.8  &    162.47929  &    58.795219  &  23.10   &  22  &  31  &  43  & 125  &  1080  & \nodata     & \nodata  &  3.42  &      II \\
\\
 SWIRE\_J104257.62+590438.9  &    160.74010  &    59.077469  &  23.39   &  21  &  23  & $<$43&  40  &   256  & $<$281      & \nodata  &  1.54  &     III \\
 SWIRE\_J104321.34+590943.0  &    160.83891  &    59.161942  &  20.03   & 184  & 239  & 345  & 492  &  2665  & $<$209      & \nodata  &  2.29  &     III\tablenotemark{c} \\
 SWIRE\_J104353.42+585316.2  &    160.97260  &    58.887840  &  22.75   &  32  &  39  &  62  &  90  &   560  & $<$103      &   0.563  &  0.69  &     III\tablenotemark{c} \\
 SWIRE\_J104359.52+590156.3  &    160.99800  &    59.032310  &  23.25   &  34  &  58  &  84  & 148  &   464  & 103$\pm$ 8  & \nodata  &  0.84  &     III \\
 SWIRE\_J104414.18+584644.6  &    161.05907  &    58.779049  &  22.26   &  63  &  81  & 126  & 268  &  1762  & 435$\pm$25  & \nodata  &  2.73  &     III \\
 SWIRE\_J104432.03+590457.7  &    161.13344  &    59.082691  &  22.80   &  56  & 111  & 192  & 339  &  1067  & 104$\pm$ 8  & \nodata  &  0.90  &     III\tablenotemark{c} \\
 SWIRE\_J104447.56+585918.7  &    161.19815  &    58.988529  &  22.90   &  29  &  38  &  66  & 127  &   660  & 120$\pm$ 4  &   2.488  &  2.44  &     III \\
 SWIRE\_J104500.71+591353.4  &    161.25297  &    59.231491  &  23.13   &  45  &  70  & 108  & 156  &   776  &  64$\pm$13  & \nodata  &  0.70  &     III\tablenotemark{c} \\
 SWIRE\_J104528.33+583651.8  &    161.36803  &    58.614391  &  22.12   &  24  &  37  &  54  &  87  &   366  & $<$325      & \nodata  &  1.54  &     III \\
 SWIRE\_J104616.93+585457.1  &    161.57056  &    58.915871  &  23.63   &  26  &  32  & $<$43&  52  &   241  &  60$\pm$ 7  & \nodata  &  0.70  &     III\tablenotemark{c} \\
 SWIRE\_J104635.87+590748.8  &    161.64946  &    59.130219  &  23.48   &  39  &  48  &  58  &  67  &   615  &  97$\pm$ 3  & \nodata  &  0.45  &     III \\
 SWIRE\_J104638.87+591931.4  &    161.66196  &    59.325378  & $>$24.4  &   9  &  16  & $<$43&  71  &   227  & $<$124      & \nodata  &  2.09  &     III \\
 SWIRE\_J104641.38+585213.9  &    161.67242  &    58.870529  &  22.75   &  42  &  84  & 204  & 541  &  2797  & 1004$\pm$8  & \nodata  &  1.10  &     III\tablenotemark{c} \\
 SWIRE\_J104644.18+590027.8  &    161.68407  &    59.007710  &  22.56   &  33  &  42  &  67  &  77  &   899  &  30$\pm$ 6  &   2.542  &  2.79  &     III\tablenotemark{c} \\
 SWIRE\_J104733.37+591200.9  &    161.88902  &    59.200241  &  22.39   &  91  & 127  & 196  & 307  &  1110  &  49$\pm$ 9  & \nodata  &  0.71  &     III \\
 SWIRE\_J104748.20+590905.7  &    161.95082  &    59.151581  &  23.48   &  25  &  33  &  50  &  63  &   375  & $<$78       & \nodata  &  3.14  &     III \\
 SWIRE\_J104749.61+584845.7  &    161.95670  &    58.812691  &  23.03   &  25  &  32  &  39  &  58  &   183  & $<$114      & \nodata  &  0.69  &     III\tablenotemark{c} \\
 SWIRE\_J104826.99+585438.8  &    162.11246  &    58.910782  &  22.89   &  32  &  53  &  84  & 156  &   519  &  87$\pm$14  & \nodata  &  1.41  &     III\tablenotemark{c} \\
 SWIRE\_J104913.37+585946.3  &    162.30569  &    58.996181  &  23.41   &  22  &  27  &  38  &  88  &   316  & $<$356      & \nodata  &  3.43  &     III\tablenotemark{c} \\
\\
 SWIRE\_J104229.75+591154.6  &    160.62396  &    59.198490  & $>$24.4  &  36  &  51  &  52  & 114  &   983  & $<$281      & \nodata  &  1.89  &      IV \\
 SWIRE\_J104241.45+591357.2  &    160.67270  &    59.232559  & $>$24.4  &  31  &  43  &  44  & 124  &   319  & $<$816      & \nodata  &  1.43  &      IV\tablenotemark{c} \\
 SWIRE\_J104254.46+591013.3  &    160.72691  &    59.170361  & $>$24.4  &  47  &  61  &  63  & 152  &   591  & $<$411      & \nodata  &  1.54  &      IV\tablenotemark{c} \\
 SWIRE\_J104414.37+584320.4  &    161.05988  &    58.722340  & $>$25.2  &  65  &  79  &  90  & 149  &  1061  & 754$\pm$22  & \nodata  &  1.78  &      IV \\
 SWIRE\_J104420.22+583948.0  &    161.08424  &    58.663342  &  24.89   &  44  &  62  &  71  & 104  &   615  & $<$346      & \nodata  &  1.70  &      IV\tablenotemark{c} \\
 SWIRE\_J104526.75+583526.0  &    161.36145  &    58.590542  & $>$23.5  & 104  & 104  & 126  & 349  &  2538  & 622$\pm$45  & \nodata  &  1.84  &      IV \\
 SWIRE\_J104528.29+591326.7  &    161.36789  &    59.224079  &  23.45   &  35  &  47  &  91  & 199  &  2410  &14379$\pm$10 & \nodata  &  0.67  &      IV\tablenotemark{c} \\
 SWIRE\_J104650.27+592602.8  &    161.70947  &    59.434120  & $>$24.4  &  31  &  48  &  67  & 203  &  1488  & $<$147      & \nodata  &  0.88  &      IV \\
 SWIRE\_J104718.10+585526.2  &    161.82542  &    58.923931  & $>$25.2  &  11  &  16  & $<$43&  29  &   314  &  60$\pm$ 5  & \nodata  &  1.95  &      IV \\
 SWIRE\_J104733.46+592108.1  &    161.88942  &    59.352242  &  23.88   &  44  &  52  &  75  & 199  &   801  &  92$\pm$30  & \nodata  &  1.11  &      IV\tablenotemark{c} \\
 SWIRE\_J104734.47+591332.9  &    161.89362  &    59.225819  & $>$25.2  &  13  &  17  & $<$43&  42  &   356  &  94$\pm$10  & \nodata  &  1.90  &      IV \\
 SWIRE\_J104736.92+591941.3  &    161.90385  &    59.328152  & $>$24.4  &  31  &  37  &  53  &  93  &   383  & $<$220      & \nodata  &  1.54  &      IV\tablenotemark{c} \\
 SWIRE\_J104754.78+590810.4  &    161.97826  &    59.136211  & $>$25.2  &  11  &  18  & $<$43&  34  &   274  & $<$26       & \nodata  &  1.83  &      IV\tablenotemark{c} \\
 SWIRE\_J104802.42+592656.5  &    162.01010  &    59.449020  & $>$24.4  &  36  &  42  &  58  &  83  &  1043  & $<$423      & \nodata  &  2.14  &      IV \\
 SWIRE\_J104813.49+590340.7  &    162.05620  &    59.061310  & $>$24.4  &  13  &  21  & $<$43&  46  &   356  &  41$\pm$10  & \nodata  &  1.82  &      IV\tablenotemark{c} \\
 SWIRE\_J104838.01+591702.6  &    162.15836  &    59.284050  & $>$24.4  &  18  &  22  &  28  &  63  &   430  & $<$438      & \nodata  &  1.79  &      IV \\
\\
 SWIRE\_J104303.50+585718.1  &    160.76460  &    58.955029  &  20.32   & 322  & 364  & 611  &1136  &  5371  & $<$244      & \nodata  &  0.60  &       V\tablenotemark{c} \\
 SWIRE\_J104351.87+584953.7  &    160.96614  &    58.831589  &  23.02   &  23  &  27  &  57  & 103  &   996  & $<$139      &   0.609  &  0.59  &       V\tablenotemark{c} \\
 SWIRE\_J104407.97+584437.0  &    161.03319  &    58.743622  &  20.32   & 655  & 858  &1148  &1595  &  7053  & 213$\pm$48  &   0.555  &  0.58  &       V\tablenotemark{c} \\
 SWIRE\_J104422.64+591304.1  &    161.09435  &    59.217819  &  21.49   & 145  & 176  & 262  & 352  &  1439  & 123$\pm$17  & \nodata  &  0.32  &       V\tablenotemark{c} \\
 SWIRE\_J104503.56+585109.9  &    161.26485  &    58.852741  &  22.83   & 133  & 176  & 237  & 344  &  1006  &  86$\pm$19  & \nodata  &  0.98  &       V\tablenotemark{c} \\
 SWIRE\_J104504.96+585947.3  &    161.27066  &    58.996479  &  21.15   &  80  &  89  & 136  & 309  &  2404  & 137$\pm$11  &   0.214  &  0.13  &       V\tablenotemark{c} \\
 SWIRE\_J104532.93+584638.6  &    161.38722  &    58.777382  &  24.37   &  34  &  45  &  64  & 124  &   345  & $<$66       & \nodata  &  1.78  &       V\tablenotemark{c} \\
 SWIRE\_J104704.33+591142.9  &    161.76802  &    59.195240  & $>$25.2  &  21  &  25  &  41  &  55  &   185  & $<$57       & \nodata  &  1.34  &       V \\
 SWIRE\_J104708.05+585026.3  &    161.78352  &    58.840641  &  20.79   &  79  & 101  & 144  & 242  &  1607  & $<$58       & \nodata  &  0.44  &       V \\
 SWIRE\_J104725.94+591025.5  &    161.85806  &    59.173759  &  21.54   & 111  & 150  & 200  & 277  &   886  &  73$\pm$ 7  & \nodata  &  0.67  &       V\tablenotemark{c} \\
 SWIRE\_J104731.84+592432.9  &    161.88266  &    59.409149  & $>$24.4  &  31  &  38  &  54  &  76  &   281  & $<$483      & \nodata  &  3.11  &       V\tablenotemark{c} \\
 SWIRE\_J104734.57+584338.7  &    161.89404  &    58.727409  &  22.68   & 105  & 119  & 180  & 302  &  1286  & 118$\pm$21  & \nodata  &  0.87  &       V \\
 SWIRE\_J104748.27+590534.7  &    161.95113  &    59.092960  & $>$25.2  &  16  &  19  &  26  &  41  & $<$230 & 199$\pm$ 7  & \nodata  &  1.04  &       V\tablenotemark{c} \\
 SWIRE\_J104826.92+592116.1  &    162.11218  &    59.354469  &  20.08   & 258  & 316  & 423  & 738  &  3632  & 450$\pm$79  & \nodata  &  0.58  &       V \\
 SWIRE\_J104829.49+591249.1  &    162.12288  &    59.213631  &  21.07   & 108  & 155  & 208  & 333  &  1117  & 125$\pm$23  & \nodata  &  0.28  &       V\tablenotemark{c} \\
 SWIRE\_J104851.80+591019.8  &    162.21584  &    59.172161  &  22.87   &  60  &  97  & 137  & 196  &   957  & $<$295      & \nodata  &  0.82  &       V \\
\\
 SWIRE\_J104344.33+585102.5  &    160.93469  &    58.850689  & $>$24.4  &   9  &  16  & $<$43& $<$40&   260  & $<$144      & \nodata  &  2.01  &      VI \\
 SWIRE\_J104407.69+585125.9  &    161.03204  &    58.857189  & $>$25.2  &   6  &   9  & $<$43& $<$40&   341  & $<$93       & \nodata  &  2.51  &      VI \\
 SWIRE\_J104408.30+590455.4  &    161.03458  &    59.082062  & $>$25.2  &   9  &  13  & $<$43& $<$40&   240  & $<$67       & \nodata  &  1.79  &      VI \\
 SWIRE\_J104530.65+585936.8  &    161.37772  &    58.993549  & $>$25.2  &  10  &  13  & $<$43& $<$40&   256  &  19$\pm$ 3  & \nodata  &  1.93  &      VI \\
 SWIRE\_J104534.22+591400.3  &    161.39259  &    59.233410  & $>$25.2  &   8  &  13  & $<$43& $<$40&   277  & $<$56       & \nodata  &  1.91  &      VI \\
 SWIRE\_J104558.79+583812.9  &    161.49496  &    58.636909  & $>$24.4  &   8  &  13  & $<$43& $<$40&   344  & $<$233      & \nodata  &  2.73  &      VI \\
 SWIRE\_J104623.23+584410.9  &    161.59677  &    58.736351  & $>$25.2  &   9  &  14  & $<$43& $<$40&   280  & $<$90       & \nodata  &  2.61  &      VI \\
 SWIRE\_J104632.92+590820.9  &    161.63716  &    59.139141  & $>$25.2  &  10  &  15  & $<$43& $<$40&   245  &  36$\pm$ 4  & \nodata  &  1.75  &      VI \\
 SWIRE\_J104703.99+585039.9  &    161.76662  &    58.844410  &  24.79   &   8  &  14  & $<$43& $<$40&   263  & $<$55       & \nodata  &  2.06  &      VI \\
 SWIRE\_J104708.15+585721.1  &    161.78397  &    58.955860  & $>$25.2  &  10  &  17  & $<$43& $<$40&   240  & $<$38       & \nodata  &  1.93  &      VI \\
 SWIRE\_J104708.92+584524.1  &    161.78716  &    58.756691  & $>$25.2  &   9  &  13  & $<$43& $<$40&   299  & $<$102      & \nodata  &  1.95  &      VI \\
 SWIRE\_J104720.43+590107.9  &    161.83514  &    59.018871  & $>$25.2  &   8  &  12  & $<$43& $<$40&   260  &  39$\pm$ 4  & \nodata  &  1.93  &      VI \\
 SWIRE\_J104728.64+585346.1  &    161.86932  &    58.896141  &  24.02   &  14  &  19  & $<$43&  41  & $<$230 & $<$57       & \nodata  &  0.58  &      VI \\
 SWIRE\_J104747.72+585903.9  &    161.94882  &    58.984421  & $>$25.2  &   6  &   9  & $<$43& $<$40&   243  & 122$\pm$ 7  & \nodata  &  2.09  &      VI \\
 SWIRE\_J104754.71+591137.2  &    161.97797  &    59.193661  & $>$25.2  &  10  &  14  & $<$43& $<$40&   254  & $<$106      & \nodata  &  1.82  &      VI \\
 SWIRE\_J104803.66+592252.5  &    162.01524  &    59.381248  & $>$24.4  &   8  &  14  & $<$43& $<$40&   259  & $<$564      & \nodata  &  2.17  &      VI \\
\enddata
\tablecomments{Typical uncertainties to the IR fluxes are $\sim$10\% of the
   measured fluxes and to the optical magnitudes are around 0.04 mag. Upper limits
correspond to 5$\sigma$ values.}
\tablenotetext{a}{IR coordinates}
\tablenotetext{b}{Class I sources are characterized by convex IR SED fitted
by a ``Torus'' template. Class II sources show power-law like optical-IR
SEDs, similar to Mrk 231 or slightly redder (A$_{\rm V}<$1). Class III
sources have power-law like optical-IR SED fitted by a reddened QSO template
(A$_{\rm V}=$0.6--1.0). Class IV sources show signatures from both a
starburst and an AGN component. Class V sources area characterized by very
red optical SEDs and power-law like IR SEDs. Class VI objects are detected
only in three bands from 3.6 to 24$\mu$m.}
\tablenotetext{c}{X-ray source}
\end{deluxetable}


\begin{deluxetable}{c c ccc r ccr cc}
\tabletypesize{\scriptsize}
\rotate
\tablecaption{X-ray properties of IR-selected AGN \label{ctagn_ir_xprop}}
\tablewidth{0pt}
\tablehead{
 \colhead{Source name}&
 \colhead{Total counts} &
 \colhead{F$_{0.3-8 keV}$\tablenotemark{a}}   &
 \colhead{F$_{0.3-2.5 keV}$\tablenotemark{a}}   &
 \colhead{F$_{2.5-8 keV}$\tablenotemark{a}}   &
 \colhead{HR} &
 \colhead{$z$} &
 \colhead{\nh$^{obs}$\tablenotemark{b}} &
 \colhead{\nh$^{rest}$\tablenotemark{c}} &
 \colhead{$Log(L)$\tablenotemark{d}} &
 \colhead{Class} \\
 \colhead{}  &
 \colhead{(0.3-8 keV)} &
 \colhead{} &
 \colhead{} &
 \colhead{} &
 \colhead{} &
 \colhead{} &
 \colhead{} &
 \colhead{} &
 \colhead{} &
 \colhead{}
} 
\startdata
 SWIRE\_J104409.95+585224.8  &      11$\pm$  5  &   19$\pm$  8  &    2$\pm$  3  &   33$\pm$ 15  &     0.85$^{+0.06}_{-0.39 }$  &   2.540\tablenotemark{f}  &   8.0  & 214.0  &  45.637  &    I \\
 SWIRE\_J104531.45+591027.0  &       7$\pm$  4  &   12$\pm$  7  &    5$\pm$  4  &   10$\pm$ 11  &  $-$0.14$^{+0.76}_{-0.84 }$  &   1.973                   &   0.7  &  11.6  &  43.986  &    I \\
 SWIRE\_J104825.76+591338.9  &      21$\pm$  7  &   34$\pm$ 12  &   19$\pm$  7  &   13$\pm$ 16  &  $-$0.61$^{+0.45}_{-0.00 }$  &   1.307                   & $<$0.1 &   0.1  &  43.497  &    I \\
\\
 SWIRE\_J104310.29+585916.0  &     118$\pm$ 13  &  202$\pm$ 22  &   84$\pm$ 12  &  162$\pm$ 30  &  $-$0.20$^{+0.20}_{-0.20 }$  &   1.997                   &   0.6  &  10.0  &  45.186  &   II \\
 SWIRE\_J104406.30+583954.1  &      22$\pm$  6  &   37$\pm$ 11  &    4$\pm$  4  &   63$\pm$ 20  &     0.61$^{+0.21}_{-0.23 }$  &   2.430\tablenotemark{f}  &   4.3  &  99.0  &  45.738  &   II \\
 SWIRE\_J104406.71+585130.8  &      10$\pm$  5  &   17$\pm$  8  &   10$\pm$  5  &    4$\pm$ 10  &  $-$0.78$^{+0.38}_{-0.33 }$  &   1.991                   &  $<$0.1&   0.2  &  43.645  &   II \\
 SWIRE\_J104525.21+585949.3  &      16$\pm$  5  &   27$\pm$  9  &   15$\pm$  6  &   10$\pm$ 12  &  $-$0.62$^{+0.34}_{-0.18 }$  &   2.045                   &  $<$0.1&   0.2  &  43.882  &   II \\
 SWIRE\_J104551.81+590345.3  &      10$\pm$  5  &   17$\pm$  8  &    5$\pm$  4  &   19$\pm$ 13  &     0.10$^{+0.56}_{-0.65 }$  &   0.999                   &   1.3  &   7.8  &  43.363  &   II \\
 SWIRE\_J104633.29+584820.3  &       6$\pm$  4  &   10$\pm$  7  &    5$\pm$  4  &    6$\pm$ 10  &  $-$0.40$^{+0.87}_{-0.00 }$  &   2.693                   &   0.2  &   6.4  &  44.102  &   II \\
 SWIRE\_J104735.93+590549.7  &      34$\pm$  7  &   55$\pm$ 11  &   21$\pm$  6  &   48$\pm$ 16  &  $-$0.12$^{+0.20}_{-0.19 }$  &   1.644                   &   0.8  &   9.4  &  44.412  &   II \\
 SWIRE\_J104836.72+585414.7  &      59$\pm$  9  &   98$\pm$ 15  &   54$\pm$  9  &   39$\pm$ 16  &  $-$0.60$^{+0.13}_{-0.11 }$  &   0.533                   &  $<$0.1&  $<$0.1&  43.002  &   II \\
 SWIRE\_J104903.66+590006.9  &      19$\pm$  7  &   32$\pm$ 12  &   21$\pm$  7  &    4$\pm$ 16  &  $-$0.87$^{+0.38}_{-0.00 }$  &   1.641                   &  $<$0.1&   0.1  &  43.716  &   II \\
\\
 SWIRE\_J104321.34+590943.0  &    1227$\pm$ 36  & 1980$\pm$ 58  &  993$\pm$ 34  & 1092$\pm$ 63  &  $-$0.45$^{+0.02}_{-0.03 }$  &   2.294                   &   0.1  &   2.9  &  46.108  &  III \\
 SWIRE\_J104353.42+585316.2  &      49$\pm$  8  &   84$\pm$ 14  &   25$\pm$  7  &   95$\pm$ 22  &     0.14$^{+0.16}_{-0.16 }$  &   0.563\tablenotemark{f}  &   1.4  &   4.5  &  43.380  &  III \\
 SWIRE\_J104432.03+590457.7  &     134$\pm$ 14  &  230$\pm$ 23  &   83$\pm$ 12  &  221$\pm$ 34  &  $-$0.04$^{+0.18}_{-0.18 }$  &   0.896                   &   0.9  &   4.9  &  44.308  &  III \\
 SWIRE\_J104500.71+591353.4  &      16$\pm$  5  &   29$\pm$  9  &    9$\pm$  5  &   32$\pm$ 15  &     0.11$^{+0.30}_{-0.30 }$  &   0.700                   &   1.3  &   5.3  &  43.161  &  III \\
 SWIRE\_J104616.93+585457.1  &      49$\pm$  8  &   84$\pm$ 15  &   37$\pm$  8  &   62$\pm$ 19  &  $-$0.27$^{+0.26}_{-0.22 }$  &   0.705                   &   0.4  &   1.8  &  43.494  &  III \\
 SWIRE\_J104641.38+585213.9  &      12$\pm$  5  &   19$\pm$  8  &    9$\pm$  5  &   14$\pm$ 12  &  $-$0.26$^{+0.48}_{-0.49 }$  &   1.095                   &   0.4  &   3.1  &  43.367  &  III \\
 SWIRE\_J104644.18+590027.8  &      33$\pm$  7  &   56$\pm$ 12  &   23$\pm$  7  &   46$\pm$ 17  &  $-$0.19$^{+0.26}_{-0.24 }$  &   2.542\tablenotemark{f}  &   0.6  &  15.7  &  44.976  &  III \\
 SWIRE\_J104749.61+584845.7  &      45$\pm$  8  &   75$\pm$ 14  &   39$\pm$  8  &   37$\pm$ 16  &  $-$0.50$^{+0.24}_{-0.20 }$  &   0.695                   &   0.05 &   0.2  &  43.237  &  III \\
 SWIRE\_J104826.99+585438.8  &      16$\pm$  5  &   27$\pm$  9  &    8$\pm$  5  &   30$\pm$ 14  &     0.10$^{+0.28}_{-0.31 }$  &   1.410                   &   1.3  &  12.8  &  44.019  &  III \\
 SWIRE\_J104913.37+585946.3  &      39$\pm$  9  &   65$\pm$ 15  &   21$\pm$  7  &   70$\pm$ 23  &     0.07$^{+0.26}_{-0.18 }$  &   3.435                   &   1.2  &  58.8  &  45.861  &  III \\
\\
 SWIRE\_J104241.45+591357.2  &      28$\pm$  7  &   45$\pm$ 12  &   23$\pm$  7  &   22$\pm$ 15  &  $-$0.51$^{+0.26}_{-0.22 }$  &   1.434                   &   0.04 &   0.4  &  43.812  &   IV \\
 SWIRE\_J104254.46+591013.3  &      32$\pm$  7  &   52$\pm$ 11  &   18$\pm$  6  &   53$\pm$ 17  &     0.03$^{+0.21}_{-0.21 }$  &   1.541                   &   1.1  &  12.3  &  44.385  &   IV \\
 SWIRE\_J104420.22+583948.0  &      56$\pm$  9  &   97$\pm$ 15  &   30$\pm$  7  &  107$\pm$ 23  &     0.11$^{+0.16}_{-0.16 }$  &   1.700                   &   1.3  &  17.5  &  44.842  &   IV \\
 SWIRE\_J104528.29+591326.7  &      18$\pm$  5  &   31$\pm$ 10  &    9$\pm$  5  &   35$\pm$ 15  &     0.16$^{+0.27}_{-0.28 }$  &   0.669                   &   1.5  &   5.5  &  43.153  &   IV \\
 SWIRE\_J104733.46+592108.1  &      40$\pm$  8  &   70$\pm$ 13  &   40$\pm$  8  &   24$\pm$ 14  &  $-$0.65$^{+0.15}_{-0.12 }$  &   1.107                   &  $<$0.1&   0.1  &  43.628  &   IV \\
 SWIRE\_J104736.92+591941.3  &       9$\pm$  4  &   15$\pm$  8  &    9$\pm$  5  &    6$\pm$ 10  &  $-$0.60$^{+0.63}_{-0.00 }$  &   1.543                   &  $<$0.1&   0.1  &  43.332  &   IV \\
 SWIRE\_J104754.78+590810.4  &      16$\pm$  5  &   25$\pm$  9  &   14$\pm$  5  &   11$\pm$ 11  &  $-$0.57$^{+0.28}_{-0.24 }$  &   1.826                   &  $<$0.1&   0.1  &  43.728  &   IV \\
 SWIRE\_J104813.49+590340.7  &      17$\pm$  5  &   28$\pm$  9  &   17$\pm$  6  &    6$\pm$ 10  &  $-$0.79$^{+0.21}_{-0.14 }$  &   1.818                   &  $<$0.1&   0.1  &  43.769  &   IV \\
\\
 SWIRE\_J104303.50+585718.1  &      14$\pm$  7  &   24$\pm$ 11  &    3$\pm$  5  &   38$\pm$ 19  &     0.60$^{+0.62}_{-0.70 }$  &   0.595                   &   3.6  &  12.0  &  43.073  &    V \\
 SWIRE\_J104351.87+584953.7  &      44$\pm$  8  &   75$\pm$ 14  &   10$\pm$  5  &  121$\pm$ 25  &     0.61$^{+0.25}_{-0.31 }$  &   0.609\tablenotemark{f}  &   3.6  &  12.5  &  43.605  &    V \\
 SWIRE\_J104407.97+584437.0  &      48$\pm$  8  &   83$\pm$ 14  &   30$\pm$  8  &   81$\pm$ 21  &  $-$0.03$^{+0.20}_{-0.20 }$  &   0.555\tablenotemark{f}  &   1.0  &   3.0  &  43.306  &    V \\
 SWIRE\_J104422.64+591304.1  &     800$\pm$ 29  & 1292$\pm$ 48  &  649$\pm$ 28  &  711$\pm$ 52  &  $-$0.45$^{+0.09}_{-0.08 }$  &   0.323                   &   0.1  &   0.3  &  43.730  &    V \\
 SWIRE\_J104503.56+585109.9  &      29$\pm$  8  &   49$\pm$ 14  &   18$\pm$  7  &   45$\pm$ 20  &  $-$0.08$^{+0.29}_{-0.28 }$  &   0.981                   &   0.8  &   4.9  &  43.729  &    V \\
 SWIRE\_J104504.96+585947.3  &      59$\pm$  9  &  102$\pm$ 16  &    3$\pm$  5  &  193$\pm$ 31  &     0.90$^{+0.06}_{-0.14 }$  &   0.214\tablenotemark{f}  &   8.7  &  14.3  &  42.736  &    V \\
 SWIRE\_J104532.93+584638.6  &     101$\pm$ 12  &  176$\pm$ 20  &   98$\pm$ 12  &   69$\pm$ 22  &  $-$0.60$^{+0.18}_{-0.15 }$  &   1.781                   &  $<$0.1&   0.1  &  44.545  &    V \\
 SWIRE\_J104725.94+591025.5  &     136$\pm$ 13  &  219$\pm$ 21  &  116$\pm$ 13  &  102$\pm$ 23  &  $-$0.54$^{+0.10}_{-0.10 }$  &   0.668                   &  $<$0.1& $<$0.1 &  43.585  &    V \\
 SWIRE\_J104731.84+592432.9  &      31$\pm$  7  &   55$\pm$ 12  &   24$\pm$  7  &   40$\pm$ 17  &  $-$0.28$^{+0.24}_{-0.22 }$  &   3.108                   &   0.4  &  16.6  &  45.174  &    V \\
 SWIRE\_J104748.27+590534.7  &      15$\pm$  5  &   24$\pm$  9  &   12$\pm$  5  &   14$\pm$ 12  &  $-$0.43$^{+0.38}_{-0.32 }$  &   1.041                   &   0.2  &   1.1  &  43.292  &    V \\
 SWIRE\_J104829.49+591249.1  &      67$\pm$ 10  &  108$\pm$ 16  &   50$\pm$  9  &   72$\pm$ 21  &  $-$0.33$^{+0.24}_{-0.22 }$  &   0.280                   &   0.3  &   0.6  &  42.577  &    V \\
\enddata
\tablenotetext{a}{X-ray flux in 10$^{-16}$\ergcm2s\ derived assuming an
 absorbed power-law model with photon index, $\Gamma$ equal to 1.7 and
 Galactic \nh\ (6$\times$10$^{19}$ \cm2). Uncertainties reflect only the
 statistical errors from the observed counts and do not include 
 uncertainties in the spectral model.}
\tablenotetext{b}{\nh\ in the observer rest-frame in 10$^{22}$ cm$^{-2}$.}
\tablenotetext{c}{\nh\ in the source rest-frame in 10$^{22}$ cm$^{-2}$.}
\tablenotetext{d}{Logarithm of the 0.3-8 keV absorption-corrected rest-frame luminosity in \ergs.}
\tablenotetext{e}{Spectroscopic $z$. Photometric $z$ are reported in
Table~\ref{ctagn_ir}.}
\end{deluxetable}


\begin{deluxetable}{lccrl}
\tabletypesize{\footnotesize}
\tablecaption{Photometric data for SW104409 \label{phot_tab1}}
\tablewidth{0pt}
\tablehead{
\colhead{Observed}& \colhead{Rest-frame}& \colhead{Magnitude} & \colhead{Flux Density} & \colhead{Instrument} \\
\colhead{Bandpass}& \colhead{Bandpass}  & \colhead{(Vega)}    & \colhead{(erg cm$^{-2}$ s$^{-1}$/$\mu$Jy)}    & \colhead{}
}
\startdata
0.3$-$2.5~keV     & 1.1$-$8.8~keV   &\nodata         & ( 2$\pm$ 3)$\times10^{-16}$ & {\it Chandra}/ACIS \\
2.5$-$8.0~keV     & 8.8$-$28.3~keV  &\nodata         & (33$\pm$15)$\times10^{-16}$ & {\it Chandra}/ACIS \\
0.3$-$8.0~keV     & 1.1$-$28.3~keV  &\nodata         & (19$\pm$ 8)$\times10^{-16}$ & {\it Chandra}/ACIS \\
$U$ 3647 \AA      & 1031 \AA        & 24.29$\pm$0.10 & 0.369$\pm$0.033                   & KPNO/Mosaic \\
\gp 4782 \AA      & 1351 \AA        & 24.00$\pm$0.07 & 0.983$\pm$0.060                   & KPNO/Mosaic \\
\rp 6288 \AA      & 1777 \AA        & 23.55$\pm$0.04 & 1.166$\pm$0.040                   & KPNO/Mosaic \\
\ip 7665 \AA      & 2166 \AA        & 23.22$\pm$0.07 & 1.279$\pm$0.071                   & KPNO/Mosaic \\
$K_s$ 2.153$\mu$m & 6085 \AA        & 19.37$\pm$0.14 & 10.55$\pm$1.42                    & Palomar/WIRC \\
3.6~$\mu$m        & 1.01~$\mu$m     &\nodata         &   65.5$\pm$0.9                    & {\it Spitzer}/IRAC \\
4.5~$\mu$m        & 1.27~$\mu$m     &\nodata         &   152$\pm$2                       & {\it Spitzer}/IRAC \\
5.8~$\mu$m        & 1.61~$\mu$m     &\nodata         &   401$\pm$5                       & {\it Spitzer}/IRAC \\
8.0~$\mu$m        & 2.25~$\mu$m     &\nodata         &  1082$\pm$6                       & {\it Spitzer}/IRAC \\
24.~$\mu$m        & 6.74~$\mu$m     &\nodata         &  4011$\pm$20                      & {\it Spitzer}/MIPS \\
1.4~GHz           & 4.95~GHz        &\nodata         &   273$\pm$15                      & VLA \\
\enddata
\end{deluxetable}

\begin{deluxetable}{lccrl}
\tabletypesize{\footnotesize}
\tablecaption{Photometric data for SW104406 \label{phot_tab2}}
\tablewidth{0pt}
\tablehead{
\colhead{Observed}& \colhead{Rest-frame}& \colhead{Magnitude} & \colhead{Flux Density} & \colhead{Instrument} \\
\colhead{Bandpass}& \colhead{Bandpass}  & \colhead{(Vega)}    & \colhead{(erg cm$^{-2}$ s$^{-1}$/$\mu$Jy)}    & \colhead{}
}
\startdata
0.3$-$2.5~keV     & 1.03$-$8.6~keV  &\nodata         & ( 4$\pm$ 4)$\times10^{-16}$ & {\it Chandra}/ACIS \\
2.5$-$8.0~keV     & 8.6$-$27.4~keV  &\nodata         & (63$\pm$20)$\times10^{-16}$ & {\it Chandra}/ACIS \\
0.3$-$8.0~keV     & 1.03$-$27.4~keV &\nodata         & (37$\pm$10)$\times10^{-16}$ & {\it Chandra}/ACIS \\
$U$ 3647 \AA      & 1063 \AA        & 24.27$\pm$0.17 & 0.376$\pm$0.033                   & KPNO/Mosaic \\
\gp 4782 \AA      & 1394 \AA        & 23.68$\pm$0.05 & 1.323$\pm$0.060                   & KPNO/Mosaic \\
\rp 6288 \AA      & 1833 \AA        & 23.35$\pm$0.06 & 1.407$\pm$0.040                   & KPNO/Mosaic \\
\ip 7665 \AA      & 2235 \AA        & 22.76$\pm$0.13 & 1.954$\pm$0.071                   & KPNO/Mosaic \\
$K_s$ 2.153$\mu$m & 6278 \AA        & 18.59$\pm$0.17 & 24.52$\pm$3.25                    & Palomar/WIRC \\
3.6~$\mu$m        & 1.04~$\mu$m     &\nodata         &   53.4$\pm$1.3                    & {\it Spitzer}/IRAC \\
4.5~$\mu$m        & 1.31~$\mu$m     &\nodata         &   67.6$\pm$1.2                    & {\it Spitzer}/IRAC \\
5.8~$\mu$m        & 1.66~$\mu$m     &\nodata         &    131$\pm$7                      & {\it Spitzer}/IRAC \\
8.0~$\mu$m        & 2.32~$\mu$m     &\nodata         &    244$\pm$5                      & {\it Spitzer}/IRAC \\
24.~$\mu$m        & 6.95~$\mu$m     &\nodata         &   1099$\pm$18                     & {\it Spitzer}/MIPS \\
1.4~GHz           & 4.95~GHz        &\nodata         &    $<$162                         & VLA \\
\enddata
\end{deluxetable}


\begin{deluxetable}{lccccc}
\tabletypesize{\footnotesize}
\tablecaption{Emission-Line Measurements of SW104409 \label{spec_tab1}}
\tablewidth{0pt}
\tablehead{
\colhead{}     & \colhead{$\lambda_{\rm obs}$} & \colhead{}         & \colhead{Flux}                  & \colhead{FWHM}   & \colhead{$W_{\rm \lambda, rest}$} \\
\colhead{Line} & \colhead{(\AA)}               & \colhead{Redshift} & \colhead{($10^{-16}$ \ergcm2s)} & \colhead{(\kms)} & \colhead{(\AA)} 
} 
\startdata
\lya\tablenotemark{a}      &  4304.5  &  2.540    &    12.5$\pm$1.3    &  1324$\pm$80    &       105$\pm$11      \\
\nv\tablenotemark{a}       &  4389.5  &  2.540    &    1.88$\pm$0.56   &  1366$\pm$300   &        16$\pm$5       \\
\civ\tablenotemark{a}      &  5479.3  &  2.537    &    2.23$\pm$0.45   &  1367$\pm$200   &        21$\pm$4       \\
\lya/\nv\tablenotemark{b}  &  \nodata & \nodata   &    23.0$\pm$2.3    &  \nodata        &       311$\pm$31      \\
\sioiv\tablenotemark{b}    &  \nodata & \nodata   &    2.34$\pm$0.47   &  \nodata        &        29$\pm$6       \\
\civ\tablenotemark{b}      &  \nodata & \nodata   &    4.15$\pm$0.62   &  \nodata        &        39$\pm$6       \\
\enddata
\tablecomments{Rest-frame equivalent widths $W_{\rm \lambda, rest}$ assume $z = 2.54$.}
\tablenotetext{a}{Gaussian fit to narrow component}
\tablenotetext{b}{Total integration.}
\end{deluxetable}

\begin{deluxetable}{lccccc}
\tabletypesize{\footnotesize}
\tablecaption{Emission-Line Measurements of SW104406 \label{spec_tab2}}
\tablewidth{0pt}
\tablehead{
\colhead{}     & \colhead{$\lambda_{\rm obs}$} & \colhead{}         & \colhead{Flux}                  & \colhead{FWHM}   & \colhead{$W_{\rm \lambda, rest}$} \\
\colhead{Line} & \colhead{(\AA)}               & \colhead{Redshift} & \colhead{($10^{-16}$ \ergcm2s)} & \colhead{(\kms)} & \colhead{(\AA)}
} 
\startdata
\lya      &  4171.3  & 2.431   &    14.9$\pm$1.5    &  1360$\pm$20   &       184$\pm$18    \\
\nv       &  4254.8  & 2.431   &    9.82$\pm$0.29   &  2037$\pm$40   &       120$\pm$4     \\
\siiv     &  4777.4  & 2.430   &    1.15$\pm$0.35   &  1684$\pm$200  &        16$\pm$5     \\
\oivp     &  4809.4  & 2.430   &    1.23$\pm$0.12   &  1120$\pm$100  &        17$\pm$2     \\
\civ      &  5310.7  & 2.428   &    7.14$\pm$0.14   &  1485$\pm$60   &        86$\pm$2     \\
\heii     &  5618.9  & 2.425   &    1.50$\pm$0.08   &  1278$\pm$100  &        20$\pm$1     \\
\enddata
\tablecomments{All parameters are derived for Gaussian fits. Rest-frame
               equivalent widths $W_{\rm \lambda, rest}$ assume $z = 2.43$.}
\end{deluxetable}

\begin{deluxetable}{lllll}
\tabletypesize{\footnotesize}
\tablecaption{X-ray Observational Details \label{xobs_log}}
\tablewidth{0pt}
\tablehead{
\colhead{Source Name} & \colhead{Sequence No.} & \colhead{OBSID} & \colhead{Date} & \colhead{Exp. Time} \\
\colhead{} & \colhead{} & \colhead{} & \colhead{} & \colhead{(ksec)} 
}
\startdata
SW104406 & 900331 & 5024 & 16 Sept 2004 & 57.97 \\
SW104409 & 900334 & 5027 & 20 Sept 2004 & 67.96 \\
\enddata
\end{deluxetable}


\begin{deluxetable}{l cccc cccc}
\tabletypesize{\scriptsize}
\rotate
\tablecaption{Luminosities and black-hole masses of SW104409 and SW104406\label{lum_tab}}
\tablewidth{0pt}
\tablehead{
 \colhead{Source Name}      &
 \colhead{$L_{Radio}$} &
 \colhead{$L(O-NIR)$\tablenotemark{a}}  &
 \colhead{$L(IR)$\tablenotemark{a}}    &
 \colhead{$L(O-IR)$\tablenotemark{a}}  &
 \colhead{$L(X)$\tablenotemark{b}}     &
 \colhead{$L_{bol}$\tablenotemark{c}}  &
 \colhead{$M_\mathrm{BH}$\tablenotemark{d}} &
 \colhead{$dM/dt$\tablenotemark{e}}  \\
 \colhead{}                 &
 \colhead{1.4 GHz}          &
 \colhead{$0.03-10\mu m$}   &
 \colhead{$3-1000\mu m$}    &
 \colhead{$0.03-1000\mu m$} &
 \colhead{$0.3-8 keV$}      &
 \colhead{}                 &
 \colhead{$10^{8}$\msun}    &
 \colhead{\msun$/yr$}
}
\startdata
 SW104406 &$<$40.07 &  45.48 &  46.08 & 46.14 & 45.74 & 46.29 (45.93) &  1.4  (0.7) & 3.4 (1.5) \\
 SW104409 &   40.32 &  47.19 &  47.53 & 47.58 & 45.64 & 47.59 (47.21) & 29.9 (12.4) &  68 (28) \\
\enddata
\tablecomments{All luminosities are in logarithmic scale and units of \ergs. See section~\ref{lum_mbh} for more details.}
\tablenotetext{a}{ Derived assuming the model shown in
 Figure~\ref{extinction} for SW104409 and in Figure~\ref{figSED2} for
 SW104406.}
\tablenotetext{b}{ Absorption-corrected X-ray luminosity between 0.3 and 8
 keV.}
\tablenotetext{c}{ Bolometric luminosity derived as the sum of $L(OIR)$,
 $L(0.3-8 keV)$ and $L(Radio)$ if available. The value in parenthesis does not include the
 luminosity emitted at $\lambda^{rest}>$10\micron\  and was obtained by replacing
 $L(O-IR)$ with $L(0-NIR)$ in the $L_{bol}$ calculation.}
\tablenotetext{d}{ Black-hole mass derived from $L_{bol}$ and assuming that
 the source is accreting at the Eddington limit. The value in parenthesis was derived using
 the $L_{bol}$ value shown in parenthesis in column 7.}
\tablenotetext{e}{ Accretion rate derived from $L_{bol}$ and assuming that
 the source is accreting with 10\%\ efficiency. The value in parethesis was derived using
 the $L_{bol}$ value shown in parenthesis in column 7.}
\end{deluxetable}

\end{document}